\documentstyle[amssymb,thmsa,a4,sw20lart]{article}

\input{tcilatex}
\begin{document}

\title{{\sc Mechanics of Continuous Media in}{\bf \ }$(\overline{L}_n,g)${\sc %
-Spaces}{\bf .}\\
{\bf II. Relative velocity and deformations}}
\author{S. Manoff \\
{\it Bulgarian Academy of Sciences,}\\
{\it \ Institute for Nuclear Research and Nuclear Energy,}\\
{\it \ Blvd. Tzarigradsko Chaussee 72,}\\
{\it \ 1784 Sofia - Bulgaria}}
\date{E-mail address: smanov@inrne.bas.bg}
\maketitle

\begin{abstract}
Basic notions of continuous media mechanics are introduced for spaces with
affine connections and metrics. The physical interpretation of the notion of
relative velocity is discussed. The notions of deformation velocity tensor,
shear velocity, rotation (vortex) velocity, and expansion velocity are
introduced. Different types of flows are considered.

\end{abstract}

\tableofcontents

\section{Introduction}

The notion of relative velocity is closely related to the notions of
deformation velocity tensor and its kinematic characteristics (shear,
rotation, and expansion). On the other side, the friction in a continuous
media could be described in analogous way as the deformation. This give rise
to considerations of friction ''velocity'' tensor and its kinematic
characteristics.

In Section 1 the introduction and the physical interpretation of the notion
of relative velocity is discussed. The notions of deformation velocity
tensor, shear velocity, rotation (vortex) velocity and vortex vector as well
as expansion velocity are introduced for $(\overline{L}_n,g)$-spaces. In
Section 2 the notions of friction velocity and it kinematic characteristics
is introduced and considered. In Section 3 different types of flows are
considered.

All considerations are given in details (even in full details) for those
readers who are not familiar with the considered problems.

{\it Remark.} The present paper is the second part of a larger research
report on the subject with the title ''Contribution to continuous media
mechanics in $(\overline{L}_n,g)$-spaces'' and with the following contents:

I. Introduction and mathematical tools.

II. Relative velocity and deformations.

III. Relative accelerations.

IV. Stress (tension) tensor.

The parts are logically self-dependent considerations of the main topics
considered in the report.

\section{Relative velocity. Deformation velocity, shear velocity, rotation
(vortex) velocity, and expansion velocity}

The notion 
\index{relative velocity@relative velocity} {\it relative velocity} vector
field (relative velocity) $_{rel}v$ can be defined (regardless of its
physical interpretation) as the orthogonal to a non-isotropic vector field $%
u $ projection of the first covariant derivative (along the same
non-isotropic vector field $u$) of (another) vector field $\xi $, i.e.

\begin{equation}
\begin{array}{c}
_{rel}v=%
\overline{g}(h_u(\nabla _u\xi ))=g^{ij}\cdot h_{\overline{j}\overline{k}%
}\cdot \xi ^k\text{ }_{;l}\cdot u^l\cdot e_i= \\ 
=g^{ij}\cdot f^m\text{ }_j\cdot f^n\text{ }_k\cdot h_{mn}\cdot \xi ^k\text{ }%
_{;l}\cdot u^l\cdot e_i\text{ ,} \\ 
e_i=\partial _i\text{ (in a co-ordinate basis),}
\end{array}
\label{5.1}
\end{equation}

\noindent where (the indices in a co-ordinate and in a non-co-ordinate basis
are written in both cases as Latin indices instead of Latin and Greek
indices)

\begin{equation}
h_u=g-\frac 1e\cdot g(u)\otimes g(u)\text{ },\text{ }h_u=h_{ij}\cdot e^i.e^j%
\text{ , }\overline{g}=g^{ij}\cdot e_i.e_j,  \label{5.2}
\end{equation}

\begin{equation}
\begin{array}{c}
\nabla _u\xi =\xi ^i\text{ }_{;j}\cdot u^j\cdot e_i\text{ } \\ 
\text{ }\xi ^i\text{ }_{;j}=e_j\xi ^i+\Gamma _{kj}^i\cdot \xi ^k\,\text{%
,\thinspace }\,\,\,\Gamma _{kj}^i\neq \Gamma _{jk}^i\text{ },
\end{array}
\label{5.3}
\end{equation}

\begin{equation}
\begin{array}{c}
g=g_{ij}\cdot e^i.e^j,\,\,\,\,\,\,\,\,\text{ }g_{ij}=g_{ji}\text{ }, \\ 
\text{ }e^i.e^j=\frac 12\cdot (e^i\otimes e^j+e^j\otimes e^i)\text{ },
\end{array}
\label{5.4}
\end{equation}

\begin{equation}
\begin{array}{c}
e=g(u,u)=g_{\overline{i}\overline{j}}\cdot u^i\cdot u^j=u_{\overline{i}%
}\cdot u^i\neq 0 \\ 
\text{ }g(u)=g_{i\overline{k}}\cdot u^k=u_i=g_{ik}\cdot u^{\overline{k}}%
\text{ , }u^{\overline{k}}=f^k\text{ }_l\cdot u^l\text{ ,} \\ 
\text{ }e_i.e_j=\frac 12\cdot (e_i\otimes e_j+e_j\otimes e_i)\text{ },\text{ 
}
\end{array}
\label{5.5}
\end{equation}

\begin{equation}
\begin{array}{c}
h_u(\nabla _u\xi )=h_{i\overline{j}}\cdot \xi ^j\text{ }_{;k}\cdot u^k\cdot
e^i \\ 
\text{ }h_{ij}=g_{ij}-\frac 1e\cdot u_i\cdot u_j\text{ .}
\end{array}
\label{5.6}
\end{equation}

In a co-ordinate basis 
\[
\begin{array}{c}
e_j\xi ^i=\xi ^i\text{ }_{,j}=\partial _j\xi ^i=\partial \xi ^i/\partial x^j,
\\ 
e^j=dx^j,\,\,\,\,\,\,\,\,e_i=\partial _i=\partial /\partial
x^i,\,\,\,\,\,\,\,\,\,\,\,\,\,\,u=u^i\cdot \partial _i,
\end{array}
\]

Every contravariant vector field $\xi $ can be written by means of its
projection along and orthogonal to $u$ in two parts - one collinear to $u$
and one - orthogonal to $u$, i.e.

\begin{equation}
\xi =\frac le\cdot u+h^u[g(\xi )]\text{ }=\frac le\cdot u+\overline{g}[%
h_u(\xi )]\text{ ,}  \label{5.7}
\end{equation}

\noindent where

\begin{equation}
\begin{array}{c}
l=g(\xi ,u)\text{ , \thinspace \thinspace \thinspace \thinspace \thinspace
\thinspace \thinspace \thinspace \thinspace \thinspace \thinspace \thinspace
\thinspace }h^u=\overline{g}-\frac 1e\cdot u\otimes u\text{ ,\thinspace
\thinspace \thinspace } \\ 
\text{ }\xi =\xi ^i\cdot \partial _i=\xi ^k\cdot e_k\text{ , \thinspace
\thinspace \thinspace \thinspace \thinspace \thinspace \thinspace \thinspace
\thinspace }h^u=h^{ij}\cdot e_i.e_j\text{ ,}
\end{array}
\label{5.8}
\end{equation}

\begin{equation}
\overline{g}(h_u)\overline{g}=h^u\text{ ,\thinspace \thinspace \thinspace
\thinspace \thinspace }h_u(\overline{g})(g)=h_u\text{ ,\thinspace \thinspace
\thinspace \thinspace \thinspace \thinspace }h^u(g)(\overline{g})=h^u\text{
,\thinspace \thinspace \thinspace \thinspace \thinspace \thinspace }%
g(h^u)g=h_u\text{ .}  \label{5.9}
\end{equation}

Therefore, $\nabla _u\xi $ can be written in the form

\begin{equation}
\nabla _u\xi =\frac{\overline{l}}e\cdot u+\overline{g}[h_u(\nabla _u\xi )]=%
\frac{\overline{l}}e\cdot u+\,_{rel}v\text{ ,\thinspace \thinspace
\thinspace \thinspace \thinspace }\overline{l}=g(\nabla _u\xi ,u)
\label{5.10}
\end{equation}

\noindent and the connection between $\nabla _u\xi $ and $_{rel}v$ is
obvious. Using the relation \cite{Yano} 
\index{Yano K.@Yano K.} between the Lie derivative $\pounds _\xi u$ and the
covariant derivative $\nabla _\xi u$

\begin{equation}
\begin{array}{c}
\pounds _\xi u=\nabla _\xi u-\nabla _u\xi -T(\xi ,u) \\ 
\text{ }T(\xi ,u)=T_{ij}\,^k\cdot \xi ^i\cdot u^j\cdot e_k\text{ ,}
\end{array}
\label{5.11}
\end{equation}

\[
T_{ij}\,^k=-T_{ji}\,^k=\Gamma _{ji}^k-\Gamma _{ij}^k-C_{ij}\text{ }^k\text{
\thinspace \thinspace \thinspace (in a non-co-ordinate basis }\{e_k\}\text{)
, } 
\]

\[
\lbrack e_i,e_j]=\pounds _{e_i}e_j\text{ }=C_{ij}\text{ }^k\cdot e_k\text{ ,}
\]

\[
T_{ij}\,^k=\Gamma _{ji}^k-\Gamma _{ij}^k\,\,\,\,\text{ (in a co-ordinate
basis }\{\partial _k\}\text{ ) ,} 
\]

\noindent one can write $\nabla _u\xi $ in the form

\begin{equation}
\nabla _u\xi =(k)g(\xi )-\pounds _\xi u\text{ }=k[g(\xi )]-\pounds _\xi u%
\text{,}  \label{5.12}
\end{equation}

\noindent or, taking into account the above expression for $\xi $, in the
form

\[
\nabla _u\xi =k[h_u(\xi )]+\frac le\cdot a-\pounds _\xi u\text{ ,} 
\]

\noindent where

\begin{equation}
\begin{array}{c}
k[g(\xi )]=\nabla _\xi u-T(\xi ,u) \\ 
\text{ }k=(u^i\text{ }_{;l}-T_{lk}\,^i\cdot u^k)\cdot g^{lj}\cdot e_i\otimes
e_j\text{ ,}
\end{array}
\label{5.13}
\end{equation}

\begin{equation}
\begin{array}{c}
k[g(u)]=k(g)u=k^{ij}\cdot g_{\overline{j}\overline{k}}\cdot u^k\cdot e_i= \\ 
=a=\nabla _uu=u^i\text{ }_{;j}\cdot u^j\cdot e_i\text{ .}
\end{array}
\label{5.14}
\end{equation}

For $h_u(\nabla _u\xi )$ it follows that

\begin{equation}
h_u(\nabla _u\xi )=h_u(\frac le\cdot a-\pounds _\xi u)+h_u(k)h_u(\xi )\text{
,}  \label{5.15}
\end{equation}

\noindent where 
\[
\begin{array}{c}
h_u(k)h_u(\xi )=h_{i\overline{k}}\cdot k^{kl}\cdot h_{\overline{l}\overline{j%
}}\cdot \xi ^j\cdot e^i, \\ 
\text{ }h_u(u)=0,\,\,\,\,\,\,\,\,\,\,\,\,u(h_u)=0, \\ 
\text{ }h_u(k)h_u(u)=0,\,\,\,\,\,\,\,\text{ }(u)h_u(k)h_u=0.
\end{array}
\]

If we introduce the abbreviation

\begin{equation}
d=h_u(k)h_u=h_{i\overline{k}}\cdot k^{kl}\cdot h_{\overline{l}j}\cdot
e^i\otimes e^j=d_{ij}\cdot e^i\otimes e^j\text{ ,}  \label{5.16}
\end{equation}

\noindent the expression for $_{rel}v$ can take the form

\[
_{rel}v=\overline{g}[h_u(\nabla _u\xi )]=\overline{g}(h_u)(\frac le\cdot
a-\pounds _\xi u)+\overline{g}[d(\xi )]= 
\]

\begin{equation}
=[g^{ik}\cdot h_{\overline{k}\overline{l}}\cdot (\frac le\cdot a^l-\pounds
_\xi u^l)+g^{ik}\cdot d_{\overline{k}\overline{l}}\cdot \xi ^l]\cdot e_i%
\text{ }=\,_{rel}v^i\cdot e_i\text{ ,}  \label{5.17}
\end{equation}

\noindent or

\begin{equation}
g(_{rel}v)=h_u(\nabla _u\xi )=h_u(\frac le\cdot a-\pounds _\xi u)+d(\xi )%
\text{ .}  \label{5.18}
\end{equation}

For the special case when the vector field $\xi $ is orthogonal to $u$, i.e. 
$\xi =\overline{g}[h_u(\xi )]$, and the Lie derivative of $u$ along $\xi $
is zero, i.e. $\pounds _\xi u=0$, then the relative velocity can be written
in the form 
\begin{equation}
g(_{rel}v)=d(\xi )  \label{5.19}
\end{equation}

\noindent or in the form 
\[
_{rel}v=\overline{g}[d(\xi )]\text{.} 
\]

{\it Remark}. All further calculations leading to a useful representation of 
$d$ are quite straightforward. The problem here was the finding out a
representation of $h_u(\nabla _u\xi )$ in the form (\ref{5.15}) which is not
a trivial task.

\subsection{Deformation velocity, shear velocity, rotation (vortex)
velocity, and expansion velocity}

The covariant tensor field $d$ is a generalization for $(\overline{L}_n,g)$%
-spaces of the well known 
\index{deformation@deformation!deformation velocity@deformation velocity} 
{\it deformation velocity }tensor for $V_n$-spaces \cite{Stephani}%
\index{Stephani H.@Stephani H.}, \cite{Kramer-2} 
\index{Kramer D.@Kramer D.} 
\index{Stephani H. (s. Kramer D.)@Stephani H. (s. Kramer D.)} 
\index{MacCallum M. (s. Kramer D.)@MacCallum M. (s. Kramer D.)} 
\index{Herlt E. (s. Kramer D.)@Herlt E. (s. Kramer D.)}. It is usually
represented by means of its three parts: the trace-free symmetric part,
called {\it shear velocity }tensor (shear), the anti-symmetric part, called 
\index{rotation@rotation!rotation velocity@rotation velocity} {\it rotation
velocity }tensor (rotation) and the trace part, in which the trace is called 
\index{expansion@expansion!expansion velocity@expansion velocity} {\it %
expansion velocity }(expansion){\it \ }invariant.

After some more complicated as for $V_n$-spaces calculations, the
deformation velocity tensor $d$ can be given in the form

\begin{equation}
\begin{array}{c}
d=h_u(k)h_u=h_u(k_s)h_u+h_u(k_a)h_u= \\ 
=\sigma +\omega +\frac 1{n-1}\cdot \theta \cdot h_u%
\text{ .}
\end{array}
\label{5.20}
\end{equation}

The symmetric trace-free tensor $\sigma $ is the 
\index{shear@shear!shear velocity@shear velocity} {\it shear velocity}
tensor (shear), 
\begin{equation}
\begin{array}{c}
\sigma =\,_sE-\,_sP=E-P-\frac 1{n-1}\cdot 
\overline{g}[E-P]\cdot h_u=\sigma _{ij}\cdot e^i.e^j= \\ 
=E-P-\frac 1{n-1}\cdot (\theta _o-\theta _1)\cdot h_u\text{ ,}
\end{array}
\label{5.21}
\end{equation}

\begin{equation}
\begin{array}{c}
_sE=E-\frac 1{n-1}\cdot \overline{g}[E]\cdot h_u\text{ ,} \\ 
\text{ }\overline{g}[E]=g^{ij}\cdot E_{\overline{i}\overline{j}}=g^{%
\overline{i}\overline{j}}\cdot E_{ij}=\theta _o\text{ ,}
\end{array}
\label{5.22}
\end{equation}

\begin{equation}
\begin{array}{c}
E=h_u(\varepsilon )h_u\text{ , \thinspace \thinspace \thinspace \thinspace }%
k_s=\varepsilon -m\text{ ,} \\ 
\text{ }\varepsilon =\frac 12\cdot (u_{\text{ };l}^i\cdot g^{lj}+u_{\text{ }%
;l}^j\cdot g^{li})\cdot e_i.e_j\text{ ,}
\end{array}
\label{5.23}
\end{equation}

\begin{equation}
m=\frac 12\cdot (T_{lk}\,^i\cdot u^k\cdot g^{lj}+T_{lk}\,^j\cdot u^k\cdot
g^{li})\cdot e_i.e_j\text{ .}  \label{5.24}
\end{equation}

The symmetric trace-free tensor $_sE$ is the {\it torsion-free shear velocity%
} 
\index{shear@shear!torsion-free shear velocity@torsion-free shear velocity}%
{\it \ }tensor, the symmetric trace-free tensor $_sP$ is the 
\index{shear@shear!shear velocity induced by the torsion@shear velocity
induced by the torsion} {\it shear velocity} tensor {\it induced by the
torsion},

\begin{equation}
\begin{array}{c}
_sP=P-\frac 1{n-1}\cdot 
\overline{g}[P]\cdot h_u\text{ ,} \\ 
\text{ }\overline{g}[P]=g^{kl}\cdot P_{\overline{k}\overline{l}}\text{ }=g^{%
\overline{k}\overline{l}}\cdot P_{kl}=\theta _1\text{,}
\end{array}
\label{5.25}
\end{equation}

\begin{equation}
\begin{array}{c}
P=h_u(m)h_u\text{ , \thinspace \thinspace \thinspace \thinspace \thinspace
\thinspace \thinspace \thinspace \thinspace \thinspace \thinspace \thinspace
\thinspace \thinspace \thinspace \thinspace \thinspace \thinspace \thinspace
\thinspace \thinspace \thinspace \thinspace \thinspace }\theta
_1=T_{kl}\,^k\cdot u^l\text{ ,} \\ 
\text{ }\theta _o=u^n\text{ }_{;n}-\frac 1{2e}\cdot (e_{,k}\cdot
u^k-g_{kl;m}\cdot u^m\cdot u^{\overline{k}}\cdot u^{\overline{l}})\text{ ,}
\end{array}
\label{5.26}
\end{equation}

\begin{equation}
e_{,k}=e_ke\text{ ,\thinspace \thinspace \thinspace \thinspace \thinspace
\thinspace \thinspace \thinspace \thinspace \thinspace \thinspace \thinspace
\thinspace \thinspace \thinspace \thinspace \thinspace \thinspace \thinspace
\thinspace \thinspace \thinspace \thinspace \thinspace \thinspace \thinspace
\thinspace \thinspace }\theta =\theta _o-\theta _1\text{ . }  \label{5.27}
\end{equation}

The invariant $\theta $ is the 
\index{expansion@expansion!expansion velocity@expansion velocity} {\it %
expansion velocity,} the invariant{\it \ $\theta _o$} is the 
\index{expansion@expansion!torsion-free expansion velocity@torsion-free
expansion velocity} {\it torsion-free expansion velocity, }the invariant $%
\theta _1$ is the {\it expansion velocity induced by the torsion,} the
antisymmetric tensor{\it \ }$\omega $ is the 
\index{rotation@rotation!rotation velocity@rotation velocity}{\it \ rotation
(vortex) velocity }tensor (rotation velocity, vortex velocity),

\begin{equation}
\omega =h_u(k_a)h_u=h_u(s)h_u-h_u(q)h_u=S-Q%
\text{ ,}  \label{5.28}
\end{equation}

\begin{equation}
\begin{array}{c}
s=\frac 12\cdot (u^k\text{ }_{;m}\cdot g^{ml}-u^l\text{ }_{;m}\cdot
g^{mk})\cdot e_k\wedge e_l\text{ ,} \\ 
\text{ }e_k\wedge e_l=\frac 12\cdot (e_k\otimes e_l-e_l\otimes e_k)\text{ , }
\end{array}
\label{5.29}
\end{equation}

\begin{equation}
\begin{array}{c}
q=\frac 12\cdot (T_{mn}\,^k\cdot g^{ml}-T_{mn}\,^l\cdot g^{mk})\cdot
u^n\cdot e_k\wedge e_l\text{ ,} \\ 
\text{ }S=h_u(s)h_u\text{ , \thinspace \thinspace \thinspace \thinspace
\thinspace \thinspace \thinspace \thinspace \thinspace \thinspace \thinspace
\thinspace \thinspace \thinspace \thinspace \thinspace \thinspace \thinspace 
}Q=h_u(q)h_u\text{ .}
\end{array}
\label{5.30}
\end{equation}

The antisymmetric tensor $S$ is the 
\index{rotation@rotation!torsion-free rotation velocity@torsion-free
rotation velocity} {\it torsion-free rotation (vortex) velocity} tensor, and
the antisymmetric tensor $Q$ is the 
\index{rotation@rotation!rotation velocity induced by the torsion@rotation
velocity induced by the torsion}{\it \ rotation (vortex) velocity }tensor 
{\it induced by the torsion.}

By means of the expressions for $\sigma $, $\omega $ and $\theta $ the
deformation velocity tensor can be written in two parts

\begin{equation}
\begin{array}{c}
d=d_o-d_1%
\text{ ,} \\ 
\text{ }d_o=\,_sE+S+\frac 1{n-1}\cdot \theta _o\cdot h_u\text{ ,} \\ 
\text{ }d_1=\,_sP+Q+\frac 1{n-1}\cdot \theta _1\cdot h_u\text{ ,}
\end{array}
\label{5.31}
\end{equation}

\noindent where $d_o$ is the 
\index{deformation@deformation!torsion-free deformation
velocity@torsion-free deformation velocity} {\it torsion-free deformation
velocity} tensor and $d_1$ is the 
\index{deformation@deformation!deformation velocity induced by the
torsion@deformation velocity induced by the torsion} {\it deformation
velocity }tensor {\it induced by the torsion. }For the case of $V_n$-spaces $%
d_1=0$ ($_sP=0$, $Q=0$, $\theta _1=0$).

If we use the explicit form of the tensor $k_s$ from (\ref{5.23}) and (\ref
{5.24}) , we can find the relations 
\begin{eqnarray}
k_s &=&\varepsilon -m=(\varepsilon ^{kl}-m^{kl})\cdot \partial _k\otimes
\partial _l%
\text{ ,}  \nonumber \\
\varepsilon ^{kl}-m^{kl} &=&\frac 12\cdot [u^k\,_{;n}\cdot
g^{nl}+u^l\,_{;n}\cdot g^{nk}-(T_{mn}\,^k\cdot g^{ml}+T_{mn}\,^l\cdot
g^{mk})\cdot u^n]=  \label{5.31a} \\
&=&\frac 12\cdot [u^k\,_{;n}\cdot g^{nl}+u^l\,_{;n}\cdot
g^{nk}-T_{mn}\,^k\cdot g^{ml}\cdot u^n-T_{mn}\,^l\cdot g^{mk}\cdot u^n]\text{
.}  \nonumber
\end{eqnarray}

On the other side, 
\begin{eqnarray}
\nabla _u\overline{g} &=&g^{ij}\,_{;k}\cdot u^k\cdot \partial _i.\partial _j%
\text{ ,\thinspace \thinspace \thinspace \thinspace \thinspace \thinspace
\thinspace \thinspace \thinspace \thinspace }\pounds _u\overline{g}=(\pounds
_ug^{ij})\cdot \partial _i.\partial _j\text{ ,}  \nonumber \\
\pounds _ug^{ij} &=&g^{ij}\,_{;k}\cdot u^k-  \nonumber \\
&&-[u^i\,_{;l}\cdot g^{lj}+u^j\,_{;l}\cdot g^{il}-T_{lk}\,^i\cdot u^k\cdot
g^{lj}-T_{lk}\,^j\cdot u^k\cdot g^{il}]\text{ ,}  \nonumber \\
\pounds _ug^{kl} &=&g^{kl}\,_{;n}\cdot u^n-  \nonumber \\
&&-[u^k\,_{;n}\cdot g^{nl}+u^l\,_{;n}\cdot g^{nk}-T_{mn}\,^k\cdot u^n\cdot
g^{ml}-T_{mn}\,^l\cdot u^n\cdot g^{km}]\text{ ,}  \nonumber \\
g^{kl}\,_{;n}\cdot u^n-\pounds _ug^{kl} &=&  \nonumber \\
&=&u^k\,_{;n}\cdot g^{nl}+u^l\,_{;n}\cdot g^{nk}-T_{mn}\,^k\cdot u^n\cdot
g^{ml}-T_{mn}\,^l\cdot u^n\cdot g^{km}\text{ .}  \label{5.31b}
\end{eqnarray}

Therefore, 
\begin{eqnarray}
\varepsilon ^{kl}-m^{kl} &=&\frac 12\cdot (g^{kl}\,_{;n}\cdot u^n-\pounds
_ug^{kl})\text{ ,}  \nonumber \\
\varepsilon -m &=&\frac 12\cdot (\nabla _u\overline{g}-\pounds _u\overline{g}%
)=k_s\text{ ,}  \label{5.31c} \\
h_u(k_s)h_u &=&\frac 12\cdot h_u(\nabla _u\overline{g}-\pounds _u\overline{g}%
)h_u\text{ .}  \nonumber
\end{eqnarray}

By the use of the last relations the shear velocity tensor $\sigma $ and the
expansion velocity invariant $\theta $ can also be written in the form

\begin{equation}
\sigma =\frac 12\cdot \{h_u(\nabla _u\overline{g}-\pounds _u\overline{g}%
)h_u-\frac 1{n-1}\cdot (h_u[\nabla _u\overline{g}-\pounds _u\overline{g}%
])\cdot h_u\}\text{ }=  \label{5.32}
\end{equation}

\begin{eqnarray}
&=&\frac 12\cdot \{h_{i\overline{k}}\cdot (g^{kl}\text{ }_{;m}\cdot
u^m-\pounds _ug^{kl})\cdot h_{\overline{l}j}-  \label{5.33} \\
&&-\frac 1{n-1}\cdot h_{\overline{k}\overline{l}}\cdot (g^{kl}\text{ }%
_{;m}\cdot u^m-\pounds _ug^{kl})\cdot h_{ij}\}\cdot e^i.e^j\text{ .} 
\nonumber
\end{eqnarray}
\begin{equation}
\begin{array}{c}
\theta =\frac 12\cdot h_u[\nabla _u\overline{g}-\pounds _u\overline{g}%
]=\frac 12\cdot [\nabla _{\overline{g}}u+T(u,\overline{g})]= \\ 
=\frac 12\cdot h_{\overline{i}\overline{j}}\cdot (g^{ij}\text{ }_{;k}\cdot
u^k-\pounds _ug^{ij})\text{ ,}
\end{array}
\label{5.33a}
\end{equation}

\noindent where 
\begin{equation}
\frac 12\cdot \overline{g}[h_u(\nabla _u\overline{g}-\pounds _u\overline{g}%
)h_u]=\frac 12\cdot h_u[\nabla _u\overline{g}-\pounds _u\overline{g}]\text{ .%
}  \label{5.33b}
\end{equation}

The main result of the above considerations can be summarized in the
following proposition:

\begin{proposition}
{\bf \ }The covariant vector field $g(_{rel}v)=h_u(\nabla _u\xi )$ can be
written in the forms:
\end{proposition}

\[
h_u(\nabla _u\xi )=h_u(\frac le\cdot a-\pounds _\xi u)+d(\xi )= 
\]

\[
=h_u(\frac le\cdot a-\pounds _\xi u)+\sigma (\xi )+\omega (\xi )+\frac
1{n-1}\cdot \theta \cdot h_u(\xi )\text{ .} 
\]

The physical interpretation of the velocity tensors $d,$ $\sigma ,$ $\omega $
and of the invariant $\theta $ for the case of $V_4$-spaces \cite{Synge}, 
\index{Synge J. L.@Synge J. L.}, \cite{Ehlers} 
\index{Ehlers J.@Ehlers J.}, can also be extended for $(%
\overline{L}_4,g)$-spaces (see Fig. 1). In this case the torsion plays an
equivalent role in the velocity tensors as the covariant derivative. It is
easy to see that the existence of some kinematic characteristics ($_sP$, $Q$%
, $\theta _1$) depends on the existence of the torsion tensor field. They
vanish if it is equal to zero (e.g. in $V_n$-spaces). On the other side, the
kinematic characteristics induced by the torsion can compensate the result
of the action of the torsion-free kinematic characteristics. For $d=0$, $%
\sigma =0$, $\omega =0$, $\theta =0$ we could have the relations $d_0=d_1$, $%
_sE=\,_sP$, $S=Q$, $\theta _0=\theta _1$ respectively leading to vanishing
the relative velocity $_{rel}v$ under the additional conditions $g(u,\xi
)=l=0$ and $\pounds _\xi u=0$.

The condition $\pounds _\xi u=0=[\xi ,u]$ for the contravariant vector
fields $\xi $ and $u$ induces a family of two dimensional sub manifolds of $%
M $. On these sub manifolds, one can choose the parameters of the integral
curves of the two vector fields $\xi $ and $u$ as co-ordinates. This
statement could be easily proved by the use of the relation 
\begin{equation}
\lbrack \xi ,u]=(\xi ^i\cdot u^j\,_{,i}-u^i\cdot \xi ^j\,_{,i})\cdot
.\partial _j=0\text{ .}  \label{5.34}
\end{equation}

\noindent For a two parametric congruence of curves (not intersecting
curves) in $M$ 
\begin{equation}
x^i=x^i(\lambda ,\tau )  \label{5.35}
\end{equation}

\noindent with 
\begin{equation}
\xi :=\frac \partial {\partial \lambda }=\frac{\partial x^i}{\partial
\lambda }\cdot \partial _i=\xi ^i\cdot \partial _i\text{ , \thinspace
\thinspace \thinspace \thinspace \thinspace \thinspace \thinspace \thinspace
\thinspace \thinspace \thinspace \thinspace }u^i:=\frac \partial {\partial
\tau }=\frac{\partial x^i}{\partial \tau }\cdot \partial _i=u^i\cdot
\partial _i\text{ ,}  \label{5.36}
\end{equation}

\noindent the integrability condition for the co-ordinates $x^i\,$along $%
\lambda $ and $\tau $ follows in the form 
\begin{equation}
u^j\,_{,i}\cdot \xi ^i=\frac{\partial ^2x^j}{\partial \lambda \partial \tau }%
=\frac{\partial ^2x^j}{\partial \tau \partial \lambda }=\xi ^j\,_{,i}\cdot
u^i\text{ .}  \label{5.37}
\end{equation}

The last expression leads to a solution of the equations for $x^i(\tau
,\lambda )$%
\begin{equation}
dx^i=\frac{\partial x^i}{\partial \lambda }\cdot d\lambda +\frac{\partial x^i%
}{\partial \tau }\cdot d\tau =\xi ^i\cdot d\lambda +u^i\cdot d\tau \text{ .}
\label{5.38}
\end{equation}

A one to one correspondence could be established between two of the
co-ordinates $x^i$ (for instance, for $i=1,2$) and the parameters $\lambda $
and $\tau $ on the basis of the relations 
\[
x^{a^{\prime }}\simeq (\lambda ,\tau )\text{ ,\thinspace \thinspace
\thinspace \thinspace \thinspace \thinspace \thinspace \thinspace \thinspace 
}x^a\simeq (x^1,x^2)\text{ , \thinspace \thinspace \thinspace \thinspace
\thinspace \thinspace }x^{a^{\prime }}=x^{a^{\prime }}(x^a)\text{
,\thinspace \thinspace \thinspace \thinspace \thinspace \thinspace
\thinspace \thinspace \thinspace \thinspace \thinspace \thinspace }%
x^a=x^a(x^{a^{\prime }})\,\text{\thinspace \thinspace ,} 
\]
\begin{equation}
dx^{a^{\prime }}=\frac{\partial x^{a^{\prime }}}{\partial x^a}\cdot dx^a%
\text{ ,\thinspace \thinspace \thinspace \thinspace \thinspace \thinspace
\thinspace \thinspace \thinspace \thinspace \thinspace \thinspace \thinspace
\thinspace \thinspace }dx^a=\frac{\partial x^a}{\partial x^{a^{\prime }}}%
\cdot dx^{a^{\prime }}\text{ .}  \label{5.39}
\end{equation}

\subsection{Physical interpretation of the notion of relative velocity}

\subsubsection{Acceleration}

Let us now consider the change of the velocity $u$ during the motion of a
material point from the point $P$ with co-ordinates $x^i(\tau _0,\lambda
_0^a)$ to the point $P_1$ with the co-ordinates $x^i(\tau _0+d\tau ,\lambda
_0^a)$. If we wish to express $u(\tau _0+d\tau ,\lambda _0^a)$ by means of $%
u(\tau _0,\lambda _0^a)$ we could use the exponent $\exp [d\tau \cdot \nabla
_u)]$ of the covariant differential operator $\nabla _u=D/d\tau $ with $%
u=d/d\tau $%
\begin{equation}
u_{(\tau _0+d\tau ,\lambda _0^a)}=u_{(\tau _0,\lambda _0^a)}+d\tau \cdot
\left( \frac{Du}{d\tau }\right) _{(\tau _0,\lambda _0^a)}+\frac 1{2!}\cdot
d\tau ^2\cdot \left( \frac{D^2u}{d\tau ^2}\right) _{(\tau _0,\lambda
_0^a)}+\cdots \text{ .}  \label{5.40}
\end{equation}

Up to the first order of $d\tau $ we have 
\begin{equation}
u_{(\tau _0+d\tau ,\lambda _0^a)}=u_{(\tau _0,\lambda _0^a)}+d\tau \cdot
\left( \frac{Du}{d\tau }\right) _{(\tau _0,\lambda _0^a)}\text{ .}
\label{5.41}
\end{equation}

Therefore, the change of the velocity $u$ from the point $P$ with $x^i(\tau
_0,\lambda _0^a)$ to the point $P_1$ with $x^i(\tau _0+d\tau ,\lambda _0^a)$
is 
\begin{equation}
u_{(\tau _0+d\tau ,\lambda _0^a)}-u_{(\tau _0,\lambda _0^a)}=d\tau \cdot
\left( \frac{Du}{d\tau }\right) _{(\tau _0,\lambda _0^a)}\text{ \thinspace
\thinspace \thinspace \thinspace .}  \label{5.42}
\end{equation}

The covariant derivative 
\begin{equation}
\left( \frac{Du}{d\tau }\right) _{(\tau _0,\lambda _0^a)}=\left( \nabla
_uu\right) _{(\tau _0,\lambda _0^a)}:=a_{(\tau _0,\lambda _0^a)}=\stackunder{%
d\tau \rightarrow 0}{\lim }\frac{u_{(\tau _0+d\tau ,\lambda _0^a)}-u_{(\tau
_0,\lambda _0^a)}}{d\tau }\text{ \thinspace \thinspace \thinspace \thinspace
,}  \label{5.43}
\end{equation}

\noindent with $u=d/d\tau =u^i\cdot \partial _i$ and $a=\nabla _uu$, can be
interpreted as the acceleration $a$ of a material point at the point $P$
with $x^i(\tau _0,\lambda _0^a)$ of the curve $x^i(\tau ,\lambda _0^a=$
const.$)$.

In analogous way, the acceleration of a material point during its motion
(transport) along a curve $x^i(\tau _0=$ const.$,\lambda ^a)$ could be found
in the form 
\begin{equation}
\left( \frac{D\xi _{(a)\perp }}{d\lambda ^a}\right) _{(\tau _0,\lambda
_0^a)}=\left( \nabla _{\xi _{(a)\perp }}\xi _{(a)\perp }\right) _{(\tau
_0,\lambda _0^a)}=\stackunder{d\lambda \rightarrow 0}{\lim }\frac{\xi
_{(a)\perp (\tau _0,\lambda _0^a+d\lambda ^a)}-\xi _{(a)\perp (\tau
_0,\lambda _0^a)}}{d\lambda ^a}\text{ \thinspace \thinspace \thinspace ,}
\label{5.44}
\end{equation}

\noindent where $\xi _{(a)\perp }=d/d\lambda ^a$. The last expression shows
the difference between the velocities along a curve $x^i(\tau _0=$ const.$%
,\lambda ^a)$ at the two different points $P_2$ with $x^i(\tau _0,\lambda
_0^a+d\lambda ^a)$ and $P$ with $x^i(\tau _0,\lambda _0^a)$. Usually, the
velocity $u$ is interpreted as the velocity of the material points
(elements) in the flow. The set of vectors $\overline{\xi }_{(a)}$ $%
(a=1,...,n-1)$ determines a cross-section of a flow in a neighborhood of a
given point. Since $\nabla _{\xi _{(a)\perp }}\xi _{(a)\perp }$ is the first
curvature vector of the curve $x^i(\tau _0,\lambda ^a)$, it could be
interpreted as a measure for the deviation of an infinitesimal cross-section
of the flow with $\nabla _{\xi _{(a)\perp }}\xi _{(a)\perp }\neq 0$ ,
orthogonal to $u$, from an auto-parallel (constructed by auto-parallel lines
of $\xi _{(a)\perp }$) infinitesimal cross-section with $\nabla _{\xi
_{(a)\perp }}\xi _{(a)\perp }=0$, orthogonal to $u$.

On the other side, we can consider

(a) the change of the vector $\xi _{(a)\perp }$ along the curve $x^i(\tau
,\lambda _0^a=$ const.$)$ and

(b) the change of the vector $u$ along the curve $x^i(\tau _0=$
const.,\thinspace $\lambda ^a)$.

\subsubsection{Relative velocity}

(a) In the first case, 
\begin{equation}
\left( \frac{D\xi _{(a)\perp }}{d\tau }\right) _{_{(\tau _0,\lambda
_0^a)}}=\left( \nabla _u\xi _{(a)\perp }\right) _{(\tau _0,\lambda _0^a)}=%
\stackunder{d\tau \rightarrow 0}{\lim }\frac{\xi _{(a)\perp (\tau _0+d\tau
,\lambda _0^a)}-\xi _{(a)\perp (\tau _0,\lambda _0^a)}}{d\tau }\text{ .}
\label{5.45}
\end{equation}

The vector $\nabla _u\xi _{(a)\perp }$ has two components with respect to
the vector $u$: one collinear to $u$ and one orthogonal to $u$, i.e. 
\begin{eqnarray}
\frac{D\xi _{(a)\perp }}{d\tau } &=&\nabla _u\xi _{(a)\perp }=\frac{%
\overline{l}_a}e\cdot u+\,_{rel}v_{(a)}\text{ ,\thinspace \thinspace }
\label{5.46} \\
\overline{l}_a &:&=g(u,\nabla _u\xi _{(a)\perp })\text{ \thinspace
\thinspace \thinspace \thinspace \thinspace ,\thinspace \thinspace
\thinspace \thinspace \thinspace \thinspace \thinspace \thinspace \thinspace
\thinspace }e=g(u,u)\neq 0\text{ \thinspace \thinspace \thinspace \thinspace
\thinspace \thinspace \thinspace \thinspace , \thinspace \thinspace } 
\nonumber \\
g(u,_{rel}v_{(a)}) &=&0\text{ ,\thinspace \thinspace \thinspace \thinspace
\thinspace }_{rel}v_{(a)}=\overline{g}[h_u(\nabla _u\xi _{(a)\perp })]=%
\overline{g}[h_u(\frac{D\xi _{(a)\perp }}{d\tau })]\text{ \thinspace
\thinspace .}  \nonumber
\end{eqnarray}

The set of the infinitesimal vectors $\{\overline{\xi }_{\left( a\right)
\perp }\}$ determines a tangential subspace at the point $x^i(\tau
_0,\lambda _0)$, orthogonal to the vector $u$. This subspace intersect the
flow in such a way that a flat cross-section appears, orthogonal to $u$ .
All material points (elements) of the flow lying at this cross-section have
one and the same proper time $\tau _0$ (if $\tau $ is interpreted as proper
time). The vector $\xi _{(a)\perp }$ was interpreted as the velocity along
the line $x^i(\tau _0,\lambda ^a)$. If we consider instead of $\xi
_{(a)\perp }$ the infinitesimal vector $\overline{\xi }_{(a)\perp
}:=d\lambda ^a\cdot \xi _{(a)\perp }$ (there is no summation over $a$) then
the flat cross-section coincides with the cross-section determined by the
points $\{x^i(\tau _0,\lambda _0^a+d\lambda ^a)$,\thinspace \thinspace
\thinspace \thinspace \thinspace $a=1,...,n-1\}$. The infinitesimal vectors $%
\overline{\xi }_{(a)\perp }$, $a=1,...,n-1$, are equal to the difference
between the co-ordinates of the point $P$ with $x^i(\tau _0,\lambda _0^a)$
and the points with co-ordinates $x^i(\tau _0,\lambda _0^a+d\lambda ^a$%
,\thinspace \thinspace $a=1,...,n-1)$. The change of the vectors $\overline{%
\xi }_{(a)\perp }$, determined by the parts $\overline{g}[h_u(\nabla _u\xi
_{(a)\perp })]$ orthogonal to $u$ and lying at the flat cross-section [when
it moves along the curve $x^i(\tau ,\lambda _0^a)$], is described by 
\begin{equation}
\left( _{rel}\overline{v}_{(a)}\right) _{(\tau _0,\lambda _0^a)}=\left( 
\overline{g}[h_u(\nabla _u\overline{\xi }_{(a)\perp })]\right) _{(\tau
_0,\lambda _0^a)}=\left( \overline{g}[h_u(\frac{D\overline{\xi }_{(a)\perp }%
}{d\tau })]\right) _{_{(\tau _0,\lambda _0^a)}}\text{ \thinspace \thinspace
\thinspace \thinspace .}  \label{5.47}
\end{equation}

Therefore, we can interpret the vector $_{rel}\overline{v}_{(a)(\tau
_0,\lambda _0^a)}$ as the {\it relative velocity vector} or {\it relative
velocity} of material points with co-ordinates $x^i(\tau _0,\lambda
_0^a+d\lambda ^a)$ with respect to the point with co-ordinates $x^i(\tau
_0,\lambda _0^a)$. Let us now determine the relation between $_{rel}%
\overline{v}$ and the total velocity $_{total}\overline{v}$ defined at the
point $P$ with $x^i(\tau _0,\lambda _0^a)$ as 
\begin{equation}
_{total}\overline{v}_{(a)(\tau _0,\lambda _0^a)}:=\stackunder{d\tau
\rightarrow 0}{\lim }\frac{x^i(\tau _0+d\tau ,\lambda _0^a+d\lambda
^a)-x^i(\tau _0,\lambda _0^a)}{d\tau }\text{ \thinspace \thinspace
\thinspace \thinspace \thinspace .}  \label{5.48}
\end{equation}

Since up to the second order of $d\tau $ and $d\lambda ^a$ we have 
\begin{eqnarray*}
x^i(\tau _0+d\tau ,\lambda _0^a+d\lambda ^a) &=&x^i(\tau _0+d\tau ,\lambda
_0^a)+d\lambda ^a\cdot \left( \frac{\partial x^i}{\partial \lambda ^a}%
\right) _{(\tau _0+d\tau ,\lambda _0^a)}= \\
&=&x^i(\tau _0+d\tau ,\lambda _0^a)+\overline{\xi }\,_{(a)\perp (\tau
_0+d\tau ,\lambda _0^a)}^{\,i}= \\
&=&x^i(\tau _0,\lambda _0^a)+d\tau \cdot \left( \frac{\partial x^i}{\partial
\tau }\right) _{(\tau _0,\lambda _0^a)}+\overline{\xi }\,_{(a)\perp (\tau
_0,\lambda _0^a)}^{\,i}+
\end{eqnarray*}
\[
+\,d\tau \cdot \left( \frac{D\overline{\xi }_{(a)\perp }^i}{d\tau }\right)
_{(\tau _0,\lambda _0^a)}= 
\]
\begin{equation}
=x^i(\tau _0,\lambda _0^a)+\overline{u}^i\,_{(\tau _0,\lambda _0^a)}+%
\overline{\xi }\,_{(a)\perp (\tau _0,\lambda _0^a)}^{\,i}+d\tau \cdot \left( 
\frac{D\overline{\xi }_{(a)\perp }^i}{d\tau }\right) _{(\tau _0,\lambda
_0^a)}\text{ ,}  \label{5.49}
\end{equation}

\noindent we can express the last term in the last relation as 
\begin{equation}
\frac{D\overline{\xi }_{(a)\perp }^i}{d\tau }=\frac{\widehat{l}_a}e\cdot
u^i+\,_{rel}\overline{v}_{(a)}^i\text{ ,\thinspace \thinspace \thinspace
\thinspace \thinspace \thinspace \thinspace \thinspace \thinspace \thinspace 
}\widehat{l}_a:=g(u,\nabla _u\overline{\xi }_{(a)\perp })\text{ \thinspace
\thinspace \thinspace \thinspace ,\thinspace \thinspace \thinspace
\thinspace \thinspace \thinspace \thinspace }g(u,\,_{rel}\overline{v}%
_{(a)})=0\text{ .}  \label{5.50}
\end{equation}

Therefore, 
\begin{eqnarray}
_{total}\overline{v}_{(\tau _0,\lambda _0^a)} &=&\stackunder{d\tau
\rightarrow 0}{\lim }\frac{x^i(\tau _0+d\tau ,\lambda _0^a+d\lambda
^a)-x^i(\tau _0,\lambda _0^a)}{d\tau }=  \nonumber \\
&=&\left[ (1+\frac{\widehat{l}_a}e)\cdot u^i\right] \,_{(\tau _0,\lambda
_0^a)}+\,_{rel}\overline{v}_{(a)(\tau _0,\lambda _0^a)}^i\text{ ,\thinspace
\thinspace \thinspace \thinspace \thinspace \thinspace \thinspace \thinspace
\thinspace \thinspace \thinspace \thinspace \thinspace \thinspace }\frac{%
d\lambda }{d\tau }=0\text{ \thinspace \thinspace \thinspace ,}  \label{5.51}
\end{eqnarray}

\noindent or for every point at a curve $x^i(\tau ,\lambda ^a)\subset $ $M$
the local total velocity is 
\begin{equation}
_{total}\overline{v}_{(a)}=(1+\frac{\widehat{l}_a}e)\cdot u+\,_{rel}%
\overline{v}_{(a)}\text{ \thinspace \thinspace \thinspace ,}  \label{5.52}
\end{equation}

\noindent and the local relative velocity $_{rel}\overline{v}_{(a)}$ of the
material point (element) in a flow is 
\begin{equation}
_{rel}\overline{v}_{(a)}=\,_{total}\overline{v}_{(a)}-(1+\frac{\widehat{l}_a}%
e)\cdot u\text{ \thinspace \thinspace \thinspace .}  \label{5.53}
\end{equation}

To find a more exact physical explanation of the structure of the relative
velocity vector $_{rel}v$ we should consider now the length of a vector
field, expressed by the use of kinematic characteristics of a flow. In the
further consideration we use the vector $u\in T(M)$ instead of the vector $%
\overline{u}=d\tau \cdot u\in T(M)$ as an infinitesimal vector at a curve $%
x^i(\tau ,\lambda _0^a)$ identified with a line segment of the curve.

\subsection{Length of a vector field, expressed by the use of the kinematic
characteristics of a flow}

Let us now consider the change of the length of a vector $\xi _{(a)\perp }$ $%
(a=1,...,n-1)$ transported from point $P$ with co-ordinates $x^i(\tau
_0,\lambda _0^a)$ to the point $P_1$ with co-ordinates $x^i(\tau _0+d\tau
,\lambda _0^a)$. The length $l_{\xi _{(a)\perp }}$ of the vector $\xi
_{(a)\perp }$ is defined as 
\begin{equation}
g(\xi _{(a)\perp },\xi _{(a)\perp })=g_{\overline{i}\overline{j}}\cdot \xi
_{(a)\perp }^i\cdot \xi _{(a)\perp }^j=\pm \,l_{\xi _{(a)\perp }}^2\text{
\thinspace \thinspace \thinspace \thinspace .}  \label{5.56}
\end{equation}

At the point $P$ with $x^i(\tau _0,\lambda _0^a)$ the vector $\xi _{(a)\perp
}$ will have the length 
\begin{equation}
\left[ g(\xi _{(a)\perp },\xi _{(a)\perp })\right] _{(\tau _0,\lambda
_0^a)}=\left[ g_{\overline{i}\overline{j}}\cdot \xi _{(a)\perp }^i\cdot \xi
_{(a)\perp }^j\right] _{(\tau _0,\lambda _0^a)}=\pm \,\left[ l_{\xi
_{(a)\perp }}^2\right] _{(\tau _0,\lambda _0^a)}\text{ \thinspace \thinspace
\thinspace \thinspace .}  \label{5.57}
\end{equation}

At the point $P_1$ with $x^i(\tau _0+d\tau ,\lambda _0^a)$ the vector $\xi
_{\perp }$ will have the length 
\begin{equation}
\left[ g(\xi _{(a)\perp },\xi _{(a)\perp })\right] _{(\tau _0+d\tau ,\lambda
_0^a)}=\left[ g(\xi _{(a)\perp },\xi _{(a)\perp })\right] _{(\tau _0,\lambda
_0^a)}+d\tau \cdot \left[ \frac{d[g(\xi _{(a)\perp },\xi _{(a)\perp })]}{%
d\tau }\right] _{_{(\tau _0,\lambda _0^a)}}\text{ \thinspace \thinspace
\thinspace \thinspace .}  \label{5.58}
\end{equation}

Since 
\begin{equation}
\frac{d[g(\xi _{(a)\perp },\xi _{(a)\perp })]}{d\tau }=\frac{D[g(\xi
_{(a)\perp },\xi _{(a)\perp })]}{d\tau }=\nabla _u[g(\xi _{(a)\perp },\xi
_{(a)\perp })]\text{ \thinspace \thinspace \thinspace ,}  \label{5.59}
\end{equation}

\noindent we can represent the last expression by the use of the kinematic
characteristics of the relative velocity. 
\begin{eqnarray}
\frac{D[g(\xi _{(a)\perp },\xi _{(a)\perp })]}{d\tau } &=&\nabla _u[g(\xi
_{(a)\perp },\xi _{(a)\perp })]=(\nabla _ug)(\xi _{(a)\perp },\xi _{(a)\perp
})+2\cdot g(\xi _{(a)\perp },\nabla _u\xi _{(a)\perp })=  \nonumber \\
&=&u[g(\xi _{(a)\perp },\xi _{(a)\perp })]=\frac d{d\tau }[g(\xi _{(a)\perp
},\xi _{(a)\perp })]\text{ \thinspace \thinspace \thinspace \thinspace .}
\label{5.60}
\end{eqnarray}

The vector $\nabla _u\xi _{(a)\perp }$ could be represented in the form 
\begin{equation}
\nabla _u\xi _{(a)\perp }=\frac{g(\nabla _u\xi _{(a)\perp },u)}{g(u,u)}\cdot
u+\,_{rel}v_{(a)}=\frac{\overline{l}_a}e\cdot u+\,_{rel}v_{(a)}\text{%
\thinspace \thinspace \thinspace \thinspace \thinspace \thinspace \thinspace
,}  \label{5.61}
\end{equation}

\noindent where 
\begin{equation}
_{rel}v_{(a)}=\overline{g}[d(\xi _{(a)\perp }]\text{ \thinspace \thinspace
\thinspace \thinspace \thinspace \thinspace .\thinspace \thinspace
\thinspace }  \label{5.62}
\end{equation}

The tensor $d$ with $d(u)=(u)(d)=0$ is the deformation velocity tensor. The
relative velocity between the material points of the flow is related to the
deformation of a cross-section of the flow (determined by $\{\xi
_{(a)}:a=1,\cdots ,n-1\}$ along a line of the flow with tangent vector $u$.

The covariant derivative of $g(\xi _{(a)\perp },\xi _{(a)\perp })$
(identical to the ordinary derivative) along the vector $u$ could be written
now as 
\begin{eqnarray}
\frac{D[g(\xi _{(a)\perp },\xi _{(a)\perp })]}{d\tau } &=&(\nabla _ug)(\xi
_{(a)\perp },\xi _{(a)\perp })+2\cdot g(\xi _{(a)\perp },\frac{\overline{l}_a%
}e\cdot u+\,_{rel}v_{(a)})=  \nonumber \\
&=&(\nabla _ug)(\xi _{(a)\perp },\xi _{(a)\perp })+2\cdot \frac{\overline{l}%
_a}e\cdot g(\xi _{(a)\perp },u)+2\cdot g(\xi _{(a)\perp },_{rel}v_{(a)})%
\text{ \thinspace \thinspace \thinspace .}  \label{5.63}
\end{eqnarray}

Since $g(\xi _{(a)\perp },u)=0$, the second term at the right side vanishes
and we have 
\begin{equation}
\frac{D[g(\xi _{(a)\perp },\xi _{(a)\perp })]}{d\tau }=(\nabla _ug)(\xi
_{(a)\perp },\xi _{(a)\perp })+2\cdot g(\xi _{(a)\perp },_{rel}v_{(a)})\text{
}\,\,\,\,\,.  \label{5.64}
\end{equation}

Let us now represent $\nabla _ug$ by means of the projective metric $h_u$
corresponding to the vector field $u$.

\subsubsection{Representation of $\nabla _ug$ by means of the projective
metric $h_u$}

Since $h_u=g-(1/e)\cdot g(u)\otimes g(u)$, the following relations can be
found: 
\begin{eqnarray}
\nabla _ug &=&g(\overline{g})(\nabla _ug)(\overline{g})g=\left( h_u+\frac
1e\cdot g(u)\otimes g(u)\right) (\overline{g})(\nabla _ug)(\overline{g})g= 
\nonumber \\
&=&h_u(\overline{g})(\nabla _ug)(\overline{g})g+\frac 1e\cdot g(u)\otimes
u(\nabla _ug)(\overline{g})g=  \nonumber \\
&=&h_u(\overline{g})(\nabla _ug)(\overline{g})\left( h_u+\frac 1e\cdot
g(u)\otimes g(u)\right) +  \nonumber \\
&&+\frac 1e\cdot g(u)\otimes u(\nabla _ug)(\overline{g})\left( h_u+\frac
1e\cdot g(u)\otimes g(u)\right)  \nonumber \\
&=&h_u(\overline{g})(\nabla _ug)(\overline{g})h_u+\frac 1e\cdot h_u(%
\overline{g})(\nabla _ug)(u)\otimes g(u)+  \nonumber \\
&&+\frac 1e\cdot g(u)\otimes u(\nabla _ug)(\overline{g})(h_u)+\frac
1{e^2}\cdot (\nabla _ug)(u,u)\cdot g(u)\otimes g(u)\text{ ,}  \label{5.65}
\end{eqnarray}

\noindent where 
\begin{eqnarray*}
g(u)\overline{g} &=&u\text{ \thinspace \thinspace \thinspace ,\thinspace
\thinspace \thinspace \thinspace \thinspace \thinspace \thinspace }h_u(%
\overline{g})(\nabla _ug)(u)=(u)(\nabla _ug)(\overline{g})h_u\text{
\thinspace \thinspace \thinspace \thinspace ,} \\
(u)(\nabla _ug)(u) &=&(\nabla _ug)(u,u)\text{ \thinspace \thinspace
\thinspace .}
\end{eqnarray*}

Therefore, 
\begin{eqnarray}
\nabla _ug &=&h_u(\overline{g})(\nabla _ug)(\overline{g})h_u+  \nonumber \\
&&+\frac 1e\cdot [h_u(\overline{g})(\nabla _ug)(u)\otimes g(u)+g(u)\otimes
h_u(\overline{g})(\nabla _ug)(u)]+  \nonumber \\
&&+\frac 1{e^2}\cdot (\nabla _ug)(u,u)\cdot g(u)\otimes g(u)\text{\thinspace
\thinspace \thinspace .}  \label{5.66}
\end{eqnarray}

\subsubsection{Explicit form of the change of the length of the vector
fields $\xi _{(a)\perp }$}

If we use the relations 
\begin{eqnarray}
\overline{g}(\nabla _ug)\overline{g} &=&-\nabla _u\overline{g}\text{%
\thinspace \thinspace \thinspace \thinspace \thinspace \thinspace \thinspace
\thinspace \thinspace \thinspace \thinspace ,\thinspace \thinspace
\thinspace \thinspace \thinspace \thinspace \thinspace \thinspace \thinspace
\thinspace \thinspace \thinspace \thinspace \thinspace \thinspace \thinspace
\thinspace \thinspace \thinspace \thinspace \thinspace \thinspace \thinspace 
}\left( g(u)\otimes g(u)\right) (\xi _{(a)\perp },\xi _{(a)\perp
})=0\,\,\,\,\,\,\text{,}  \nonumber \\
&&\left( h_u(\overline{g})(\nabla _ug)(u)\otimes g(u)\right) (\xi _{(a)\perp
},\xi _{(a)\perp })  \nonumber \\
&=&\left( h_u(\overline{g})(\nabla _ug)(u)\right) (\xi _{(a)\perp })\cdot
g(u,\xi _{(a)\perp })=0\text{ \thinspace \thinspace \thinspace \thinspace ,}
\label{5.67} \\
&&\left( g(u)\otimes h_u(\overline{g})(\nabla _ug)(u)\right) (\xi _{(a)\perp
},\xi _{(a)\perp })  \nonumber \\
&=&g(u,\xi _{(a)\perp })\cdot \left( h_u(\overline{g})(\nabla _ug)(u)\right)
(\xi _{(a)\perp })=0\text{ ,}  \nonumber
\end{eqnarray}

\noindent we can find some useful expressions leading to the explicit form
of the change of the length of the vector fields $\xi _{(a)\perp }$.

For $(\nabla _ug)(\xi _{(a)\perp },\xi _{(a)\perp })$, we obtain 
\begin{eqnarray}
(\nabla _ug)(\xi _{(a)\perp },\xi _{(a)\perp }) &=&\left( h_u(\overline{g}%
)(\nabla _ug)(\overline{g})h_u\right) (\xi _{(a)\perp },\xi _{(a)\perp })= 
\nonumber \\
&=&-\text{ }\left( h_u(\nabla _u\overline{g})h_u\right) (\xi _{(a)\perp
},\xi _{(a)\perp })\text{\thinspace \thinspace \thinspace \thinspace .}
\label{5.68}
\end{eqnarray}

On the other side, 
\begin{eqnarray}
\overline{g}\left[ h_u(\nabla _u\overline{g})h_u\right] &=&g^{\overline{i}%
\overline{j}}\cdot h_{ik}\cdot g^{\overline{k}\overline{l}}\,_{;n}\cdot
u^n\cdot h_{lj}=  \nonumber \\
&=&g^{\overline{i}\overline{j}}\cdot h_{i\overline{k}}\cdot
g^{kl}\,_{;n}\cdot u^n\cdot h_{\overline{l}j}=  \nonumber \\
&=&h_{ki}\cdot g^{\overline{i}\overline{j}}\cdot h_{jl}\cdot g^{\overline{k}%
\overline{l}}\,_{;n}\cdot u^n=  \nonumber \\
&=&\left( h_u(\overline{g})h_u\right) [\nabla _u\overline{g}]\text{
\thinspace \thinspace \thinspace \thinspace \thinspace \thinspace ,}
\label{5.69}
\end{eqnarray}
\begin{eqnarray}
\lbrack g(u)]\overline{g} &=&\overline{g}[g(u)]=u\text{ \thinspace
\thinspace \thinspace ,\thinspace \thinspace \thinspace \thinspace
\thinspace \thinspace \thinspace \thinspace \thinspace \thinspace \thinspace
\thinspace \thinspace }h_u(\overline{g})h_u=g(\overline{g})h_u=h_u\text{
\thinspace \thinspace ,}  \nonumber \\
\overline{g}[h_u(\nabla _u\overline{g})h_u] &=&h_u[\nabla _u\overline{g}]=h_{%
\overline{i}\overline{j}}\cdot g^{ij}\,_{;n}\cdot u^n\text{ \thinspace
\thinspace \thinspace .}  \label{5.70}
\end{eqnarray}

It follows now for $(\nabla _ug)(\xi _{(a)\perp },\xi _{(a)\perp })$%
\begin{eqnarray}
(\nabla _ug)(\xi _{(a)\perp },\xi _{(a)\perp }) &=&-[h_u(\nabla _u\overline{g%
})h_u-\frac 1{n-1}\cdot \left( h_u[\nabla _u\overline{g}]\right) \cdot
h_u](\xi _{(a)\perp },\xi _{(a)\perp })-  \nonumber \\
&&-\frac 1{n-1}\cdot \left( h_u[\nabla _u\overline{g}]\right) \cdot h_u(\xi
_{(a)\perp },\xi _{(a)\perp })\text{ \thinspace \thinspace \thinspace .}
\label{5.71}
\end{eqnarray}

After introducing the abbreviations: 
\begin{eqnarray}
_\nabla \sigma &:&=\frac 12\cdot [h_u(\nabla _u\overline{g})h_u-\frac
1{n-1}\cdot \left( h_u[\nabla _u\overline{g}]\right) \cdot h_u]\text{
\thinspace \thinspace \thinspace ,}  \label{5.72a} \\
_\nabla \theta &:&=\frac 12\cdot h_u[\nabla _u\overline{g}]\text{ \thinspace
\thinspace \thinspace ,}  \label{5.72b}
\end{eqnarray}

\noindent we obtain for $(\nabla _ug)(\xi _{(a)\perp },\xi _{(a)\perp })$%
\begin{equation}
(\nabla _ug)(\xi _{(a)\perp },\xi _{(a)\perp })=-2\cdot \,_\nabla \sigma
(\xi _{(a)\perp },\xi _{(a)\perp })-\frac 2{n-1}\cdot _\nabla \theta \cdot
\,h_u(\xi _{(a)\perp },\xi _{(a)\perp })\text{ \thinspace \thinspace
\thinspace .}  \label{5.73}
\end{equation}

We can now find the explicit form of $\,\nabla _u[g(\xi _{(a)\perp },\xi
_{(a)\perp })]$ by the use of the last relations 
\begin{eqnarray}
\frac D{d\tau }[g(\xi _{(a)\perp },\xi _{(a)\perp })] &=&(\nabla _ug)(\xi
_{(a)\perp },\xi _{(a)\perp })+2\cdot g(\xi _{(a)\perp },_{rel}v_{(a)})= 
\nonumber \\
&=&-2\cdot \,_\nabla \sigma (\xi _{(a)\perp },\xi _{(a)\perp })-\frac
2{n-1}\cdot _\nabla \theta \cdot \,h_u(\xi _{(a)\perp },\xi _{(a)\perp })%
\text{ }+  \nonumber \\
&&+2\cdot g(\xi _{(a)\perp },_{rel}v_{(a)})\text{ .}  \label{5.74}
\end{eqnarray}

On the other side, 
\begin{eqnarray}
g(\xi _{(a)\perp },_{rel}v_{(a)}) &=&g(\xi _{(a)\perp },\,\overline{g}[d(\xi
_{(a)\perp })=  \nonumber \\
&=&g_{\overline{i}\overline{j}}\cdot \xi _{(a)\perp }^i\cdot g^{jk}\cdot d_{%
\overline{k}\overline{l}}\cdot \xi _{(a)\perp }^l=  \nonumber \\
&=&g_{\overline{i}\overline{j}}\cdot g^{jk}\cdot \xi _{(a)\perp }^i\cdot d_{%
\overline{k}\overline{l}}\cdot \xi _{(a)\perp }^l=  \nonumber \\
&=&d_{\overline{k}\overline{l}}\cdot \xi _{(a)\perp }^k\cdot \xi _{(a)\perp
}^l=d(\xi _{(a)\perp }^k,\,\xi _{(a)\perp }^l)\text{ \thinspace \thinspace
\thinspace \thinspace .}  \label{5.75}
\end{eqnarray}

Therefore, 
\begin{eqnarray}
\frac D{d\tau }[g(\xi _{(a)\perp },\xi _{(a)\perp })] &=&-2\cdot \,_\nabla
\sigma (\xi _{(a)\perp },\xi _{(a)\perp })-\frac 2{n-1}\cdot _\nabla \theta
\cdot \,h_u(\xi _{(a)\perp },\xi _{(a)\perp })\text{ }+  \nonumber \\
&&+2\cdot d(\xi _{(a)\perp }^k,\,\xi _{(a)\perp }^l)\text{ .}  \label{5.76}
\end{eqnarray}

The deformation velocity tensor $d$ could be given in its explicit form as 
\cite{Manoff-9}: 
\begin{equation}
d=\sigma +\omega +\frac 1{n-1}\cdot \theta \cdot h_u\text{ ,}  \label{5.77}
\end{equation}

\noindent where 
\begin{eqnarray}
\sigma &=&\frac 12\cdot \{h_u(\nabla _u\overline{g})h_u-h_u(\pounds _u%
\overline{g})h_u-  \nonumber \\
&&-\frac 1{n-1}\cdot \left( h_u[\nabla _u\overline{g}]\right) \cdot
h_u+\frac 1{n-1}\cdot \left( h_u[\pounds _u\overline{g}]\right) \cdot h_u\}%
\text{ \thinspace \thinspace \thinspace \thinspace ,}  \label{5.78}
\end{eqnarray}
\begin{equation}
\theta =\frac 12\cdot h_u[\nabla _u\overline{g}]-\frac 12\cdot h_u[\pounds _u%
\overline{g}]\text{ \thinspace \thinspace \thinspace \thinspace .}
\label{5.79}
\end{equation}

If we introduce the abbreviations 
\begin{eqnarray}
_{\pounds }\sigma &=&\frac 12\cdot \{h_u(\pounds _u\overline{g})h_u-\frac
1{n-1}\cdot \left( h_u[\pounds _u\overline{g}]\right) \cdot h_u\}\text{
\thinspace \thinspace ,}  \label{5.80a} \\
_{\pounds }\theta &=&\frac 12\cdot h_u[\pounds _u\overline{g}]\text{
\thinspace \thinspace \thinspace ,}  \label{5.80b}
\end{eqnarray}

\noindent it follows for the shear velocity tensor $\sigma $ and for the
expansion velocity invariant $\theta $ the expressions 
\begin{equation}
\sigma =\,_\nabla \sigma -\,_{\pounds }\sigma \text{ \thinspace \thinspace
\thinspace \thinspace \thinspace ,\thinspace \thinspace \thinspace
\thinspace \thinspace \thinspace \thinspace \thinspace }\theta =\,_\nabla
\theta -\,_{\pounds }\theta \text{ \thinspace \thinspace \thinspace
\thinspace \thinspace \thinspace \thinspace .}  \label{5.81}
\end{equation}

Then we obtain the deformation velocity tensor $d$ in the form 
\begin{equation}
d=\,_\nabla \sigma -\,_{\pounds }\sigma +\omega +\frac 1{n-1}\cdot (_\nabla
\theta -\,_{\pounds }\theta )\cdot h_u\text{\thinspace \thinspace \thinspace
\thinspace \thinspace \thinspace ,}  \label{5.82}
\end{equation}

\noindent and $\nabla _u[g(\xi _{(a)\perp },\xi _{(a)\perp })]$ could be
written in the form 
\begin{eqnarray}
\frac D{d\tau }[g(\xi _{(a)\perp },\xi _{(a)\perp })] &=&\frac d{d\tau
}[g(\xi _{(a)\perp },\xi _{(a)\perp })]=  \nonumber \\
&=&-2\cdot \,_\nabla \sigma (\xi _{(a)\perp },\xi _{(a)\perp })-\frac
2{n-1}\cdot \,_\nabla \theta \cdot \,h_u(\xi _{(a)\perp },\xi _{(a)\perp })%
\text{ }+  \nonumber \\
&&+2\cdot _\nabla \sigma (\xi _{(a)\perp },\xi _{(a)\perp })-2\cdot
\,_{\pounds }\sigma (\xi _{(a)\perp },\xi _{(a)\perp })+  \nonumber \\
&&+2\cdot \omega (\xi _{(a)\perp },\xi _{(a)\perp })+\frac 2{n-1}\cdot
\,_\nabla \theta \cdot h_u(\xi _{(a)\perp },\xi _{(a)\perp })-  \nonumber \\
&&-\frac 2{n-1}\cdot \,_{\pounds }\theta \cdot h_u(\xi _{(a)\perp },\xi
_{(a)\perp })  \nonumber \\
\frac D{d\tau }[g(\xi _{(a)\perp },\xi _{(a)\perp })] &=&-2\cdot [_{\pounds
}\sigma (\xi _{(a)\perp },\xi _{(a)\perp })+\frac 1{n-1}\cdot \,_{\pounds
}\theta \cdot h_u(\xi _{(a)\perp },\xi _{(a)\perp })]\text{ \thinspace
\thinspace ,}  \label{5.83}
\end{eqnarray}

\noindent where $\omega (\xi _{(a)\perp },\xi _{(a)\perp })=0$. Therefore,
the change of the square $\pm \,l_{\xi _{(a)\perp }}^2=$ $g(\xi _{(a)\perp
},\xi _{(a)\perp })=g_{\overline{i}\overline{j}}\cdot \xi _{(a)\perp
}^i\cdot \xi _{(a)\perp }^j$ of the length $l_{\xi _{(a)\perp }}$ along the
curve $x^i(\tau ,\lambda _0^a)$ is depending only on the shear and expansion
velocity induced by the dragging along $u$ and not on the transport along $u$%
.

Now we can determine the difference between the length of the vector $\xi
_{(a)\perp }$ at the point $P_1$ with $x^i(\tau _0+d\tau ,\lambda _0^a)$ and
the length of the vector $\xi _{(a)\perp }$ at the point $P$ with $x^i(\tau
_0,\lambda _0^a)$: 
\[
\left[ g(\xi _{(a)\perp },\xi _{(a)\perp })\right] _{(\tau _0+d\tau ,\lambda
_0^a)}=\left[ g(\xi _{(a)\perp },\xi _{(a)\perp })\right] _{(\tau _0,\lambda
_0^a)}- 
\]
\begin{equation}
-2\cdot d\tau \cdot \,\left[ _{\pounds }\sigma (\xi _{(a)\perp },\xi
_{(a)\perp })+\frac 1{n-1}\cdot \,_{\pounds }\theta \cdot h_u(\xi _{(a)\perp
},\xi _{(a)\perp })\right] _{_{(\tau _0,\lambda _0^a)}}\text{ .}
\label{5.84}
\end{equation}

Since $h_u(\xi _{(a)\perp },\xi _{(a)\perp })=g(\xi _{(a)\perp },\xi
_{(a)\perp })$, we obtain 
\begin{eqnarray}
\left[ g(\xi _{(a)\perp },\xi _{(a)\perp })\right] _{(\tau _0+d\tau ,\lambda
_0^a)} &=&(1-\frac{2\cdot d\tau }{n-1}\cdot \,_{\pounds }\theta )\cdot
\left[ g(\xi _{(a)\perp },\xi _{(a)\perp })\right] _{(\tau _0,\lambda _0^a)}-
\nonumber \\
&&-2\cdot d\tau \cdot \left[ \,_{\pounds }\sigma (\xi _{(a)\perp },\xi
_{(a)\perp })\right] _{(\tau _0,\lambda _0^a)}\text{ \thinspace \thinspace
\thinspace \thinspace \thinspace ,}  \label{5.85}
\end{eqnarray}
\[
\left\{ \frac D{d\tau }[g(\xi _{(a)\perp },\xi _{(a)\perp })]\right\}
_{(\tau _0,\lambda _0^a)}= 
\]
\[
\stackunder{d\tau \rightarrow 0}{=\lim }\frac 1{d\tau }\cdot \left\{ \left[
g(\xi _{(a)\perp },\xi _{(a)\perp })\right] _{(\tau _0+d\tau ,\lambda
_0^a)}-\left[ g(\xi _{(a)\perp },\xi _{(a)\perp })\right] _{(\tau _0,\lambda
_0^a)}\right\} = 
\]
\begin{equation}
=-2\cdot \left[ _{\pounds }\sigma (\xi _{(a)\perp },\xi _{(a)\perp })+\frac
1{n-1}\cdot \,_{\pounds }\theta \cdot g(\xi _{(a)\perp },\xi _{(a)\perp
})\right] _{_{(\tau _0,\lambda _0^a)}}\text{ \thinspace \thinspace
\thinspace \thinspace \thinspace \thinspace \thinspace \thinspace .}
\label{5.86}
\end{equation}

If we introduce the set of unit vectors $\{n_{(a)}\}$ $(a=1,...,n-1)$%
\begin{equation}
n_{(a)}=\frac{\xi _{(a)\perp }}{l_{\xi _{(a)\perp }}}\text{ \thinspace
\thinspace \thinspace \thinspace ,\thinspace \thinspace \thinspace
\thinspace \thinspace \thinspace \thinspace \thinspace \thinspace \thinspace
\thinspace \thinspace \thinspace \thinspace }g(n_{(a)},n_{(a)})=\frac
1{l_{\xi _{(a)\perp }}^2}\cdot g(\xi _{(a)\perp },\xi _{(a)\perp })=\pm 1%
\text{ \thinspace ,}  \label{5.87}
\end{equation}

\noindent then 
\begin{equation}
\frac{Dl_{\xi _{(a)\perp }}^2}{d\tau }=\frac{dl_{\xi _{(a)\perp }}^2}{d\tau }%
=-2\cdot \left[ _{\pounds }\sigma (l_{\xi _{(a)\perp }}\cdot n_{(a)},l_{\xi
_{(a)\perp }}\cdot n_{(a)})+\frac 1{n-1}\cdot \,_{\pounds }\theta \cdot
l_{\xi _{(a)\perp }}^2\cdot \text{\thinspace }g(n_{(a)},n_{(a)})\right] 
\text{ ,}  \label{5.88}
\end{equation}
\begin{eqnarray}
\frac 1{l_{\xi _{(a)\perp }}^2}\cdot \frac{dl_{\xi _{(a)\perp }}^2}{d\tau }
&=&2\cdot \frac 1{l_{\xi _{(a)\perp }}}\cdot \frac{dl_{\xi _{(a)\perp }}}{%
d\tau }=  \nonumber \\
&=&-2\cdot \left[ _{\pounds }\sigma (n_{(a)},n_{(a)})+\frac 1{n-1}\cdot
\,_{\pounds }\theta \cdot \text{\thinspace }g(n_{(a)},n_{(a)})\right] = 
\nonumber \\
&=&-2\cdot \left[ _{\pounds }\sigma (n_{(a)},n_{(a)})\pm \frac 1{n-1}\cdot
\,_{\pounds }\theta \right] \text{ }\,\,\,\,\,\text{,}  \label{5.89}
\end{eqnarray}
\begin{equation}
\frac 1{l_{\xi _{(a)\perp }}}\cdot \frac{dl_{\xi _{(a)\perp }}}{d\tau }%
=-\left[ _{\pounds }\sigma (n_{(a)},n_{(a)})\pm \frac 1{n-1}\cdot
\,_{\pounds }\theta \right] \text{ \thinspace \thinspace \thinspace
\thinspace \thinspace .}  \label{5.90}
\end{equation}

The invariant $_{\pounds }\theta $ is called {\it expansion velocity induced
by the Lie differential operator} $\pounds _u$ acting on the contravariant
metric $\overline{g}$. The invariant $_\nabla \theta $ is called {\it %
expansion velocity induced by the covariant differential operator} $\nabla
_u $ acting on $\overline{g}$. It follows from the last two expressions that
the length of the vector $\xi _{(a)\perp }$ does not depend on $_\nabla
\theta $. The invariant $\theta =\,_\nabla \theta -\,_{\pounds }\theta $ is
called expansion velocity. This notion is related to the fact that the
change of the invariant volume element $d\omega $ along a given vector field 
$u$ after a transport or a dragging along $u$ could be expressed by means of 
$_\nabla \theta $ and $_{\pounds }\theta $ respectively. For $\nabla
_u(d\omega )$ and $\pounds _u(d\omega )$ we have 
\begin{equation}
\nabla _u(d\omega )=\frac 12\cdot \overline{g}[\nabla _ug]\cdot d\omega 
\text{ \thinspace \thinspace \thinspace \thinspace \thinspace ,\thinspace
\thinspace \thinspace \thinspace \thinspace }\pounds _u(d\omega )=\frac
12\cdot \overline{g}[\pounds _ug]\cdot d\omega \text{\thinspace \thinspace
\thinspace \thinspace \thinspace .\thinspace }  \label{5.91}
\end{equation}

On the other side, if we use the relations 
\begin{eqnarray}
\nabla _u\overline{g} &=&-\overline{g}(\nabla _ug)\overline{g}\text{%
\thinspace \thinspace \thinspace \thinspace \thinspace \thinspace ,} 
\nonumber \\
g[\overline{g}(\nabla _ug)\overline{g}] &=&\overline{g}[\nabla _ug]\text{%
\thinspace \thinspace \thinspace \thinspace \thinspace \thinspace \thinspace
\thinspace ,}  \label{5.92} \\
\left( g(u)\otimes g(u)\right) [\overline{g}(\nabla _ug)\overline{g}]
&=&(\nabla _ug)(u,u)\text{ \thinspace \thinspace \thinspace \thinspace ,} 
\nonumber
\end{eqnarray}

\noindent we can find the expression for $_\nabla \theta $ in the form 
\begin{eqnarray}
_\nabla \theta &=&\frac 12\cdot h_u[\nabla _u\overline{g}]=-\frac 12\cdot
h_u[\overline{g}(\nabla _ug)\overline{g}]=  \nonumber \\
&=&-\frac 12\cdot \overline{g}[\nabla _ug]+\frac 1{2\cdot e}\cdot (\nabla
_ug)(u,u)\text{ ,}  \label{5.93}
\end{eqnarray}

\noindent and therefore, 
\begin{equation}
\frac 12\cdot \overline{g}[\nabla _ug]=[\frac 1{2\cdot e}\cdot (\nabla
_ug)(u,u)-\,_\nabla \theta ]\text{ \thinspace \thinspace \thinspace
\thinspace .}  \label{5.94}
\end{equation}

It follows for the change $\nabla _u(d\omega )$ of the invariant volume
element $d\omega $%
\begin{eqnarray}
\nabla _u(d\omega ) &=&\frac 12\cdot \overline{g}[\nabla _ug]\cdot d\omega =
\nonumber \\
&=&[-\,_\nabla \theta +\frac 1{2\cdot e}\cdot (\nabla _ug)(u,u)]\cdot
d\omega \text{ \thinspace \thinspace \thinspace \thinspace .}  \label{5.95}
\end{eqnarray}

{\it Special case:} (Pseudo) Riemannian spaces with or without torsion ($U_n$
or $V_n$-spaces): metric transport $\nabla _ug:=0$. 
\begin{equation}
\nabla _u(d\omega )=-\,_\nabla \theta \cdot d\omega =0\text{ \thinspace
\thinspace \thinspace .}  \label{5.96}
\end{equation}

This means that the invariant volume element $d\omega $ does not change
under a transport of $d\omega $ along $u$ in (pseudo) Riemannian spaces with
or without torsion. This result follows directly from $\nabla _u(d\omega
)=(1/2)\cdot \overline{g}[\nabla _ug]\cdot d\omega $.

In an analogous way, for the expansion velocity $_{\pounds }\theta $ induced
by the Lie differential operator the relations are valid: 
\begin{eqnarray}
_{\pounds }\theta &=&\frac 12\cdot h_u[\pounds _u\overline{g}]=-\frac
12\cdot h_u[\overline{g}(\pounds _ug)\overline{g}]=  \nonumber \\
&=&-\frac 12\cdot \overline{g}[\pounds _ug]+\frac 1{2\cdot e}\cdot (\pounds
_ug)(u,u)\text{ ,}  \label{5.97}
\end{eqnarray}
\begin{eqnarray}
\pounds _u(d\omega ) &=&\frac 12\cdot \overline{g}[\pounds _ug]\cdot d\omega
=  \nonumber \\
&=&[-\,_{\pounds }\theta +\frac 1{2\cdot e}\cdot (\pounds _ug)(u,u)]\cdot
d\omega \text{ \thinspace \thinspace \thinspace \thinspace \thinspace .}
\label{5.98}
\end{eqnarray}

If we use further the explicit form of $\pounds _ug$ for spaces with affine
connections and metrics 
\begin{eqnarray}
\pounds _ug &=&(\pounds _ug_{ij})\cdot dx^i.dx^j=  \nonumber \\
&=&[g_{ij;k}\cdot u^k+g_{kj}\cdot u^{\overline{k}}\,_{;\underline{i}%
}+g_{ik}\cdot u^{\overline{k}}\,_{;\underline{j}}+  \nonumber \\
&&+(g_{kj}\cdot T_{l\underline{i}}\,^{\overline{k}}+g_{ik}\cdot T_{l%
\underline{j}}\,^{\overline{k}})\cdot u^l]\cdot dx^i.dx^j\text{ \thinspace
\thinspace \thinspace ,}  \label{5.99}
\end{eqnarray}

\noindent and the relation 
\begin{equation}
(\pounds _ug)(u,u)=\nabla _u[g(u,u)]=ue\text{ ,}  \label{5.100}
\end{equation}

\noindent we obtain 
\begin{eqnarray}
\frac 12\cdot \overline{g}[\pounds _ug] &=&(-\,_{\pounds }\theta +\frac{ue}{%
2\cdot e})\text{ ,}  \label{5.101} \\
\pounds _u(d\omega ) &=&\frac 12\cdot \overline{g}[\pounds _ug]\cdot d\omega
=(-\,_{\pounds }\theta +\frac{ue}{2\cdot e})\cdot d\omega \text{\thinspace
\thinspace \thinspace \thinspace \thinspace \thinspace .}  \label{5.101a}
\end{eqnarray}

For a normalized vector field $u$ with $e=\,$const. and therefore, $ue=0$,
it follows 
\begin{equation}
\pounds _u(d\omega )=-\,_{\pounds }\theta \cdot d\omega \text{\thinspace
\thinspace \thinspace \thinspace \thinspace \thinspace .}  \label{5.102}
\end{equation}

The last expression leads to the interpretation of $_{\pounds }\theta $ as
expansion velocity induced by the Lie differential operator $\pounds _u$ on
the covariant metric tensor $g$. In other words, $_{\pounds }\theta $ is the
expansion velocity induced by a dragging of $d\omega $ along a vector field $%
u$.

Since 
\begin{eqnarray}
\nabla _u(d\omega )-\pounds _u(d\omega ) &=&\{\frac 12\cdot \overline{g}[%
\nabla _ug]-\frac 12\cdot \overline{g}[\pounds _ug]\}\cdot d\omega = 
\nonumber \\
&=&\{\frac 12\cdot \overline{g}[\nabla _ug-\pounds _ug]\}\cdot d\omega = 
\nonumber \\
&=&[-\,_\nabla \theta +\frac 1{2\cdot e}\cdot (\nabla _ug)(u,u)+\,_{\pounds
}\theta -\frac{ue}{2\cdot e}]\cdot d\omega =  \nonumber \\
&=&\{\,_{\pounds }\theta -\,_\nabla \theta +\frac 1{2\cdot e}\cdot [(\nabla
_ug)(u,u)-ue]\}.d\omega =  \nonumber \\
&=&\{-\theta +\frac 1{2\cdot e}\cdot [(\nabla _ug)(u,u)-ue]\}.d\omega
\label{5.103}
\end{eqnarray}

\noindent and 
\begin{eqnarray}
(\nabla _ug)(u,u)-ue &=&(\nabla _ug)(u,u)-\nabla _u[g(u,u)]=  \nonumber \\
&=&-2\cdot g(u,\nabla _uu)=-2\cdot g(u,a)\text{ \thinspace \thinspace
\thinspace \thinspace \thinspace ,\thinspace \thinspace \thinspace
\thinspace \thinspace \thinspace \thinspace \thinspace \thinspace \thinspace
\thinspace \thinspace }\nabla _uu=a\text{ ,}  \label{5.104}
\end{eqnarray}

\noindent it follows that 
\begin{equation}
\nabla _u(d\omega )-\pounds _u(d\omega )=-[\theta +\frac 1e\cdot
g(u,a)]\cdot d\omega \text{ \thinspace \thinspace \thinspace .}
\label{5.105}
\end{equation}

If the vector field $u$ is an auto-parallel vector field $(\nabla _uu=a=0)$,
then 
\begin{equation}
\nabla _u(d\omega )-\pounds _u(d\omega )=-\theta \cdot d\omega \text{%
\thinspace \thinspace \thinspace \thinspace \thinspace \thinspace \thinspace
\thinspace .}  \label{5.106}
\end{equation}

Therefore, the expansion velocity $\theta $ determines the difference
between the change of the invariant volume element $d\omega $ after a
transport along an auto-parallel curve and a dragging along the vector field 
$u$.

The trace-free symmetric tensor field $\sigma $ is called shear velocity
tensor (shear velocity, shear) 
\begin{equation}
\sigma =\,_\nabla \sigma -\,_{\pounds }\sigma \text{ \thinspace \thinspace
\thinspace \thinspace ,\thinspace \thinspace \thinspace \thinspace
\thinspace \thinspace \thinspace \thinspace \thinspace \thinspace \thinspace
\thinspace \thinspace }\overline{g}[\sigma ]=0\text{ \thinspace \thinspace
\thinspace \thinspace ,\thinspace \thinspace \thinspace \thinspace
\thinspace \thinspace \thinspace \thinspace \thinspace }\overline{g}[_\nabla
\sigma ]=0\text{ \thinspace \thinspace \thinspace \thinspace \thinspace
,\thinspace \thinspace \thinspace \thinspace \thinspace \thinspace
\thinspace \thinspace }\overline{g}[_{\pounds }\sigma ]=0\text{ .}
\label{5.107}
\end{equation}

The tensor field $_\nabla \sigma $ is called shear velocity induced by a
transport along the vector field $u$, $_{\pounds }\sigma $ is called shear
velocity induced by a dragging along $u$. The change of the length of the
vector $\xi _{(a)\perp }$ along a curve $x^i(\tau ,\lambda _0)$ is
determined only by $_{\pounds }\sigma $ and not by $_\nabla \sigma $. The
shear velocity tensor $_{\pounds }\sigma $ determines the change of the
length of the vector $\xi _{(a)\perp }$ together with the expansion velocity 
$_{\pounds }\theta $. The shear velocity $_{\pounds }\sigma $ does not
contain the part, responding for the change of the invariant volume element.
This is because $_{\pounds }\sigma $ is constructed by a tensor, which trace 
$_{\pounds }\theta $, determining the change of the invariant volume element 
$d\omega $, is substracted from it. Thus the shear velocity tensor does not
change a volume in a space with affine connections and metrics. It generates
a volume-preserving shape deformation. This means that a sphere could be
deformed to an ellipsoid under keeping its volume if $_{\pounds }\sigma \neq
0$.

We can now resume, that the expansion velocity and the shear velocity have
their corresponding physical meaning in the continuum media mechanics in
spaces with affine connections and metrics.

The tensor field $\omega $, with $\omega (u)=-(u)(\omega )=0$, is called
rotation (vortex) velocity tensor (rotation velocity, rotation, vortex). It
does not change the length of a vector field $\xi _{(a)\perp }$ and
therefore, it changes only the direction of $\xi _{(a)\perp }$ causing its
rotation [$\omega (\xi _{(a)\perp })$], with $(u)(\omega (\xi _{(a)\perp
})=0 $, in the $n-1$ dimensional subspace, orthogonal to the vector $u$.

\subsection{Relative velocity and contravariant vector fields}

The kinematic characteristics related to the notion of relative velocity can
be used in finding out their influence on the rate of change of the length
of a contravariant vector field as well as the rate of change of the cosine
between two contravariant vector fields.

\subsubsection{Relative velocity and change of the length of a contravariant
vector field}

Let we now consider the influence of the kinematic characteristics related
to the relative velocity upon the change of the length of a contravariant
vector field.

Let $l_\xi =\,\mid g(\xi ,\xi )\mid ^{\frac 12}$ be the length of a
contravariant vector field $\xi $. The rate of change $ul_\xi $ of $l_\xi $
along a contravariant vector field $u$ can be expressed in the form $\pm
\,2\cdot l_\xi \cdot (ul_\xi )=(\nabla _ug)(\xi ,\xi )+2\cdot g(\nabla _u\xi
,\xi )$. By the use of the projections of $\xi $ and $\nabla _u\xi $ along
and orthogonal to $u$ (see the section about kinematic characteristics and
relative velocity) we can find the relations 
\[
\begin{array}{c}
2\cdot g(\nabla _u\xi ,\xi )=2\cdot \frac le\cdot g(\nabla _u\xi ,u)+2\cdot
g(_{rel}v,\xi _{\perp })\text{ ,} \\ 
(\nabla _ug)(\xi ,\xi )=(\nabla _ug)(\xi _{\perp },\xi _{\perp })+2\cdot
\frac le\cdot (\nabla _ug)(\xi _{\perp },u)+\frac{l^2}{e^2}\cdot (\nabla
_ug)(u,u)\text{ .}
\end{array}
\]

Then, it follows for $\pm \,2\cdot l_\xi \cdot (ul_\xi )$ the expression 
\begin{equation}
\begin{array}{c}
\pm \,2\cdot l_\xi \cdot (ul_\xi )=(\nabla _ug)(\xi _{\perp },\xi _{\perp
})+2\cdot \frac le\cdot [(\nabla _ug)(\xi _{\perp },u)+g(\nabla _u\xi ,u)]+
\\ 
+\,\frac{l^2}{e^2}\cdot (\nabla _ug)(u,u)+2\cdot g(_{rel}v,\xi _{\perp })%
\text{ ,}
\end{array}
\label{5.108}
\end{equation}

\noindent where 
\begin{equation}
g(_{rel}v,\xi _{\perp })=\frac le\cdot h_u(a,\xi _{\perp })+h_u(\pounds
_u\xi ,\xi _{\perp })+d(\xi _{\perp },\xi _{\perp })\text{ ,}  \label{5.109}
\end{equation}
\begin{equation}
d(\xi _{\perp },\xi _{\perp })=\sigma (\xi _{\perp },\xi _{\perp })+\frac
1{n-1}\cdot \theta \cdot l_{\xi _{\perp }}^2\text{ .}  \label{5.110}
\end{equation}

For finding out the last two expressions the following relations have been
used: 
\begin{equation}
g(\overline{g}(h_u)a,\xi _{\perp })=h_u(a,\xi _{\perp })\text{ ,\thinspace
\thinspace \thinspace \thinspace \thinspace \thinspace \thinspace }g(%
\overline{g}(h_u)(\pounds _u\xi ),\xi _{\perp })=h_u(\pounds _u\xi ,\xi
_{\perp })\text{ ,}  \label{5.111}
\end{equation}
\begin{equation}
g(\overline{g}[d(\xi )],\xi _{\perp })=d(\xi _{\perp },\xi _{\perp })\text{
,\thinspace \thinspace \thinspace \thinspace \thinspace \thinspace
\thinspace \thinspace }d(\xi )=d(\xi _{\perp })\text{ .}  \label{5.112}
\end{equation}

{\it Special case}: $g(u,\xi )=l:=0:\xi =\xi _{\perp }$. 
\begin{equation}
\pm \,2\cdot l_{\xi _{\perp }}\cdot (ul_{\xi _{\perp }})=(\nabla _ug)(\xi
_{\perp },\xi _{\perp })+2\cdot g(_{rel}v,\xi _{\perp })\text{ .}
\label{5.113}
\end{equation}

{\it Special case}: $V_n$-spaces: $\nabla _\eta g=0$ for $\forall \eta \in
T(M)$ ($g_{ij;k}=0$), $g(u,\xi )=l:=0:\xi =\xi _{\perp }$. 
\begin{equation}
\pm \,l_{\xi _{\perp }}\cdot (ul_{\xi _{\perp }})=g(_{rel}v,\xi _{\perp })%
\text{ .}  \label{5.114}
\end{equation}

In $(\overline{L}_n,g)$-spaces as well as in $(L_n,g)$-spaces the covariant
derivative $\nabla _ug$ of the metric tensor field $g$ along $u$ can be
decomposed in its trace free part $^s\nabla _ug$ and its trace part $\frac
1n\cdot Q_u\cdot g$ as 
\[
\nabla _ug=\,^s\nabla _ug+\frac 1n\cdot Q_u\cdot g\text{ ,\thinspace
\thinspace \thinspace \thinspace \thinspace \thinspace \thinspace \thinspace
\thinspace }\dim M=n\text{ ,} 
\]

\noindent where 
\[
\overline{g}[^s\nabla _ug]=0\text{ ,\thinspace \thinspace \thinspace
\thinspace \thinspace \thinspace \thinspace }Q_u=\overline{g}[\nabla _ug]=g^{%
\overline{k}\overline{l}}\cdot g_{kl;j}\cdot u^j=Q_j\cdot u^j\text{ ,
\thinspace \thinspace \thinspace \thinspace }Q_j=g^{\overline{k}\overline{l}%
}\cdot g_{kl;j}\text{ .} 
\]

The covariant vector $\overline{Q}=\frac 1n\cdot Q=\frac 1n\cdot Q_j\cdot
dx^j=\frac 1n\cdot Q_\alpha \cdot e^\alpha $ is called 
\index{Weyl's vector field@Weyl's vector field} {\it Weyl's covariant vector
field}. The operator $\nabla _u=\,^s\nabla _u+\frac 1n\cdot Q_u$ is called 
{\it trace free covariant operator}.

If we use now the decomposition of $\nabla _ug$ in the expression for $\pm
\,2\cdot l_\xi \cdot (ul_\xi )$ we find the relation 
\begin{equation}
\begin{array}{c}
\pm \,2\cdot l_\xi \cdot (ul_\xi )=(^s\nabla _ug)(\xi ,\xi )\pm \frac
1n\cdot Q_u\cdot l_\xi ^2+2\cdot g(\nabla _u\xi ,\xi )= \\ 
=(^s\nabla _ug)(\xi _{\perp },\xi _{\perp })+ \\ 
+\frac le\cdot [2\cdot (^s\nabla _ug)(\xi _{\perp },u)+2\cdot g(\nabla _u\xi
,u)+\frac le\cdot (^s\nabla _ug)(u,u)]+ \\ 
+\frac 1n\cdot Q_u\cdot (\pm \,\,l_{\xi _{\perp }}^2+%
\frac{l^2}e)+2\cdot g(_{rel}v,\xi _{\perp })\text{ ,}
\end{array}
\label{5.115}
\end{equation}

\noindent where $\pm \,l_{\xi _{\perp }}^2=g(\xi _{\perp },\xi _{\perp })$, $%
l=g(\xi ,u)$.

For $l_\xi \neq 0:$%
\begin{equation}
ul_\xi =\pm \frac 1{2\cdot l_\xi }\cdot (^s\nabla _ug)(\xi ,\xi )+\frac
1{2\cdot n}\cdot Q_u\cdot l_\xi \pm \frac 1{l_\xi }\cdot g(\nabla _u\xi ,\xi
)\text{ .}  \label{5.116}
\end{equation}

In the case of a parallel transport ($\nabla _u\xi =0$) of $\xi $ along $u$
the change $ul_\xi $ of the length\thinspace \thinspace $l_\xi $ is 
\begin{equation}
ul_\xi =\pm \frac 1{2.l_\xi }\cdot (^s\nabla _ug)(\xi ,\xi )+\frac 1{2\cdot
n}\cdot Q_u\cdot l_\xi \text{ .\thinspace \thinspace \thinspace \thinspace
\thinspace \thinspace \thinspace \thinspace \thinspace \thinspace \thinspace
\thinspace \thinspace \thinspace \thinspace }  \label{5.117}
\end{equation}

{\it Special case}: $\nabla _u\xi =0$ and $^s\nabla _ug=0$. 
\begin{equation}
ul_\xi =\frac 1{2\cdot n}\cdot Q_u\cdot l_\xi \text{ .\thinspace \thinspace }
\label{5.118a}
\end{equation}

If $u=\frac d{ds}=u^i\cdot \partial _i=(dx^i/ds)\cdot \partial _i$, then 
\begin{eqnarray}
l_\xi (s+ds) &\approx &l_\xi (s)+\frac{dl_\xi }{ds}\cdot ds=l_\xi (s)+\frac
1{2\cdot n}\cdot Q_u(s)\cdot l_\xi (s)\cdot ds=  \nonumber \\
&=&(1+\frac 1{2\cdot n}\cdot Q_u(s)\cdot ds)\cdot l_\xi (s)=\triangle
_u(s)\cdot l_\xi (s)\text{ ,\thinspace \thinspace \thinspace \thinspace
\thinspace \thinspace \thinspace }  \nonumber \\
\,\,\,\triangle _u(s) &=&1+\frac 1{2\cdot n}\cdot Q_u(s)\cdot ds\text{ ,}
\label{5.118b} \\
\frac{dl_\xi }{ds} &=&\lim_{ds\rightarrow \,0}\frac{l_\xi (s+ds)-l_\xi (s)}{%
ds}=+\frac 1{2\cdot n}\cdot Q_u(s)\cdot l_\xi (s)  \label{5.118c}
\end{eqnarray}

Therefore, the rate of change of $l_\xi $ along $u$ is linear to $l_\xi $.

{\it Special case}: $g(u,\xi )=l:=0:\xi =\xi _{\perp }$. 
\[
\pm 2\cdot l_{\xi _{\perp }}\cdot (ul_{\xi _{\perp }})=(^s\nabla _ug)(\xi
_{\perp },\xi _{\perp })\pm \frac 1n\cdot Q_u\cdot l_{\xi _{\perp
}}^2+2\cdot g(_{rel}v,\xi _{\perp })\text{ .} 
\]
\begin{equation}
ul_{\xi _{\perp }}=\pm \frac 1{2\cdot l_{\xi _{\perp }}}\cdot (^s\nabla
_ug)(\xi _{\perp },\xi _{\perp })+\frac 1{2\cdot n}\cdot Q_u\cdot l_{\xi
_{\perp }}\pm \frac 1{l_{\xi _{\perp }}}\cdot g(_{rel}v,\xi _{\perp })\text{
,\thinspace \thinspace \thinspace \thinspace \thinspace \thinspace
\thinspace \thinspace \thinspace \thinspace }l_{\xi _{\perp }}\neq 0\text{ .}
\label{5.119}
\end{equation}

{\it Special case}: Quasi-metric transports: $\nabla _ug:=2\cdot g(u,\eta
)\cdot g$, \thinspace \thinspace \thinspace $u$, $\eta \in T(M)$. 
\begin{equation}
\pm 2\cdot l_\xi \cdot (ul_\xi )=2\cdot g(u,\eta )\cdot (\pm \,l_{\xi
_{\perp }}^2+\frac{l^2}e)+2\cdot [\frac le\cdot g(\nabla _u\xi
,u)+g(_{rel}v,\xi _{\perp })]\text{ .}  \label{5.120}
\end{equation}

\subsubsection{Relative velocity and change of the cosine between two
contravariant vector fields}

The cosine between two contravariant vector fields $\xi $ and $\eta $ has
been defined as $g(\xi ,\eta )=l_\xi \cdot l_\eta \cdot \cos (\xi ,\eta )$.
The rate of change of the cosine along a contravariant vector field $u$ can
be found in the form 
\begin{equation}
\begin{array}{c}
l_\xi \cdot l_\eta \cdot \{u[\cos (\xi ,\eta )]\}=(\nabla _ug)(\xi ,\eta
)+g(\nabla _u\xi ,\eta )+g(\xi ,\nabla _u\eta )- \\ 
-[l_\eta \cdot (ul_\xi )+l_\xi \cdot (ul_\eta )]\cdot \cos (\xi ,\eta )\text{
.}
\end{array}
\label{5.120a}
\end{equation}

{\it Special case}: $\nabla _u\xi =0$, $\nabla _u\eta =0$, $^s\nabla _ug=0$. 
\[
l_\xi \cdot l_\eta \cdot \{u[\cos (\xi ,\eta )]\}=\frac 1n\cdot Q_u\cdot
g(\xi ,\eta )-[l_\eta \cdot (ul_\xi )+l_\xi \cdot (ul_\eta )]\cdot \cos (\xi
,\eta )\text{ .} 
\]

Since $g(\xi ,\eta )=l_\xi \cdot l_\eta \cdot \cos (\xi ,\eta )$, it follows
from the last relation 
\[
l_\xi \cdot l_\eta \cdot \{u[\cos (\xi ,\eta )]\}=\{\frac 1n\cdot Q_u\cdot
l_\xi \cdot l_\eta -[l_\eta \cdot (ul_\xi )+l_\xi \cdot (ul_\eta )]\}\cdot
\cos (\xi ,\eta )\text{ .} 
\]

Therefore, if $\cos (\xi ,\eta )=0$ between two parallel transported along $%
u $ vector fields $\xi $ and $\eta $, then the right angle between them
[determined by the condition $\cos (\xi ,\eta )=0$] does not change along
the contravariant vector field $u$. In the cases, when $\cos (\xi ,\eta
)\neq 0$, the rate of change of the cosine of the angle between two vector
fields $\xi $ and $\eta $ is linear to $\cos (\xi ,\eta )$.

By the use of the definitions and the relations: 
\begin{equation}
_{rel}v_\xi :=\overline{g}[h_u(\nabla _u\xi )]=\,_{rel}v\text{ ,\thinspace
\thinspace \thinspace \thinspace \thinspace \thinspace \thinspace \thinspace
\thinspace \thinspace \thinspace \thinspace \thinspace \thinspace \thinspace
\thinspace \thinspace }_{rel}v_\eta :=\overline{g}[h_u(\nabla _u\eta )]\text{
,}  \label{5.121}
\end{equation}
\begin{equation}
\begin{array}{c}
g(\nabla _u\xi ,\eta )=\frac 1e\cdot g(u,\eta )\cdot g(\nabla _u\xi
,u)+g(_{rel}v_\xi ,\eta )\text{ ,} \\ 
g(\nabla _u\eta ,\xi )=\frac 1e\cdot g(u,\xi )\cdot g(\nabla _u\eta
,u)+g(_{rel}v_\eta ,\xi )\text{ ,}
\end{array}
\label{5.122}
\end{equation}
\begin{equation}
(\nabla _ug)(\xi ,\eta )=(^s\nabla _ug)(\xi ,\eta )+\frac 1n\cdot Q_u\cdot
g(\xi ,\eta )\text{ ,}  \label{5.123}
\end{equation}
\begin{equation}
\begin{array}{c}
(^s\nabla _ug)(\xi ,\eta )=(^s\nabla _ug)(\xi _{\perp },\eta _{\perp
})+\frac le\cdot (^s\nabla _ug)(u,\eta _{\perp })+\frac{\overline{l}}e\cdot
(^s\nabla _ug)(\xi _{\perp },u)+ \\ 
+\frac le\cdot \frac{\overline{l}}e\cdot (^s\nabla _ug)(u,u)\text{
,\thinspace \thinspace \thinspace \thinspace \thinspace \thinspace
\thinspace \thinspace }\overline{l}=g(u,\eta )\text{ ,\thinspace \thinspace
\thinspace \thinspace \thinspace \thinspace \thinspace \thinspace }\eta
_{\perp }=\overline{g}[h_u(\eta )]\text{ ,\thinspace \thinspace \thinspace
\thinspace \thinspace }l=g(u,\xi )\text{ ,}
\end{array}
\label{5.124}
\end{equation}
\begin{equation}
\begin{array}{c}
(\nabla _ug)(\xi ,\eta )=(^s\nabla _ug)(\xi ,\eta )+\frac 1n\cdot Q_u\cdot
g(\xi ,\eta )= \\ 
=(^s\nabla _ug)(\xi _{\perp },\eta _{\perp })+\frac le\cdot (^s\nabla
_ug)(u,\eta _{\perp })+\frac{\overline{l}}e\cdot (^s\nabla _ug)(\xi _{\perp
},u)+ \\ 
+\frac le\cdot \frac{\overline{l}}e\cdot (^s\nabla _ug)(u,u)+\frac 1n\cdot
Q_u\cdot [\frac{l\cdot \overline{l}}e+g(\xi _{\perp },\eta _{\perp })]\text{
,}
\end{array}
\label{5.125}
\end{equation}

\noindent the expression of $l_\xi \cdot l_\eta \cdot \{u[\cos (\xi ,\eta
)]\}$ follows in the form 
\begin{equation}
\begin{array}{c}
l_\xi \cdot l_\eta \cdot \{u[\cos (\xi ,\eta )]\}=(^s\nabla _ug)(\xi _{\perp
},\eta _{\perp })+\frac le\cdot [(^s\nabla _ug)(u,\eta _{\perp })+g(\nabla
_u\eta ,u)]+ \\ 
+\frac{\overline{l}}e\cdot [(^s\nabla _ug)(\xi _{\perp },u)+g(\nabla _u\xi
,u)]+\frac{l\cdot \overline{l}}{e^2}\cdot (^s\nabla _ug)(u,u)+ \\ 
+\,\frac 1n\cdot Q_u\cdot [\frac{l\cdot \overline{l}}e+g(\xi _{\perp },\eta
_{\perp })]+g(_{rel}v_\xi ,\eta )+g(_{rel}v_\eta ,\xi )- \\ 
-[l_\eta \cdot (ul_\xi )+l_\xi \cdot (ul_\eta )]\cdot \cos (\xi ,\eta )\text{
.}
\end{array}
\label{5.126}
\end{equation}

{\it Special case}: $g(u,\xi )=l:=0$, $g(u,\eta )=\overline{l}:=0:\xi =\xi
_{\perp }$, $\eta =\eta _{\perp }$. 
\begin{equation}
\begin{array}{c}
l_{\xi _{\perp }}\cdot l_{\eta _{\perp }}\cdot \{u[\cos (\xi _{\perp },\eta
_{\perp })]\}=(^s\nabla _ug)(\xi _{\perp },\eta _{\perp })+\,\frac 1n\cdot
Q_u\cdot l_{\xi _{\perp }}\cdot l_{\eta _{\perp }}\cdot \cos (\xi _{\perp
},\eta _{\perp })+ \\ 
+\,g(_{rel}v_{\xi _{\perp }},\eta _{\perp })+g(_{rel}v_{\eta _{\perp }},\xi
_{\perp })-[l_{\eta _{\perp }}\cdot (ul_{\xi _{\perp }})+l_{\xi _{\perp
}}\cdot (ul_{\eta _{\perp }})]\cdot \cos (\xi _{\perp },\eta _{\perp })\text{
,}
\end{array}
\label{5.127}
\end{equation}

\noindent where $g(\xi _{\perp },\eta _{\perp })=l_{\xi _{\perp }}\cdot
l_{\eta _{\perp }}\cdot \cos (\xi _{\perp },\eta _{\perp })$.

The kinematic characteristics related to the relative velocity and used in
considerations of the rate of change of the length of a contravariant vector
field as well as the change of the angle between two contravariant vector
fields could also be useful for description of the motion of physical
systems in $(\overline{L}_n,g)$-spaces.

\subsection{Expansion velocity and variation of the invariant volume element}

From the explicit form of the expansion velocity $\theta $%
\begin{equation}
\begin{array}{c}
\theta =\frac 12\cdot h_u[\nabla _u\overline{g}-\pounds _u\overline{g}]= \\ 
=\frac 12\cdot \{\overline{g}[\pounds _ug]-\overline{g}[\nabla _ug]+\frac
1e\cdot (\nabla _ug)(u,u)-\frac 1e\cdot (\pounds _ug)(u,u)\}\text{ ,}
\end{array}
\label{5.128}
\end{equation}

\noindent where 
\begin{equation}
\lbrack g(u)\otimes g(u)][\nabla _u\overline{g}]=-(\nabla _ug)(u,u)\text{ ,}
\label{5.129}
\end{equation}
\begin{equation}
\lbrack g(u)\otimes g(u)][\pounds _u\overline{g}]=-(\pounds _ug)(u,u)\text{ ,%
}  \label{5.130}
\end{equation}

\noindent one can draw the conclusion that the variation of the invariant
volume element $d\omega $ \cite{Manoff-9} is connected with the expansion
velocity $\theta $. From 
\[
\nabla _u(d\omega )=\frac 12\cdot \overline{g}[\nabla _ug]\cdot d\omega 
\text{ ,\thinspace \thinspace \thinspace \thinspace \thinspace \thinspace
\thinspace \thinspace \thinspace \thinspace \thinspace \thinspace \thinspace
\thinspace }\pounds _\xi (d\omega )=\frac 12\cdot \overline{g}[\pounds _ug%
]\cdot d\omega 
\]

\noindent and (\ref{5.128}) the following relations are fulfilled 
\begin{equation}
\theta \cdot d\omega =\pounds _u(d\omega )-\nabla _u(d\omega )+\frac
1{2\cdot e}\cdot [(\nabla _ug)(u,u)-(\pounds _ug)(u,u)]\cdot d\omega \text{ ,%
}  \label{5.131}
\end{equation}
\begin{equation}
\begin{array}{c}
\pounds _u(d\omega )-\nabla _u(d\omega )=\frac 12\cdot \overline{g}[\pounds
_ug-\nabla _ug]\cdot d\omega = \\ 
=[\theta +\frac 1{2e}\cdot (\pounds _ug-\nabla _ug)(u,u)]\cdot d\omega \text{
,}
\end{array}
\label{5.132}
\end{equation}
\begin{equation}
\pounds _u(d\omega )=[\theta +\frac 12\cdot \overline{g}[\nabla _ug]+\frac
1{2\cdot e}\cdot (\pounds _ug-\nabla _ug)(u,u)]\cdot d\omega \text{ ,}
\label{5.133}
\end{equation}
\begin{equation}
\nabla _u(d\omega )=[-\theta +\frac 12\cdot \overline{g}[\pounds _ug]+\frac
1{2\cdot e}\cdot (\nabla _ug-\pounds _ug)(u,u)]\cdot d\omega \text{ ,}
\label{5.134}
\end{equation}

\noindent where 
\begin{equation}
\frac 12\cdot g[\nabla _u\overline{g}]=\frac 12\cdot h_u[\nabla _u\overline{g%
}]-\frac 1{2\cdot e}\cdot (\nabla _ug)(u,u)=  \label{5.135}
\end{equation}
\begin{equation}
=\theta +\frac 12\cdot h_u[\pounds _u\overline{g}]-\frac 1{2\cdot e}\cdot
(\nabla _ug)(u,u)\text{ ,}  \label{5.136}
\end{equation}
\begin{equation}
\begin{array}{c}
\nabla _u(d\omega )=\frac 12\cdot \overline{g}[\nabla _ug]\cdot d\omega
=-\frac 12\cdot g[\nabla _u\overline{g}]\cdot d\omega = \\ 
=\{\frac 1{2\cdot e}\cdot (\nabla _ug)(u,u)-\frac 12\cdot h_u[\nabla _u%
\overline{g}]\}\cdot d\omega = \\ 
=\{-\theta -\frac 12\cdot h_u[\pounds _u\overline{g}]+\frac 1{2\cdot e}\cdot
(\nabla _ug)(u,u)\}\cdot d\omega \text{ .}
\end{array}
\label{5.137}
\end{equation}

{\it Special case}: Metric transports ($\nabla _ug=0$): 
\begin{equation}
\theta =-\frac 12\cdot h_u[\pounds _u\overline{g}]=\frac 12\cdot \{\overline{%
g}[\pounds _ug]-\frac 1e\cdot (\pounds _ug)(u,u)\},  \label{5.138}
\end{equation}
\begin{equation}
\nabla _u(d\omega )=0,  \label{5.139}
\end{equation}
\begin{equation}
\pounds _u(d\omega )=[\theta +\frac 1{2\cdot e}\cdot (\pounds
_ug)(u,u)]\cdot d\omega \text{ .}  \label{5.140}
\end{equation}

If the additional condition $g(u,u)=e=\,$const. is fulfilled, then $\pounds
_u[g(u,u)]=0=(\pounds _ug)(u,u)$ and 
\[
\pounds _u(d\omega )=\theta \cdot d\omega \text{ .} 
\]

{\it Special case}: Metric transports ($\nabla _ug=0$) and isometric
draggings-along (motions) ($\pounds _ug=0$): 
\begin{equation}
\theta =0\text{ , }\nabla _u(d\omega )=0\text{ , }\pounds _u(d\omega )=0%
\text{ . }  \label{5.141}
\end{equation}

At the same time, 
\begin{equation}
\begin{array}{c}
\pounds _u(d\omega )=\frac 12\cdot \overline{g}[\pounds _ug]\cdot d\omega
=-\frac 12\cdot g[\pounds _u\overline{g}]\cdot d\omega = \\ 
=\{\frac 1{2\cdot e}\cdot (\pounds _ug)(u,u)-\frac 12\cdot h_u[\pounds _u%
\overline{g}]\}\cdot d\omega = \\ 
=\{\theta -\frac 12\cdot h_u[\nabla _u\overline{g}]+\frac 1{2\cdot e}\cdot
(\pounds _ug)(u,u)\}\cdot d\omega \text{ .}
\end{array}
\label{5.142}
\end{equation}

After introducing the abbreviations 
\begin{equation}
_l\theta _u=\frac 12\cdot \overline{g}[\pounds _ug]\text{ , \thinspace
\thinspace \thinspace \thinspace \thinspace \thinspace \thinspace \thinspace
\thinspace \thinspace \thinspace \thinspace \thinspace \thinspace }_c\theta
_u=\frac 12\cdot \overline{g}[\nabla _ug]\text{ ,}  \label{5.143}
\end{equation}
\begin{equation}
_r\theta _u=\frac 1{2\cdot e}\cdot (\pounds _ug-\nabla _ug)(u,u)\text{ ,}
\label{5.144}
\end{equation}

$\theta $, $\nabla _u(d\omega )$, $\pounds _u(d\omega )$ and $\pounds
_ug-\nabla _ug$ can be written in the form 
\begin{equation}
\theta =\,_l\theta _u-\,_c\theta _u-\,_r\theta _u\text{ ,}  \label{5.145}
\end{equation}
\begin{equation}
\nabla _u(d\omega )=\,_c\theta _u\cdot d\omega =(-\theta +\,_l\theta
_u-\,_r\theta _u)\cdot d\omega \text{ ,}  \label{5.146}
\end{equation}
\begin{equation}
\pounds _u(d\omega )=\,_l\theta _u\cdot d\omega =(\theta +\,_c\theta
_u+\,_r\theta _u)\cdot d\omega \text{ ,}  \label{5.147}
\end{equation}
\begin{equation}
\pounds _u(d\omega )-\nabla _u(d\omega )=(_l\theta _u-\,_c\theta _u)\cdot
d\omega \text{ .}  \label{5.148}
\end{equation}

The variation of the invariant volume element along a contravariant vector
field, orthogonal to the contravariant vector field $u$ can be found by
means of the projections of a contravariant vector field $\xi $ along $u$%
\[
\xi =\frac le\cdot u+\xi _{\perp }\text{ , \thinspace \thinspace \thinspace
\thinspace \thinspace \thinspace \thinspace \thinspace \thinspace \thinspace
\thinspace }\nabla _\xi =\frac le\cdot \nabla _u+\nabla _{\xi _{\perp }}%
\text{ ,} 
\]
\begin{equation}
\nabla _\xi (d\omega )=\frac le\cdot \nabla _u(d\omega )+\nabla _{\xi
_{\perp }}(d\omega )\text{ ,}  \label{5.149}
\end{equation}
\begin{equation}
\nabla _{\xi _{\bot }}(d\omega )=\frac 12\cdot \overline{g}[\nabla _{\xi
_{\perp }}g]\cdot d\omega \text{ .}  \label{5.150}
\end{equation}

By means of the relations (\ref{5.137}) and (\ref{5.142}) the following
propositions can be proved:

\begin{proposition}
The necessary and sufficient condition for the existence of the covariant
derivative $\nabla _u(d\omega )$ of the invariant volume element $d\omega $
in the form 
\begin{equation}
\nabla _u(d\omega )=-\theta \cdot d\omega  \label{5.151}
\end{equation}
is the condition 
\begin{equation}
h_u[\pounds _u\overline{g}]=\frac 1e\cdot (\nabla _ug)(u,u)\text{ .}
\label{5.152}
\end{equation}
\end{proposition}

Proof: 1. Sufficiency. From (\ref{5.152}) and (\ref{5.137}) 
\[
\nabla _u(d\omega )=\{-\theta -\frac 12\cdot h_u[\pounds _u\overline{g}%
]+\frac 1{2e}\cdot (\nabla _ug)(u,u)\}\cdot d\omega \text{ ,} 
\]

\noindent it follows 
\[
\nabla _u(d\omega )=-\theta \cdot d\omega \text{ .} 
\]

2. Necessity. From (\ref{5.151}) and (\ref{5.137}), it follows 
\[
\{-h_u[\pounds _u\overline{g}]+\frac 1e\cdot (\nabla _ug)(u,u)\}\cdot
d\omega =0\text{ ,} 
\]

\noindent from where, for $d\omega \neq 0$, (\ref{5.152}) follows.

\begin{proposition}
The necessary and sufficient condition for the existence of the Lie
derivative $\pounds _u(d\omega )$ of the invariant volume element $d\omega $
in the form 
\begin{equation}
\pounds _u(d\omega )=\theta \cdot d\omega  \label{5.153}
\end{equation}
is the condition 
\begin{equation}
h_u[\nabla _u\overline{g}]=\frac 1e\cdot (\pounds _ug)(u,u)\text{ .}
\label{5.154}
\end{equation}
\end{proposition}

Proof: 1. Sufficiency. From (\ref{5.154}) and (\ref{5.142}) 
\[
\pounds _u(d\omega )=\{\theta -\frac 12\cdot h_u[\nabla _u\overline{g}%
]+\frac 1{2\cdot e}\cdot (\pounds _ug)(u,u)\}\cdot d\omega 
\]

\noindent it follows 
\[
\pounds _u(d\omega )=\theta \cdot d\omega \text{ .} 
\]

2. Necessity. From (\ref{5.153}) and (\ref{5.142}), it follows 
\[
\{-h_u[\nabla _u\overline{g}]+\frac 1e\cdot (\pounds _ug)(u,u)\}\cdot
d\omega =0\text{ , } 
\]

\noindent from where, for $d\omega \neq 0$, (\ref{5.154}) follows.

{\it Special case}: Quasi-projective non-metric transports 
\[
\begin{array}{c}
\nabla _ug=\frac 12\cdot [p\otimes g(u)+g(u)\otimes p]\text{ ,} \\ 
\nabla _u\overline{g}=\frac 12\cdot (v\otimes u+u\otimes v)\text{ , }v=-%
\overline{g}(p)\text{ ,}
\end{array}
\]
\begin{equation}
h_u[\nabla _u\overline{g}]=0\text{ ,}  \label{5.155}
\end{equation}
\begin{equation}
(\nabla _ug)(u,u)=e\cdot p(u)\text{ ,\thinspace \thinspace \thinspace
\thinspace \thinspace \thinspace \thinspace \thinspace \thinspace \thinspace
\thinspace \thinspace \thinspace \thinspace \thinspace \thinspace }%
e=g(u,u)\neq 0\text{ ,}  \label{5.156}
\end{equation}
\begin{equation}
\nabla _u(d\omega )=\frac 12\cdot p(u)\cdot d\omega =\frac 1{2\cdot e}\cdot
(\nabla _ug)(u,u)\cdot d\omega \text{ ,}  \label{5.157}
\end{equation}
\begin{equation}
\pounds _u(d\omega )=[\theta +\frac 1{2\cdot e}\cdot (\pounds
_ug)(u,u)]\cdot d\omega \text{ .}  \label{5.158}
\end{equation}

The variation of the invariant volume element and its preservation is
connected with the structures of a Lagrangian theory of tensor fields over $(%
\overline{L}_n,g)$-spaces \cite{Manoff-01}.

\subsection{Rotation (vortex) velocity}

The tensor $\omega $ is called rotation (vortex) velocity tensor (rotation
velocity, vortex velocity, rotation, vortex). It does not change the length
of a vector field $\xi _{\perp }$ and, therefore, $\omega $ changes only the
direction of $\xi _{\perp }$, causing its rotation in the $n-1$ dimensional
sub space, orthogonal to $u$ [because of the relation $[\omega (\xi _{\perp
})](u)=(u)(\omega )(\xi _{\perp })=\omega (u,\xi _{\perp })=0$]. By the use
of the Levi-Civita symbols or the star operator $*$ we can define the
corresponding rotation velocity vector.

\subsubsection{Definition of the rotation (vortex) velocity vector}

For a differentiable manifold $M$ with $\dim M=4$, a vector $\overline{%
\omega }$ corresponding to the rotation velocity tensor $\omega $ could be
defined by the use of $\omega $, $u$, and the Hodge (star) operator $*$%
\begin{equation}
\overline{\omega }:=\overline{g}(*\,\,(g(u)\wedge \omega ))\text{
,\thinspace \thinspace \thinspace \thinspace \thinspace \thinspace
\thinspace \thinspace \thinspace \thinspace \thinspace }\overline{\omega }=%
\overline{\omega }^i\cdot \partial _i\text{ \thinspace \thinspace \thinspace
\thinspace , \thinspace \thinspace \thinspace \thinspace \thinspace
\thinspace \thinspace }\overline{\omega }\in T(M)\text{ .}  \label{5.159}
\end{equation}

Let us find now the explicitly form of the rotation (vortex) velocity vector 
$\overline{\omega }$. For this purpose, we should write $\omega $,\thinspace 
$g(u)$, and $\overline{g}$ in a co-ordinate (or non-co-ordinate) basis 
\begin{equation}
\overline{g}=g^{ij}\cdot \partial _i.\partial _j\text{ ,\thinspace
\thinspace \thinspace \thinspace \thinspace \thinspace }g(u)=g_{i\overline{k}%
}\cdot u^k\cdot dx^i\text{ ,\thinspace \thinspace \thinspace \thinspace
\thinspace \thinspace \thinspace }\omega =\omega _{ij}\cdot dx^i\wedge dx^j%
\text{ .}  \label{5.160}
\end{equation}

Then $g(u)\wedge \omega $ will have the form 
\begin{eqnarray}
g(u)\wedge \omega &=&g_{im}\cdot u^{\overline{m}}\cdot \omega _{jk}\cdot
dx^i\wedge dx^j\wedge dx^k=  \nonumber \\
&=&\,_aA_{[ijk]}\cdot dx^i\wedge dx^j\wedge dx^k=\,_aA\,\text{ ,}
\label{5.161} \\
_aA &:&=g(u)\wedge \omega \text{ \thinspace \thinspace ,}  \nonumber
\end{eqnarray}

\noindent where 
\begin{eqnarray}
_aA_{[ijk]} &=&u^{\overline{m}}\cdot g_{m[i}\omega _{jk]}\text{ ,}
\label{5.162a} \\
_a\overline{A} &=&A^{[ijk]}\cdot \partial _i\wedge \partial _j\wedge
\partial _k\text{ \thinspace \thinspace ,}  \nonumber \\
A^{[ijk]} &=&g^{i\overline{l}}\cdot g^{j\overline{m}}\cdot g^{k\overline{n}%
}\cdot \,_aA_{[lmn]}=  \label{5.162b} \\
&=&g^{i\overline{l}}\cdot g^{j\overline{m}}\cdot g^{k\overline{n}}\cdot u^{%
\overline{r}}\cdot g_{r[l}\omega _{mn]}\text{ \thinspace \thinspace
\thinspace .}  \nonumber
\end{eqnarray}

For $\dim M=4$, $k=3$, we have 
\begin{eqnarray}
\ast \,(_aA) &=&[*\,(_aA)]_s\cdot dx^s=\frac 1{3!}\cdot \sqrt{-d_g}\cdot
\varepsilon _{sijk}\cdot g^{\overline{i}\overline{l}}\cdot g^{\overline{j}%
\overline{m}}\cdot g^{\overline{k}\overline{n}}\cdot u^{\overline{r}}\cdot
g_{r[l}\omega _{mn]}\cdot dx^s\text{ ,}  \label{5.163a} \\
g_{r[l}\omega _{mn]} &=&\frac 1{3!}\cdot (g_{rl}\cdot \omega
_{mn}+g_{rn}\cdot \omega _{ml}+g_{rn}\cdot \omega _{lm}+g_{rm}\cdot \omega
_{nl}-g_{rl}\cdot \omega _{nm}-g_{rm}\cdot \omega _{nl})=  \nonumber \\
&=&\frac 1{3!}\cdot 2\cdot g_{rl}\cdot \omega _{mn}=\frac 13\cdot
g_{rl}\cdot \omega _{mn}\text{ \thinspace \thinspace \thinspace ,\thinspace
\thinspace \thinspace \thinspace \thinspace \thinspace \thinspace \thinspace
\thinspace \thinspace }\omega _{lm}=-\omega _{ml}\text{ ,}  \label{5.163b}
\end{eqnarray}
\begin{eqnarray}
g^{\overline{i}\overline{l}}\cdot g^{\overline{j}\overline{m}}\cdot g^{%
\overline{k}\overline{n}}\cdot u^{\overline{r}}\cdot g_{r[l}\omega _{mn]}
&=&\frac 13\cdot g^{\overline{i}\overline{l}}\cdot g_{rl}\cdot g^{\overline{j%
}\overline{m}}\cdot g^{\overline{k}\overline{n}}\cdot u^{\overline{r}}\cdot
\omega _{mn}=  \nonumber \\
&=&\frac 13\cdot g_r^i\cdot u^{\overline{r}}\cdot g^{\overline{j}\overline{m}%
}\cdot g^{\overline{k}\overline{n}}\cdot \omega _{mn}=  \nonumber \\
&=&\frac 13\cdot u^{\overline{i}}\cdot \omega ^{\overline{j}\overline{k}}%
\text{ ,}  \label{5.164}
\end{eqnarray}
\begin{eqnarray}
\ast \,(g(u)\wedge \omega ) &=&*\,(_aA)=\frac 1{3!}\cdot \sqrt{-d_g}\cdot
\varepsilon _{sijk}\cdot \frac 13\cdot u^{\overline{i}}\cdot \omega ^{%
\overline{j}\overline{k}}\cdot dx^s=  \nonumber \\
&=&\frac 1{18}\cdot \sqrt{-d_g}\cdot \varepsilon _{ijkl}\cdot u^{\overline{j}%
}\cdot \omega ^{\overline{k}\overline{l}}\cdot dx^i=\overline{\omega }%
_i\cdot dx^i\text{ ,}  \label{6.165a} \\
\overline{\omega }_i &=&\frac 1{18}\cdot \sqrt{-d_g}\cdot \varepsilon
_{ijkl}\cdot u^{\overline{j}}\cdot \omega ^{\overline{k}\overline{l}}\text{
\thinspace ,}  \label{5.165b}
\end{eqnarray}
\begin{eqnarray}
\overline{\omega } &=&\overline{g}(*\,(g(u)\wedge \omega ))=g^{i\overline{m}%
}\cdot \overline{\omega }_m\cdot \partial _i=\overline{\omega }^i\cdot
\partial _i\text{ ,}  \label{5.166} \\
&&  \nonumber \\
\overline{\omega }^i &=&g^{i\overline{m}}\cdot \overline{\omega }_m\cdot
=\frac 1{18}\cdot \sqrt{-d_g}\cdot \varepsilon _{mjkl}\cdot g^{i\overline{m}%
}\cdot u^{\overline{j}}\cdot \omega ^{\overline{k}\overline{l}}\text{
\thinspace \thinspace \thinspace .}  \nonumber
\end{eqnarray}

All further results for $\overline{\omega }$ are specialized for $\dim M=4$.

\subsubsection{Properties of the rotation (vortex) velocity vector}

By the use of the above expressions, we can find some important relations
and properties of the rotation (vortex) vector.

1. The rotation (vortex) velocity vector $\overline{\omega }$ is orthogonal
to the vector $u$ 
\begin{equation}
g(u,\overline{\omega })=g_{\overline{i}\overline{j}}\cdot u^i\cdot \overline{%
\omega }^j=0\text{\thinspace \thinspace \thinspace \thinspace \thinspace
\thinspace .}  \label{5.167}
\end{equation}

Proof: 
\begin{eqnarray}
g(u,\overline{\omega }) &=&g_{\overline{i}\overline{j}}\cdot u^i\cdot 
\overline{\omega }^j=g_{\overline{i}\overline{j}}\cdot u^i\cdot \frac
1{18}\cdot \sqrt{-d_g}\cdot \varepsilon _{mnkl}\cdot g^{j\overline{m}}\cdot
u^{\overline{n}}\cdot \omega ^{\overline{k}\overline{l}}=  \nonumber \\
&=&\frac 1{18}\cdot \sqrt{-d_g}\cdot g_{ij}\cdot g^{\overline{j}\overline{m}%
}\cdot u^{\overline{i}}\cdot u^{\overline{n}}\cdot \varepsilon _{mnkl}\cdot
\omega ^{\overline{k}\overline{l}}=  \nonumber \\
&=&\frac 1{18}\cdot \sqrt{-d_g}\cdot g_i^m\cdot \varepsilon _{mnkl}\cdot u^{%
\overline{i}}\cdot u^{\overline{n}}\cdot \omega ^{\overline{k}\overline{l}}=
\nonumber \\
&=&\frac 1{18}\cdot \sqrt{-d_g}\cdot \varepsilon _{mnkl}\cdot u^{\overline{m}%
}\cdot u^{\overline{n}}\cdot \omega ^{\overline{k}\overline{l}}=0\text{ .}
\label{5.167a}
\end{eqnarray}

2. Representation of the rotation velocity tensor $\omega $ by means of the
rotation velocity vector $\overline{\omega }$.

We can use the definition of the rotation velocity vector $\overline{\omega }
$ to express the rotation velocity tensor $\omega $. From $\overline{\omega }%
=\overline{g}(*\,(g(u)\wedge \omega ))$ we obtain 
\begin{equation}
g(\overline{\omega })=*\,(g(u)\wedge \omega )\text{ \thinspace \thinspace
\thinspace \thinspace \thinspace ,\thinspace \thinspace \thinspace
\thinspace \thinspace \thinspace \thinspace \thinspace \thinspace \thinspace
\thinspace \thinspace \thinspace \thinspace }*\,(g(\overline{\omega }%
))=*\,(*\,(g(u)\wedge \omega ))\text{ ,}  \label{5.168}
\end{equation}
\begin{eqnarray}
\ast \,(g(\overline{\omega })) &=&\varepsilon \cdot (-1)^{3\cdot (4-3)}\cdot 
\frac{4!}{(4-3)!3!}\cdot g(u)\wedge \omega =  \nonumber \\
&=&\varepsilon \cdot (-1)\cdot 4\cdot g(u)\wedge \omega \text{ }=  \nonumber
\\
&=&-4\cdot \varepsilon \cdot g(u)\wedge \omega \text{ \thinspace \thinspace
\thinspace \thinspace .}  \label{5.169}
\end{eqnarray}

On the other side the following relations are valid

(a) $S(u,g(u))=g(u,u)=e$.

Proof:

\begin{eqnarray}
S(u,g(u)\wedge \omega ) &=&S(u,g(u))\wedge \omega +(-1)^1\cdot g(u)\cdot
S(u,\omega )\text{ \thinspace ,}  \label{5.170a} \\
S(u,g(u)) &=&S(u^i\cdot \partial _i,\,g_{kl}\cdot u^{\overline{l}}\cdot
dx^k)=  \nonumber \\
&=&u^i\cdot g_{kl}\cdot u^{\overline{l}}\cdot S(\partial _i,dx^k)=u^i\cdot
g_{kl}\cdot u^{\overline{l}}\cdot f^k\,_i=  \nonumber \\
&=&g_{kl}\cdot u^{\overline{k}}\cdot u^{\overline{l}}=g(u,u)=e\text{
\thinspace ,}  \label{5.170b}
\end{eqnarray}

(b) $S(u,\omega )=-\omega (u)=(u)(\omega )=0$.

Proof:

\begin{eqnarray}
S(u,\omega ) &=&S(u^i\cdot \partial _i,\omega _{kl}\cdot dx^k\wedge dx^l)= 
\nonumber \\
&=&u^i\cdot \omega _{kl}\cdot \frac 12\cdot S(\partial _i,dx^k\otimes
dx^l-dx^l\otimes dx^k)=  \nonumber \\
&=&\frac 12\cdot u^i\cdot \omega _{kl}\cdot (f^k\,_i\cdot dx^l-f^l\,_i\cdot
dx^k)=  \nonumber \\
&=&\frac 12\cdot u^i\cdot (\omega _{kl}\cdot f^k\,_i\cdot dx^l-\omega
_{kl}\cdot f^l\,_i\cdot dx^k)=  \nonumber \\
&=&\frac 12\cdot u^i\cdot (\omega _{lk}\cdot f^l\,_i\cdot dx^k-\omega
_{kl}\cdot f^l\,_i\cdot dx^k)=  \nonumber \\
&=&\frac 12\cdot u^i\cdot f^l\,_i\cdot (\omega _{lk}-\omega _{kl})\cdot
dx^k=-u^i\cdot f^l\,_i\cdot \omega _{kl}\cdot dx^k=  \nonumber \\
&=&-\omega _{kl}\cdot u^{\overline{l}}\cdot dx^k=-\omega (u)=(u)(\omega )=0%
\text{ ,}  \label{5.171}
\end{eqnarray}

(c) $S(u,g(u)\wedge \omega )=e\cdot \omega $.

(d) $S(u,\,*\,[g(\overline{\omega })])=-\,4\cdot \varepsilon \cdot e\cdot
\omega $.

Proof: 
\begin{eqnarray}
S(u,\,*\,[g(\overline{\omega })]) &=&S(u,\varepsilon \cdot (-1)^{3\cdot
(4-3)}\cdot \frac{4!}{(4-3)!3!}\cdot g(u)\wedge \omega )=  \nonumber \\
&=&-\,4\cdot \varepsilon \cdot S(u,g(u)\wedge \omega )=-\,4\cdot \varepsilon
\cdot e\cdot \omega \text{ .}  \label{5.172}
\end{eqnarray}

From the last expression, it follows that 
\begin{equation}
\omega =-\,\frac 1{4\cdot \varepsilon \cdot e}\cdot S(u,\,*\,[g(\overline{%
\omega })])=\omega _{kl}\cdot dx^k\wedge dx^l\text{ .}  \label{5.173}
\end{equation}

On the other side, we have $*\,[g(\overline{\omega })]=\sqrt{-d_g}\cdot
\varepsilon _{ijkl}\cdot \overline{\omega }^{\overline{l}}\cdot dx^i\wedge
dx^j\wedge dx^k$.

Proof:

\begin{eqnarray}
\ast \,[g(\overline{\omega })] &=&\sqrt{-d_g}\cdot \varepsilon _{ijkl}\cdot
g^{\overline{l}\overline{m}}\cdot g_{m\overline{n}}\cdot \overline{\omega }%
^n\cdot dx^i\wedge dx^j\wedge dx^k=  \nonumber \\
&=&\sqrt{-d_g}\cdot \varepsilon _{ijkl}\cdot g^{\overline{l}\overline{m}%
}\cdot g_{mn}\cdot \overline{\omega }^{\overline{n}}\cdot dx^i\wedge
dx^j\wedge dx^k=  \nonumber \\
&=&\sqrt{-d_g}\cdot \varepsilon _{ijkl}\cdot g_n^l\cdot \overline{\omega }^{%
\overline{n}}\cdot dx^i\wedge dx^j\wedge dx^k=  \nonumber \\
&=&\sqrt{-d_g}\cdot \varepsilon _{ijkl}\cdot \overline{\omega }^{\overline{l}%
}\cdot dx^i\wedge dx^j\wedge dx^k\text{ .}  \label{5.174}
\end{eqnarray}

Then 
\begin{equation}
S(u,\,*\,[g(\overline{\omega })])=3\cdot \sqrt{-d_g}\cdot \varepsilon
_{ijkl}\cdot u^{\overline{i}}\cdot \overline{\omega }^{\overline{l}}\cdot
dx^j\wedge dx^k\text{ .}  \label{5.175}
\end{equation}

Proof:

\begin{eqnarray}
S(u,\,*\,[g(\overline{\omega })]) &=&S(u^m\cdot \partial _m,\sqrt{-d_g}\cdot
\varepsilon _{ijkl}\cdot \overline{\omega }^{\overline{l}}\cdot dx^i\wedge
dx^j\wedge dx^k)=  \nonumber \\
&=&\sqrt{-d_g}\cdot \varepsilon _{ijkl}\cdot \overline{\omega }^{\overline{l}%
}\cdot u^m\cdot S(\partial _m,dx^i\wedge dx^j\wedge dx^k)\text{ ,}
\label{5.175a}
\end{eqnarray}

$\blacktriangle \,$For $S(\partial _m,dx^i\wedge dx^j\wedge dx^k)$ we obtain

\begin{equation}
S(\partial _m,dx^i\wedge dx^j\wedge dx^k)=f^i\,_m\cdot dx^j\wedge
dx^k-f^j\,_m\cdot dx^i\wedge dx^k+f^k\,_m\cdot dx^i\wedge dx^j\text{ .}
\label{5.176}
\end{equation}

Proof: 
\[
S(\partial _m,dx^i\wedge dx^j\wedge dx^k)=f^i\,_m\cdot dx^j\wedge
dx^k+(-1)\cdot dx^i\wedge S(\partial _m,dx^j\wedge dx^k)= 
\]
\begin{eqnarray*}
&=&f^i\,_m\cdot dx^j\wedge dx^k-dx^i\wedge (f^j\,_m\cdot dx^k-dx^j\cdot
S(\partial _m,dx^k))= \\
&=&f^i\,_m\cdot dx^j\wedge dx^k-f^j\,_m\cdot dx^i\wedge dx^k+f^k\,_m\cdot
dx^i\wedge dx^j\text{ .\thinspace \thinspace \thinspace }\blacktriangle
\end{eqnarray*}

Therefore, 
\begin{eqnarray*}
S(u,\,*\,[g(\overline{\omega })]) &=&\sqrt{-d_g}\cdot \varepsilon
_{ijkl}\cdot \overline{\omega }^{\overline{l}}\cdot u^m\cdot \\
&&\cdot (f^i\,_m\cdot dx^j\wedge dx^k-f^j\,_m\cdot dx^i\wedge
dx^k+f^k\,_m\cdot dx^i\wedge dx^j)
\end{eqnarray*}
\begin{eqnarray*}
&=&\sqrt{-d_g}\cdot (\varepsilon _{ijkl}\cdot \overline{\omega }^{\overline{l%
}}\cdot u^m\cdot f^i\,_m\cdot dx^j\wedge dx^k- \\
&&-\varepsilon _{ijkl}\cdot \overline{\omega }^{\overline{l}}\cdot u^m\cdot
f^j\,_m\cdot dx^i\wedge dx^k+ \\
&&+\varepsilon _{ijkl}\cdot \overline{\omega }^{\overline{l}}\cdot u^m\cdot
f^k\,_m\cdot dx^i\wedge dx^j)\text{ ,}
\end{eqnarray*}
\begin{eqnarray}
S(u,\,*\,[g(\overline{\omega })]) &=&\sqrt{-d_g}\cdot u^m\cdot \overline{%
\omega }^{\overline{l}}\cdot f^i\,_m\cdot (\varepsilon _{ijkl}-\varepsilon
_{jikl}+\varepsilon _{jkil})\cdot dx^j\wedge dx^k=  \nonumber \\
&=&3\cdot \sqrt{-d_g}\cdot \varepsilon _{ijkl}\cdot u^{\overline{i}}\cdot 
\overline{\omega }^{\overline{l}}\cdot dx^j\wedge dx^k\text{.}  \label{5.177}
\end{eqnarray}

For the rotation (vortex) velocity $\omega $, we obtain 
\begin{eqnarray}
\omega &=&-\,\frac 1{4\cdot \varepsilon \cdot e}\cdot S(u,\,*\,[g(\overline{%
\omega })])=\omega _{jk}\cdot dx^j\wedge dx^k=  \nonumber \\
&=&-\,\frac 3{4\cdot \varepsilon \cdot e}\cdot \sqrt{-d_g}\cdot \varepsilon
_{ijkl}\cdot u^{\overline{i}}\cdot \overline{\omega }^{\overline{l}}\cdot
dx^j\wedge dx^k\text{ \thinspace \thinspace \thinspace ,}  \label{5.178}
\end{eqnarray}
\begin{eqnarray}
\omega _{jk} &=&-\,\frac 3{4\cdot \varepsilon \cdot e}\cdot \sqrt{-d_g}\cdot
\varepsilon _{ijkl}\cdot u^{\overline{i}}\cdot \overline{\omega }^{\overline{%
l}}=  \nonumber \\
&=&-\,\frac 3{4\cdot \varepsilon \cdot e}\cdot \sqrt{-d_g}\cdot \varepsilon
_{iljk}\cdot u^{\overline{i}}\cdot \overline{\omega }^{\overline{l}}\text{%
\thinspace \thinspace \thinspace \thinspace \thinspace .}  \label{5.178a}
\end{eqnarray}

3. The rotation velocity tensor $\omega $ is orthogonal to $u$ and $%
\overline{\omega }$. The first property $\omega (u)=-(u)(\omega )=0$ follows
from the construction of $\omega $. The second property $\omega (\overline{%
\omega })=-(\overline{\omega })(\omega )=0$ can be proved.

Proof: 
\begin{eqnarray}
\omega (\overline{\omega }) &=&\frac 12\cdot \omega _{jk}\cdot (dx^j\otimes
dx^k-dx^k\otimes dx^j)(\overline{\omega }^l\cdot \partial _l)=  \nonumber \\
&=&\frac 12\cdot \omega _{jk}\cdot \overline{\omega }^l\cdot (f^k\,_l\cdot
dx^j-f^j\,_l\cdot dx^k)=  \nonumber \\
&=&\frac 12\cdot (\omega _{jk}\cdot \overline{\omega }^l\cdot f^k\,_l\cdot
dx^j-\omega _{jk}\cdot \overline{\omega }^l\cdot f^j\,_l\cdot dx^k)= 
\nonumber \\
&=&\frac 12\cdot (\omega _{jk}\cdot \overline{\omega }^{\overline{k}}\cdot
dx^j-\omega _{jk}\cdot \overline{\omega }^{\overline{j}}\cdot dx^k)= 
\nonumber \\
&=&\frac 12\cdot (\omega _{jk}\cdot \overline{\omega }^{\overline{k}}\cdot
dx^j-\omega _{kj}\cdot \overline{\omega }^{\overline{k}}\cdot dx^j)= 
\nonumber \\
&=&\frac 12\cdot (\omega _{jk}+\omega _{jk})\cdot \overline{\omega }^{%
\overline{k}}\cdot dx^j=  \nonumber \\
&=&\omega _{jk}\cdot \overline{\omega }^{\overline{k}}\cdot dx^j=  \nonumber
\\
&=&-\,\frac 3{4\cdot \varepsilon \cdot e}\cdot \sqrt{-d_g}\cdot \varepsilon
_{iljk}\cdot u^{\overline{i}}\cdot \overline{\omega }^{\overline{l}}\cdot 
\overline{\omega }^{\overline{k}}\cdot dx^j=0\text{ .}  \label{5.179}
\end{eqnarray}

Therefore, the rotation (vortex) velocity vector $\overline{\omega }$ is
orthogonal to the velocity vector $u$ and to the rotation velocity tensor $%
\omega $.

{\it Special case}: $(\overline{L}_n,g)$-space admitting the conditions $%
\sigma =0$, $\theta =0$.

Since $g(u,\overline{\omega })=0$, we can chose $u$ and $\overline{\omega }$
as tangent vectors to the two of the co-ordinate lines, i.e. $u$ and $%
\overline{\omega }$ could fulfil the condition $\pounds _u\overline{\omega }%
=0$. Then 
\begin{equation}
h_u(\nabla _u\overline{\omega })=\omega (\overline{\omega })=0\text{
,\thinspace \thinspace \thinspace \thinspace \thinspace \thinspace
\thinspace \thinspace }_{rel}v=\overline{g}[h_u(\nabla _u\overline{\omega })%
]=0\text{ \thinspace \thinspace \thinspace .}  \label{5.180}
\end{equation}

On the other side, 
\begin{equation}
\nabla _u\overline{\omega }=\frac 1e\cdot g(u,\nabla _u\overline{\omega }%
)\cdot u+\,_{rel}v=\frac 1e\cdot g(u,\nabla _u\overline{\omega })\cdot u%
\text{ ,}  \label{5.181}
\end{equation}
\begin{eqnarray}
g(u,\nabla _u\overline{\omega }) &=&\nabla _u[g(u,\overline{\omega }%
)]-g(\nabla _uu,\overline{\omega })-(\nabla _ug)(u,\overline{\omega })\text{
,\thinspace \thinspace \thinspace \thinspace \thinspace \thinspace
\thinspace }g(u,\overline{\omega })=0\text{ \thinspace ,\thinspace
\thinspace \thinspace }  \nonumber \\
\nabla _u[g(u,\overline{\omega })] &=&u[g(u,\overline{\omega })]=0\text{ ,}
\label{5.182}
\end{eqnarray}
\begin{eqnarray}
\nabla _u\overline{\omega } &=&-\,\frac 1e\cdot [g(\nabla _uu,\overline{%
\omega })+(\nabla _ug)(u,\overline{\omega })]\cdot u\text{ \thinspace
,\thinspace \thinspace \thinspace \thinspace \thinspace \thinspace
\thinspace }\nabla _uu=a\text{ \thinspace \thinspace \thinspace ,}  \nonumber
\\
\text{\thinspace }\nabla _u\overline{\omega } &=&-\,\frac 1e\cdot [g(a,%
\overline{\omega })+(\nabla _ug)(u,\overline{\omega })]\cdot u\text{
\thinspace \thinspace .}  \label{5.183}
\end{eqnarray}

The rotation vector $\overline{\omega }$ does not change in directions,
orthogonal to $u$. Its change along $u$ is collinear to $u$ and depends on $%
\nabla _ug$ and on the acceleration $a$.

{\it Special case}: Weyl's spaces. In Weyl's spaces (with or without
torsion), where $\nabla _\xi g=(1/n)\cdot Q_\xi \cdot g$ for $\forall \xi
\in T(M)$ and $(\nabla _ug)(u,\overline{\omega })=(1/n)\cdot Q_\xi \cdot g(u,%
\overline{\omega })=0$, the rotation vector $\overline{\omega }$ does not
change along $u$ if $u$ is a tangent vector on an auto-parallel curve, i.e.
if $\nabla _uu=a=0$, 
\begin{equation}
\nabla _u\overline{\omega }=0\text{ ,\thinspace \thinspace \thinspace
\thinspace \thinspace \thinspace \thinspace \thinspace \thinspace }a:=0\text{
\thinspace \thinspace ,}  \label{5.184}
\end{equation}

\noindent i.e. $\overline{\omega }$ is transported parallel along $u$.

4. Change of the velocity vector $u$ along the rotation (vortex) velocity
vector $\overline{\omega }$.

The change of the velocity $u$ along $\overline{\omega }$ could be
represented in the form 
\begin{eqnarray}
\nabla _{\overline{\omega }}u &=&\overline{g}(_sE)(\overline{\omega })+%
\overline{g}(S)(\overline{\omega })+\frac 1{n-1}\cdot \overline{g}[E]\cdot 
\overline{g}[h_u(\overline{\omega })]+  \nonumber \\
&&+\,\frac 1{2\cdot e}\cdot [\overline{\omega }e-(\nabla _{\overline{\omega }%
}g)(u,u)]\cdot u\text{ .}  \label{5.185}
\end{eqnarray}

Since $\overline{g}[h_u(\overline{\omega })]=\overline{\omega }$ and $%
\overline{g}[E]=\theta _o$, we can also write 
\begin{eqnarray}
\nabla _{\overline{\omega }}u &=&\overline{g}(_sE)(\overline{\omega })+%
\overline{g}(S)(\overline{\omega })+\frac 1{n-1}\cdot \theta _o\cdot 
\overline{\omega }+  \nonumber \\
&&+\,\frac 1{2\cdot e}\cdot [\overline{\omega }e-(\nabla _{\overline{\omega }%
}g)(u,u)]\cdot u\text{ .}  \label{5.186}
\end{eqnarray}

{\it Special case}: Shear-free ($\sigma =\,_sE=0$) and expansion-free ($%
\overline{g}[E]=\theta =0$) $\overline{V}_n$-spaces with $\omega =S$ and $%
\nabla _\xi g=0$ for $\forall \xi \in T(M)$, $e:=$const.$\neq 0$. For these
types of spaces $\nabla _{\overline{\omega }}u=0$. Therefore, in $\overline{U%
}_n$- and $\overline{V}_n$-spaces the velocity $u$ does not change along the
rotation (vortex) velocity vector $\overline{\omega }$. This means that all
particles (material points, material elements) lying on an axis, collinear
to $\overline{\omega }$, have one and the same velocity which remains
unchanged along this axis 
\begin{equation}
\nabla _{\overline{\omega }}[g(u,u)]=\overline{\omega }e=(\nabla _{\overline{%
\omega }}g)(u,u)+2\cdot g(\nabla _{\overline{\omega }}u,u)=0\text{
\thinspace \thinspace \thinspace \thinspace .}  \label{5.187}
\end{equation}

\noindent and $u$ is parallel transported along $u$. This fact is related to
the physical interpretation of $\overline{\omega }$ as a rotation axis.

In $(\overline{L}_n,g)$- and $(L_n,g)$-spaces, the rotation velocity vector $%
\overline{\omega }$ changes in general along the velocity $u$. The same is
valid for the velocity $u$ along the rotation (vortex) vector $\overline{%
\omega }$. This means that every material point in the flow could have its
own rotation (vortex) velocity vector $\overline{\omega }$ different from
that of the other material points in its neighborhoods and even different
from the rotation vector of the points lying on the $\overline{\omega }$
itself. In general, we have the relation 
\begin{equation}
\pounds _u\overline{\omega }=\nabla _u\overline{\omega }-\nabla _{\overline{%
\omega }}u-T(u,\overline{\omega })\text{ .}  \label{5.188}
\end{equation}

On the other side, we can calculate the change of the velocity vector $u$ by
its transport along the rotation (vortex) velocity vector $\overline{\omega }
$. From the relations 
\begin{equation}
\nabla _u\overline{\omega }=\frac 1e\cdot g(u,\nabla _u\overline{\omega }%
)\cdot u+\,_{rel}v\text{ \thinspace ,}  \label{5.189}
\end{equation}
\begin{equation}
g(u,\nabla _u\overline{\omega })=-[g(a,\overline{\omega })+(\nabla _ug)(%
\overline{\omega },u)]\text{ ,}  \label{5.190}
\end{equation}
\begin{eqnarray}
_{rel}v &=&\overline{g}[h_u(\nabla _u\overline{\omega })]=\overline{g}%
(h_u)(\frac le\cdot a-\pounds _{\overline{\omega }}u)+\overline{g}[d(%
\overline{\omega })]=  \nonumber \\
&=&\overline{g}(h_u)(\pounds _u\overline{\omega })+\overline{g}[d(\overline{%
\omega })]\text{\thinspace \thinspace \thinspace \thinspace \thinspace
,\thinspace \thinspace \thinspace \thinspace \thinspace \thinspace
\thinspace \thinspace }l=g(u,\overline{\omega })=0\text{ ,}  \label{5.191}
\end{eqnarray}
\begin{eqnarray}
\nabla _u\overline{\omega } &=&-\frac 1e\cdot [g(a,\overline{\omega }%
)+(\nabla _ug)(\overline{\omega },u)]\cdot u+\overline{g}(h_u)(\pounds _u%
\overline{\omega })+\overline{g}[d(\overline{\omega })]\text{\thinspace }= 
\nonumber \\
&=&-\frac 1e\cdot [g(a,\overline{\omega })+(\nabla _ug)(\overline{\omega }%
,u)]\cdot u+\overline{g}(h_u)(\pounds _u\overline{\omega })+  \nonumber \\
&&+\overline{g}[\sigma +\omega +\frac 1{n-1}\cdot \theta \cdot h_u](%
\overline{\omega })\text{ \thinspace \thinspace \thinspace \thinspace ,
\thinspace \thinspace }  \label{5.192}
\end{eqnarray}
\begin{equation}
\text{\thinspace \thinspace \thinspace }h_u(\overline{\omega })=g(\overline{%
\omega })\text{ , \thinspace \thinspace \thinspace \thinspace \thinspace
\thinspace \thinspace \thinspace }\omega (\overline{\omega })=0\text{
\thinspace \thinspace ,\thinspace \thinspace \thinspace \thinspace
\thinspace \thinspace \thinspace \thinspace \thinspace \thinspace \thinspace 
}h_u(\overline{g})h_u=(h_u)\overline{g}(h_u)=h_u\text{ ,}  \label{5.193}
\end{equation}

\noindent it follows for $\nabla _u\overline{\omega }$%
\begin{eqnarray}
\nabla _u\overline{\omega } &=&-\frac 1e\cdot [g(a,\overline{\omega }%
)+(\nabla _ug)(\overline{\omega },u)]\cdot u+(\pounds _u\overline{\omega }%
)_{\perp }+  \nonumber \\
&&+\overline{g}(\sigma )(\overline{\omega })+\frac 1{n-1}\cdot \theta \cdot 
\overline{\omega }\text{ \thinspace \thinspace \thinspace \thinspace
,\thinspace \thinspace \thinspace \thinspace \thinspace \thinspace }
\label{5.194} \\
\text{\thinspace \thinspace }(\pounds _u\overline{\omega })_{\perp } &:&=%
\overline{g}(h_u)(\pounds _u\overline{\omega })  \nonumber
\end{eqnarray}

If we substitute now the expressions for $\nabla _u\overline{\omega }$ and
for $\nabla _{\overline{\omega }}u$ in $\pounds _u\overline{\omega }=\nabla
_u\overline{\omega }-\nabla _{\overline{\omega }}u-T(u,\overline{\omega })$
then we will obtain the explicit form of the torsion vector $T(u,\overline{%
\omega })$%
\begin{eqnarray}
T(u,\overline{\omega }) &=&-\frac 1e\cdot \{g(a,\overline{\omega })+(\nabla
_ug)(\overline{\omega },u)+g(u,\pounds _u\overline{\omega })+  \nonumber \\
&&+\frac 12\cdot [\overline{\omega }e-(\nabla _{\overline{\omega }%
}g)(u,u)]\}\cdot u-  \nonumber \\
&&-[\overline{g}(\sigma _1)(\overline{\omega })+\overline{g}(\omega _1)(%
\overline{\omega })+\frac 1{n-1}\cdot \theta _1\cdot \overline{\omega }]%
\text{ ,}  \label{5.195}
\end{eqnarray}

\noindent where 
\begin{eqnarray*}
\omega &=&\omega _o-\omega _1\text{ \thinspace \thinspace \thinspace
\thinspace ,\thinspace \thinspace \thinspace \thinspace \thinspace
\thinspace }\omega _o=S\text{ ,\thinspace \thinspace \thinspace \thinspace
\thinspace \thinspace \thinspace \thinspace }\omega _1=Q\text{ ,\thinspace
\thinspace \thinspace \thinspace \thinspace \thinspace \thinspace \thinspace
\thinspace \thinspace \thinspace \thinspace }\omega (\overline{\omega }%
)=0=\omega _o(\overline{\omega })-\omega _1(\overline{\omega })\text{ ,} \\
\sigma &=&\sigma _o-\sigma _1\text{ \thinspace \thinspace \thinspace
\thinspace \thinspace , }\sigma _o=\,_sE\text{ ,\thinspace \thinspace
\thinspace \thinspace \thinspace }\sigma _1=\,_sP\text{ .}
\end{eqnarray*}

{\it Special case}: If $u$ and $\overline{\omega }$ are tangent vectors to
co-ordinate lines in $M$, then $\pounds _u\overline{\omega }=0$ and we can
find a representation of the torsion vector $T(u,\overline{\omega })$ in the
form 
\begin{equation}
T(u,\overline{\omega })=\nabla _u\overline{\omega }-\nabla _{\overline{%
\omega }}u\text{ .}  \label{5.196}
\end{equation}

{\it Special case}: $\overline{U}_n$-spaces: $\nabla _\xi g=0$ for $\forall
\xi \in T(M)$, $n=4$, $\pounds _u\overline{\omega }=0$. 
\begin{equation}
T(u,\overline{\omega })=-\frac 1e\cdot [g(a,\overline{\omega })+\frac
12\cdot \overline{\omega }e]\cdot u-\overline{g}[d_1(\overline{\omega })]%
\text{ \thinspace \thinspace .}  \label{5.197}
\end{equation}

If $T(u,\overline{\omega }):=0$, then 
\begin{equation}
\overline{g}[d_1(\overline{\omega })]=-\frac 1e\cdot [g(a,\overline{\omega }%
)+\frac 12\cdot \overline{\omega }e]\cdot u\text{ \thinspace \thinspace
\thinspace .}  \label{5.198}
\end{equation}

Since $d_1(u)=0$ and $[g(u)](\overline{g})[d_1(\overline{\omega })]=d_1(u,%
\overline{\omega })=0$, we obtain 
\begin{equation}
g(a,\overline{\omega })+\frac 12\cdot \overline{\omega }e=0\,\text{%
\thinspace \thinspace \thinspace \thinspace \thinspace \thinspace \thinspace
,}\,\,\,\,\overline{\omega }e=-2\cdot \,g(a,\overline{\omega })\text{
\thinspace \thinspace \thinspace \thinspace .}\,\,\,\,\,  \label{5.199}
\end{equation}

In this special case [$U_n$-space, $n=4$, $\pounds _u\overline{\omega }=0$, $%
T(u,\overline{\omega })=0$] the velocity $u$ will change along the axis $%
\overline{\omega }$ only if $g(a,\overline{\omega })=0$. This means that the
condition $\,\overline{\omega }e=0$ will be fulfilled only if the
acceleration $a$ is orthogonal to $\overline{\omega }$ [$g(a,\overline{%
\omega })=0$] or $a=0$ (if $\overline{\omega }\neq 0$). The last condition
is fulfilled if the material points are moving on auto-parallel trajectories
and the same time having vortex velocity $\overline{\omega }\neq 0$.

{\it Special case:} $V_n$-spaces. $n=4$, $\pounds _u\overline{\omega }=0$, $%
T(\xi ,\eta ):=0$ for $\forall \xi $, $\eta \in T(M)$. 
\begin{equation}
\,\,\overline{\omega }e=-2\cdot \,g(a,\overline{\omega })\text{ .}
\label{5.200}
\end{equation}

If further $e:=\,$const.$\,\neq 0:g(a,\overline{\omega })=0$.

In the Einstein theory of gravitation (where $e:=\,$const.$\,\neq 0$) the
vortex velocity vector $\overline{\omega }$ is always orthogonal to the
acceleration $a$ or the acceleration $a$ is equal to zero (auto-parallel,
geodesic trajectories). Since in the general case $g(a,u)=0$, and $g(u,%
\overline{\omega })=0$, the vectors $u$, $\overline{\omega }$, and $a$
construct a triad (3-Bein), where $u$ is a time-like vector, where $%
\overline{\omega }$ and $a$ are space-like vectors.

We can introduce abbreviations for the following invariants: 
\begin{eqnarray}
g(\overline{\omega },\overline{\omega }) &:&=\overline{\omega }^2=\pm \,\,l_{%
\overline{\omega }}^2\text{\thinspace \thinspace \thinspace \thinspace
\thinspace \thinspace \thinspace \thinspace \thinspace \thinspace \thinspace
,\thinspace \thinspace \thinspace \thinspace \thinspace \thinspace
\thinspace \thinspace \thinspace }l_{\overline{\omega }}=\,\mid g(\overline{%
\omega },\overline{\omega })\mid ^{1/2}\text{ \thinspace \thinspace
\thinspace \thinspace \thinspace ,}  \nonumber \\
\overline{g}[\omega (\overline{g})\omega ] &=&\omega _{ij}\cdot g^{\overline{%
j}\overline{k}}\cdot \omega _{kl}\cdot g^{\overline{i}\overline{l}}:=\omega
^2\text{ \thinspace \thinspace \thinspace \thinspace \thinspace \thinspace
\thinspace \thinspace ,}  \nonumber \\
\overline{g}[\sigma (\overline{g})\sigma ] &=&\sigma _{ik}\cdot g^{\overline{%
k}\overline{l}}\cdot \sigma _{lj}\cdot g^{\overline{i}\overline{j}}:=\sigma
^2\,\,\,\,\,\,\,\,\,\,\,\text{,}  \label{5.201} \\
\overline{g}[\sigma (\overline{g})\sigma (\overline{g})\sigma ] &=&\sigma
_{ik}\cdot \overline{g}^{\overline{k}\overline{l}}\cdot \sigma _{lm}\cdot 
\overline{g}^{\overline{m}\overline{n}}\cdot \sigma _{nj}\cdot g^{\overline{i%
}\overline{j}}:=\sigma ^3\text{ \thinspace \thinspace \thinspace ,} 
\nonumber \\
g(a,a) &:&=a^2=\,\pm l_a^2\text{ \thinspace \thinspace \thinspace
,\thinspace \thinspace \thinspace \thinspace \thinspace \thinspace
\thinspace }l_a=\,\mid g(a,a)\mid ^{1/2}\text{ \thinspace \thinspace .} 
\nonumber
\end{eqnarray}

Let us now consider the change of the vector $u$ along the curve $x^i(\tau
_0=$ const.,\thinspace $\lambda ^a)$.

\section{Friction velocity. Deformation friction velocity, shear friction
velocity, rotation (vortex) friction velocity, and expansion friction
velocity}

\subsection{Friction velocity}

In the case of change of the vector $u$ along the curve $x^i(\tau _0=$
const.,\thinspace $\lambda ^a)$, we have 
\begin{equation}
\left( \frac{Du}{d\lambda ^a}\right) _{(\tau _0,\lambda _0^a)}=\left( \nabla
_{\xi _{(a)\perp }}u\right) _{(\tau _0,\lambda _0^a)}=\stackunder{d\lambda
^a\rightarrow 0}{\lim }\frac{u_{(\tau _0,\lambda _0^a+d\lambda ^a)}-u_{(\tau
_0,\lambda _0^a)}}{d\lambda ^a}\text{ \thinspace \thinspace \thinspace
\thinspace , \thinspace \thinspace \thinspace \thinspace \thinspace }\xi
_{(a)\perp }=\frac d{d\lambda ^a}\text{ \thinspace \thinspace \thinspace .}
\label{5.202}
\end{equation}

The vector $\nabla _{\xi _{(a)\perp }}u$ describes the change of the
velocity $u$ of material points along the orthogonal to $u$ curves $x^i(\tau
_0,\lambda ^a)$. This means that $\nabla _{\xi _{(a)\perp }}u$ shows how the
velocity of the material elements changes at a cross-section of a flow.
Usually, the change of the velocity of material points in a direction,
orthogonal to the velocity, is related to the existence of inner friction
(viscosity of the media) between the different current lines of the flow.

In an analogous way as for $\nabla _u\xi _{(a)\perp }$ the vector $\nabla
_{\xi _{(a)\perp }}u$ can be decomposed in two parts: one collinear to $\xi
_{(a)\perp }$ and one orthogonal to $\xi _{(a)\perp }$, i.e. 
\begin{eqnarray}
\frac{Du}{d\lambda ^a} &=&\nabla _{\xi _{(a)\perp }}u=\frac{g(\nabla _{\xi
_{(a)\perp }}u,\xi _{(a)\perp })}{g(\xi _{(a)\perp },\xi _{(a)\perp })}\cdot
\xi _{(a)\perp }+R_{u(a)}=  \nonumber \\
&=&\frac{\widetilde{l}_{(a)}}{\pm l_{\xi (a)\perp }^2}\cdot \xi _{(a)\perp
}+R_{u(a)}\text{ \thinspace \thinspace \thinspace \thinspace \thinspace
,\thinspace \thinspace }  \label{5.203} \\
\widetilde{l}_{(a)} &=&g(\nabla _{\xi _{(a)\perp }}u,\xi _{(a)\perp })\text{
\thinspace \thinspace \thinspace ,\thinspace \thinspace \thinspace
\thinspace \thinspace \thinspace }\pm l_{\xi (a)\perp }^2=\text{\thinspace }%
g(\xi _{(a)\perp },\xi _{(a)\perp })\text{ \thinspace \thinspace \thinspace ,%
}  \nonumber \\
R_{u(a)} &=&\overline{g}[h_{\xi _{(a)\perp }}(\nabla _{\xi _{(a)\perp }}u)]=%
\overline{g}[h_{\xi _{(a)\perp }}(\frac{Du}{d\lambda ^a})]\text{ \thinspace
\thinspace \thinspace ,}  \nonumber \\
g(R_{u(a)},\xi _{(a)\perp }) &=&0\text{ \thinspace \thinspace \thinspace
,\thinspace \thinspace \thinspace \thinspace \thinspace \thinspace
\thinspace }h_{\xi _{(a)\perp }}=g-g(\xi _{(a)\perp })\otimes g(\xi
_{(a)\perp })\text{ \thinspace \thinspace \thinspace \thinspace .}  \nonumber
\end{eqnarray}

The vector $R_{u(a)}$ is called {\it friction velocity vector} or {\it %
friction velocity.} It could be consider as a measure for the friction
between the layers of a flow. We will consider later the structure of the
friction vector $R_{u(a)}$.

\subsection{Deformation friction velocity, shear friction velocity, rotation
(vortex) friction velocity, and expansion friction velocity}

Let us now consider a vector field $\xi _{\perp }$ with $g(u,\xi _{\perp
})=0 $, $\pounds _{\xi _{\perp }}u=-\pounds _u\xi _{\perp }=0$. Then we have
the relations 
\begin{eqnarray}
\nabla _{\xi _{\perp }}u &=&\nabla _u\xi _{\perp }-\pounds _u\xi _{\perp
}-T(u,\xi _{\perp })=(_\xi k)g(u)-\pounds _u\xi _{\perp }\,\,\,\,\,\text{,} 
\nonumber \\
(_\xi k)g(u) &=&\,_\xi k[g(u)]=\nabla _u\xi _{\perp }-T(u,\xi _{\perp
})=(\xi _{\perp ;j}^i\cdot u^j-T_{kl}\,^i\cdot u^k\cdot \xi _{\perp
}^l)\cdot \partial _i=  \nonumber \\
&=&(\xi _{\perp ;l}^i-T_{lk}\,^i\cdot \xi _{\perp }^k)\cdot u^l\cdot
\partial _i=(\xi _{\perp ;l}^i-T_{lk}\,^i\cdot \xi _{\perp }^k)\cdot
g^{lj}\cdot g_{\overline{j}\overline{m}}\cdot u^m\cdot \partial _i= 
\nonumber \\
&=&\,_\xi k^{ij}\cdot g_{\overline{j}\overline{m}}\cdot u^m\cdot \partial
_i=\,_\xi k(g)(u)=\,(_\xi k)g(u)=\,_\xi k[g(u)]\,\,\,\,\,\text{,\thinspace
\thinspace \thinspace }  \label{5.204} \\
_\xi k &=&\,_\xi k^{ij}\cdot \partial _i\otimes \partial _j\text{ \thinspace
\thinspace \thinspace \thinspace ,\thinspace \thinspace \thinspace
\thinspace \thinspace \thinspace \thinspace \thinspace \thinspace \thinspace
\thinspace }_\xi k^{ij}=(\xi _{\perp ;l}^i-T_{lk}\,^i\cdot \xi _{\perp
}^k)\cdot g^{lj}\cdot \partial _i\otimes \partial _j\text{ \thinspace
\thinspace \thinspace ,}  \nonumber \\
_\xi k(g)(u) &=&\,_\xi k(h_{\xi _{\perp }}+\frac 1{g(\xi _{\perp },\xi
_{\perp })}\cdot g(\xi _{\perp })\otimes g(\xi _{\perp }))(u)=\,(_\xi
k)h_{\xi _{\perp }}(u)\,\,\,\,\,\,\text{,}  \nonumber \\
\nabla _{\xi _{\perp }}u &=&-\pounds _u\xi _{\perp }+(_\xi k)h_{\xi _{\perp
}}(u)\,\,\,\,\,\text{.}  \nonumber
\end{eqnarray}

On the other side, from the relations

\begin{eqnarray}
\nabla _{\xi _{\perp }}u &=&\frac{g(\nabla _{\xi _{\perp }}u,u)}{g(\xi
_{\perp },\xi _{\perp })}\cdot \xi _{\perp }+R_u\text{\thinspace \thinspace
\thinspace \thinspace \thinspace \thinspace \thinspace \thinspace \thinspace
,\thinspace \thinspace \thinspace \thinspace \thinspace \thinspace
\thinspace \thinspace \thinspace \thinspace \thinspace \thinspace \thinspace
\thinspace \thinspace }R_u=\overline{g}[h_{\xi _{\perp }}(\nabla _{\xi
_{\perp }}u)]\,\,\,\,\text{,}  \label{5.205} \\
h_{\xi _{\perp }}(\xi _{\perp }) &=&0\text{ \thinspace ,\thinspace
\thinspace \thinspace \thinspace \thinspace \thinspace \thinspace \thinspace
\thinspace \thinspace }h^{\xi _{\perp }}=\overline{g}-\frac 1{g(\xi _{\perp
},\xi _{\perp })}\cdot \xi _{\perp }\otimes \xi _{\perp }\text{\thinspace
\thinspace \thinspace \thinspace \thinspace \thinspace \thinspace
,\thinspace \thinspace \thinspace \thinspace \thinspace \thinspace
\thinspace \thinspace }  \label{5.206} \\
g(R_u) &=&h_{\xi _{\perp }}(\nabla _{\xi _{\perp }}u)\,\,\,\,\,\,\text{,}
\label{5.207}
\end{eqnarray}

\noindent it follows the expression 
\begin{eqnarray}
h_{\xi _{\perp }}(\nabla _{\xi _{\perp }}u) &=&-h_{\xi _{\perp }}(\pounds
_u\xi _{\perp })+h_{\xi _{\perp }}(_\xi k)h_{\xi _{\perp }}(u)=  \nonumber \\
&=&-h_{\xi _{\perp }}(\pounds _u\xi _{\perp })+R(u)\,\,\,\,\,\,\,\,\text{,}
\label{5.208} \\
R(u) &=&h_{\xi _{\perp }}(_\xi k)h_{\xi _{\perp }}(u)\text{ \thinspace
\thinspace \thinspace \thinspace \thinspace \thinspace ,}  \label{5.209}
\end{eqnarray}

[compare with $h_u(\nabla _u\xi )=h_u(\frac le\cdot a-\pounds _\xi
u)+h_u(k)h_u(\xi )$ for $l=0$].

The tensor of second rank $R$ is called {\it friction deformation velocity
tensor}. It can be represented in the form analogous of the form of the
deformation velocity tensor 
\begin{equation}
R=\,_\sigma R+\,_\omega R+\frac 1{n-1}\cdot \,_\theta R\cdot h_{\xi _{\perp
}}\text{ \thinspace \thinspace \thinspace .}  \label{5.210}
\end{equation}

The tensors $_\sigma R$, $_\omega R$, and the invariant $_\theta R$ could be
found in analogous way as the tensors $\sigma $, $\omega $, and the
invariant $\theta $.

The symmetric trace-free tensor $_\sigma R$ has the form 
\begin{eqnarray}
_\sigma R &=&\,_{\xi s}E-\,_{\xi s}P=\,_\xi E-\,_\xi P-\frac 1{n-1}\cdot 
\overline{g}[_\xi E-\,_\xi P]\cdot h_{\xi _{\perp }}=\,_\sigma R_{ij}\cdot
dx^i.dx^j=  \nonumber \\
&=&\,_\sigma R_{ij}\cdot e^i.e^j\text{\thinspace \thinspace \thinspace
\thinspace ,}  \nonumber \\
\,_{\xi s}E &=&\,_\xi E-\frac 1{n-1}\cdot \overline{g}[_\xi E]\cdot h_{\xi
_{\perp }}\text{ \thinspace \thinspace \thinspace ,\thinspace \thinspace
\thinspace \thinspace \thinspace \thinspace \thinspace \thinspace \thinspace
\thinspace \thinspace \thinspace \thinspace \thinspace \thinspace \thinspace
\thinspace \thinspace \thinspace }\overline{g}[_\xi E]=g^{ij}\cdot \,_\xi E_{%
\overline{i}\overline{j}}=g^{\overline{i}\overline{j}}\cdot \,_\xi
E_{ij}=\,_{\theta _o}R\text{ \thinspace \thinspace \thinspace ,}  \nonumber
\\
_{\theta _o}R &=&\xi _{\perp ;n}^n-\frac 1{2\cdot e_{\xi _{\perp }}}\cdot
(e_{\xi _{\perp },k}\cdot \xi _{\perp }^k-g_{kl;m}\cdot \xi _{\perp }^m\cdot
\xi _{\perp }^{\overline{k}}\cdot \xi _{\perp }^{\overline{l}})\text{
\thinspace \thinspace \thinspace \thinspace \thinspace ,\thinspace
\thinspace \thinspace \thinspace \thinspace }  \label{5.211} \\
\text{\thinspace }e_{\xi _{\perp }} &=&g(\xi _{\perp },\xi _{\perp })=\pm
l_{\xi _{\perp }}^2\text{\thinspace \thinspace \thinspace \thinspace
,\thinspace \thinspace }  \nonumber \\
\,_\xi E &=&h_{\xi _{\perp }}(_\xi \varepsilon )h_{\xi _{\perp }}\text{
\thinspace \thinspace \thinspace \thinspace \thinspace ,\thinspace
\thinspace \thinspace \thinspace \thinspace \thinspace \thinspace \thinspace
\thinspace \thinspace \thinspace \thinspace \thinspace }_\xi k_s=\,_\xi
\varepsilon -\,_\xi m\text{ \thinspace \thinspace \thinspace \thinspace
\thinspace \thinspace \thinspace ,\thinspace \thinspace \thinspace
\thinspace \thinspace \thinspace \thinspace \thinspace \thinspace \thinspace
\thinspace \thinspace }e_{\xi _{\perp },k}=\partial _ke_{\xi _{\perp
}}=e_k(e_{\xi _{\perp }})\,\,\,\,\,\,\text{,}  \nonumber \\
\text{\thinspace \thinspace \thinspace \thinspace \thinspace \thinspace }%
_\xi \varepsilon &=&\frac 12\cdot (\xi _{\perp ;l}^i\cdot g^{lj}+\xi _{\perp
;l}^j\cdot g^{li})\cdot \partial _i.\partial j\text{ \thinspace \thinspace
\thinspace \thinspace ,}  \nonumber \\
_\xi m &=&\frac 12\cdot (T_{lk}\,^i\cdot \xi _{\perp }^k\cdot
g^{lj}+T_{lk}\,^j\cdot \xi _{\perp }^k\cdot g^{li})\cdot \partial
_i.\partial j\text{ \thinspace \thinspace \thinspace \thinspace .}  \nonumber
\end{eqnarray}

The symmetric trace-free tensor $_{\xi s}E$ is the torsion-free shear
friction velocity tensor (shear friction), the symmetric trace-free tensor $%
_{\xi s}P$ is the shear friction velocity tensor induced by the torsion, 
\begin{eqnarray}
_{\xi s}P &=&\,_\xi P-\frac 1{n-1}\cdot \overline{g}[_\xi P]\cdot h_{\xi
_{\perp }}\,\,\,\,\,\,\,\text{,\thinspace \thinspace \thinspace \thinspace
\thinspace \thinspace \thinspace \thinspace \thinspace \thinspace \thinspace 
}\overline{g}[_\xi P]=g^{ij}\cdot \,_\xi P_{\overline{i}\overline{j}}=g^{%
\overline{i}\overline{j}}\cdot \,_\xi P_{ij}=\,_{\theta _1}R\text{
\thinspace \thinspace \thinspace ,}  \label{5.212} \\
_\xi P &=&h_{\xi _{\perp }}(_\xi m)h_{\xi _{\perp }}\text{ \thinspace
\thinspace \thinspace \thinspace \thinspace ,\thinspace \thinspace
\thinspace \thinspace \thinspace \thinspace \thinspace \thinspace \thinspace
\thinspace \thinspace \thinspace \thinspace \thinspace \thinspace \thinspace
\thinspace \thinspace \thinspace \thinspace \thinspace \thinspace \thinspace
\thinspace \thinspace \thinspace \thinspace \thinspace \thinspace \thinspace
\thinspace \thinspace \thinspace \thinspace \thinspace \thinspace \thinspace
\thinspace \thinspace \thinspace \thinspace \thinspace \thinspace \thinspace
\thinspace \thinspace \thinspace \thinspace \thinspace \thinspace \thinspace
\thinspace \thinspace }_{\theta _1}R=T_{kl}\,^k\cdot \xi _{\perp }^l\text{%
\thinspace \thinspace \thinspace \thinspace \thinspace \thinspace \thinspace
\thinspace \thinspace \thinspace \thinspace \thinspace ,\thinspace
\thinspace \thinspace }  \label{5.213} \\
_\theta R &=&\,_{\theta _0}R\,-\,_{\theta _1}R\,\,\,\,\,\,\,\text{.}
\label{5.214}
\end{eqnarray}

The invariant $_\theta R$ is the expansion friction velocity (expansion
friction), the invariant$_{\theta _o}R$ is the torsion-free expansion
friction velocity, the invariant $_{\theta _1}R$ is the expansion friction
velocity induced by the torsion.

The antisymmetric tensor of second rank $_\omega R$ is the rotation (vortex)
friction velocity tensor (rotation friction) 
\begin{eqnarray}
_\omega R &=&h_{\xi _{\perp }}(_\xi k_a)h_{\xi _{\perp }}=h_{\xi _{\perp
}}(_\xi s)h_{\xi _{\perp }}-h_{\xi _{\perp }}(_\xi q)h_{\xi _{\perp
}}=\,_\xi S-\,_\xi Q\,\,\,\,\,\,\text{,}  \label{5.215} \\
_\xi S &=&h_{\xi _{\perp }}(_\xi s)h_{\xi _{\perp }}\,\,\,\,\,\,\text{%
,\thinspace \thinspace \thinspace \thinspace \thinspace \thinspace
\thinspace \thinspace \thinspace }_\xi Q=\text{\thinspace }h_{\xi _{\perp
}}(_\xi q)h_{\xi _{\perp }}\,\,\,\,\,\,\text{,}  \label{5.216} \\
\text{\thinspace }_\xi s &=&\frac 12\cdot (\xi _{\perp ;m}^k\cdot g^{ml}-\xi
_{\perp ;m}^l\cdot g^{mk})\cdot \partial _i\wedge \partial j\text{
\thinspace \thinspace \thinspace \thinspace ,\thinspace \thinspace }
\label{5.217} \\
_\xi q &=&\frac 12\cdot (T_{mn}\,^k\cdot \xi _{\perp }^n\cdot
g^{ml}-T_{mn}\,^l\cdot \xi _{\perp }^n\cdot g^{mk})\cdot \partial _i\wedge
\partial j\text{ \thinspace \thinspace \thinspace \thinspace .}  \nonumber
\end{eqnarray}

The antisymmetric tensor $_\xi S$ is the torsion-free rotation (vortex)
friction velocity tensor, the antisymmetric tensor $_\xi Q$ is the rotation
(vortex) friction velocity tensor induced by the torsion.

By means of the expressions for $_\sigma R$, $_\omega R$, and $_\theta R$
the friction deformation velocity tensor $R$ could be written in the form 
\begin{equation}
R=\,_0R-\,_TR\text{ ,}  \label{5.218}
\end{equation}

\noindent where 
\begin{eqnarray}
_0R &=&\,_{\xi s}E+\,_\xi S+\frac 1{n-1}\cdot \,_{\theta _o}R\cdot h_{\xi
_{\perp }}\text{ ,}  \label{5.219} \\
_TR &=&\,_{\xi s}P+\,_\xi Q+\frac 1{n-1}\cdot \,_{\theta _1}R\cdot h_{\xi
_{\perp }}\,\,\,\,\,\,\text{. }  \label{5.220}
\end{eqnarray}

The tensor $_0R$ is the torsion-free friction deformation velocity tensor
and the tensor $_TR$ is the friction deformation velocity induced by the
torsion. for the case of $V_n$-spaces, $_TR=0$ ($\,_{\xi s}P=0$, $_\xi Q=0$, 
$_{\theta _1}R=0$).

In an analogous way as in the case of the shear velocity tensor $\sigma $
and the expansion velocity invariant $\theta $, the shear friction velocity
tensor $_\sigma R$ and the expansion friction invariant $_\theta R$ could be
represented in the forms 
\begin{eqnarray}
_\sigma R &=&\frac 12\cdot \{h_{\xi _{\perp }}(\nabla _{\xi _{\perp }}%
\overline{g}-\pounds _{\xi _{\perp }}\overline{g})h_{\xi _{\perp }}-\frac
1{n-1}\cdot (h_{\xi _{\perp }}[\nabla _{\xi _{\perp }}\overline{g}-\pounds
_{\xi _{\perp }}\overline{g}])h_{\xi _{\perp }}\,\,\,\,\,\text{,}
\label{5.221} \\
_\theta R &=&\frac 12\cdot h_{\xi _{\perp }}[\nabla _{\xi _{\perp }}%
\overline{g}-\pounds _{\xi _{\perp }}\overline{g}]\,\,\,\,\,\,\,\text{.}
\label{5.222}
\end{eqnarray}

\subsection{Representation of the friction velocity by the use of the
kinematic characteristics of the relative velocity}

The relative velocity tensor and the friction velocity tensor can be related
to each other on the basis of the relation $\pounds _u\xi =\nabla _u\xi
-\nabla _\xi u-T(u,\xi )$. Let us now consider the representation of $\nabla
_\xi u$ by the use of the corresponding to $u$ projective metrics $h_u$ and $%
h^u$. We can write for $\nabla _\xi u$%
\begin{eqnarray}
\nabla _\xi u &=&u^i\,_{;j}\cdot u^j\cdot \partial _i=u^\alpha \,_{/\beta
}\cdot \xi ^\beta \cdot e_\alpha =  \nonumber \\
&=&(\varepsilon +s)g(\xi )=(\varepsilon +s)[g(\xi )]=  \nonumber \\
&=&(\varepsilon +s)[h_u+\frac 1e\cdot g(u)\otimes g(u)](\xi )=  \nonumber \\
&=&(\varepsilon +s)h_u(\xi )+\frac 1e\cdot (\varepsilon +s)[g(u)]\otimes
[g(u)](\xi )=  \nonumber \\
&=&\{(\varepsilon +s)h_u+\frac 1e\cdot (\varepsilon +s)[g(u)]\otimes
[g(u)]\}(\xi )\text{ \thinspace \thinspace \thinspace ,}  \label{5.223}
\end{eqnarray}
\begin{eqnarray}
g(\nabla _\xi u) &=&\{g(\varepsilon +s)h_u+\frac 1e\cdot g(\varepsilon
+s)[g(u)]\otimes [g(u)]\}(\xi )=  \nonumber \\
&=&\{[h_u+\frac 1e\cdot g(u)\otimes g(u)](\varepsilon +s)h_u+\frac 1e\cdot
g(\varepsilon +s)[g(u)]\otimes [g(u)]\}(\xi )=  \nonumber \\
&=&\{h_u(\varepsilon +s)h_u+\frac 1e\cdot g(u)\otimes [g(u)](\varepsilon
+s)h_u+  \nonumber \\
&&+\frac 1e\cdot g(\varepsilon +s)[g(u)]\otimes [g(u)]\}(\xi )\,\,\,\,\,\,\,.
\label{5.224}
\end{eqnarray}

Since 
\begin{eqnarray*}
h_u(\varepsilon +s)h_u &=&h_u(\varepsilon )h_u+h_u(s)h_u=E+S= \\
&=&\,_sE+S+\frac 1{n-1}\cdot \overline{g}[E]\cdot h_u= \\
&=&\,_sE+S+\frac 1{n-1}\cdot \theta _o\cdot h_u=d_o\text{ \thinspace
\thinspace \thinspace ,} \\
(\varepsilon +s)[g(u)] &=&a=\nabla _uu\,\,\,\,\,\text{,} \\
\lbrack g(u)](\varepsilon +s)h_u &=&[g(u)](\varepsilon +s)g-\frac 1e\cdot
g(u,a)\cdot g(u)\,\,\,\,\,\,\text{,} \\
\lbrack g(u)](\varepsilon +s)g(\xi ) &=&\frac 12\cdot [\xi e-(\nabla _\xi
g)(u,u)]\,\,\,\,\,\,\,\text{,}
\end{eqnarray*}

\noindent we obtain for $g(\nabla _\xi u)$ and $\nabla _\xi u$ respectively 
\begin{eqnarray}
g(\nabla _\xi u) &=&\{d_o+\frac 1e\cdot g(a)\otimes g(u)-  \nonumber \\
&&-\frac 1{2\cdot e^2}\cdot [ue-(\nabla _ug)(u,u)]\cdot g(u)\otimes
g(u)\}(\xi )+  \nonumber \\
&&+\frac 1{2\cdot e}\cdot [\xi e-(\nabla _\xi g)(u,u)]\cdot
g(u)\,\,\,\,\,\,\,\text{,}  \label{5.227} \\
g(\nabla _\xi u) &=&\{\frac 1e\cdot g(a)\otimes g(u)+\,_sE+S+\frac
1{n-1}\cdot \overline{g}[E]\cdot h_u\}(\xi )+  \nonumber \\
&&+\frac 1{2\cdot e}\cdot [\xi e-(\nabla _\xi g)(u,u)]\cdot g(u)-  \nonumber
\\
&&-\frac 1{2\cdot e^2}\cdot \{[ue-(\nabla _ug)(u,u)]\cdot g(u)\otimes
g(u)\}(\xi )  \nonumber \\
\nabla _\xi u &=&\{\overline{g}(d_o)+\frac 1e\cdot a\otimes g(u)-\frac
1{2\cdot e^2}\cdot [ue-(\nabla _ug)(u,u)]\cdot u\otimes g(u)\}(\xi )+ 
\nonumber \\
&&+\frac 1{2\cdot e}\cdot [\xi e-(\nabla _\xi g)(u,u)]\cdot u\,\,\,\,\,\,%
\text{,}  \label{5.228} \\
\nabla _\xi u &=&\{\frac 1e\cdot a\otimes g(u)+\overline{g}(\,_sE)+\overline{%
g}(S)+\frac 1{n-1}\cdot \overline{g}[E]\cdot \overline{g}(h_u)\}(\xi )+ 
\nonumber \\
&&+\frac 1{2\cdot e}\cdot [\xi e-(\nabla _\xi g)(u,u)]\cdot u-  \nonumber \\
&&-\frac 1{2\cdot e^2}\cdot \{[ue-(\nabla _ug)(u,u)]\cdot u\otimes
g(u)\}(\xi )\,\,\,\,\,\,\text{.}  \label{5.229}
\end{eqnarray}

The relations for $g(\nabla _\xi u)$ and $\nabla _\xi u$ are valid for an
arbitrary given contravariant vector field $\xi \in T(M)$. For an orthogonal
to $u$ vector field $\xi _{\perp }$,$\,[g(u,\xi _{\perp })=0]$, we obtain
the representations for $g(\nabla _\xi u)$ and $\nabla _\xi u$ in the forms 
\begin{eqnarray}
g(\nabla _{\xi _{\perp }}u) &=&d_o(\xi _{\perp })+\frac 1{2\cdot e}\cdot
[\xi _{\perp }e-(\nabla _{\xi _{\perp }}g)(u,u)]\cdot g(u)\,\,\,\,\,\,\text{,%
}  \label{5.230} \\
\nabla _{\xi _{\perp }}u &=&\overline{g}(d_o)(\xi _{\perp })+\frac 1{2\cdot
e}\cdot [\xi _{\perp }e-(\nabla _{\xi _{\perp }}g)(u,u)]\cdot u\,\,\,\,\,%
\text{.}  \nonumber
\end{eqnarray}

Therefore, for $\nabla _{\xi _{\perp }}u$ and $h_{\xi _{\perp }}(\nabla
_{\xi _{\perp }}u)$ we have the relations 
\begin{eqnarray}
\nabla _{\xi _{\perp }}u &=&\frac{g(\nabla _{\xi _{\perp }}u,u)}{g(\xi
_{\perp },\xi _{\perp })}\cdot \xi _{\perp }+R_u=  \nonumber \\
&=&\overline{g}(d_o)(\xi _{\perp })+\frac 1{2\cdot e}\cdot [\xi _{\perp
}e-(\nabla _{\xi _{\perp }}g)(u,u)]\cdot u\text{ \thinspace \thinspace
\thinspace \thinspace \thinspace ,}  \label{5.231} \\
h_{\xi _{\perp }}(\nabla _{\xi _{\perp }}u) &=&h_{\xi _{\perp }}(\overline{g}%
)(d_o)(\xi _{\perp })+\frac 1{2\cdot e}\cdot [\xi _{\perp }e-(\nabla _{\xi
_{\perp }}g)(u,u)]\cdot g(u)\,\,\,\,\text{,}  \label{5.231a} \\
h_{\xi _{\perp }}(u) &=&g(u)-\frac 1{e_{\xi _{\perp }}}\cdot g(\xi _{\perp
},u)\cdot g(\xi _{\perp })=g(u)  \nonumber \\
R_u &=&\overline{g}[h_{\xi _{\perp }}(\nabla _{\xi _{\perp }}u)]=(\overline{g%
})h_{\xi _{\perp }}(\overline{g})(d_o)(\xi _{\perp })+\frac 1{2\cdot e}\cdot
[\xi _{\perp }e-(\nabla _{\xi _{\perp }}g)(u,u)]\cdot u=  \nonumber \\
&=&h^{\xi _{\perp }}\,(d_o)(\xi _{\perp })+\frac 1{2\cdot e}\cdot [\xi
_{\perp }e-(\nabla _{\xi _{\perp }}g)(u,u)]\cdot u\,\,\,\,\,\,\text{,}
\label{5.232} \\
h^{\xi _{\perp }} &=&(\overline{g})h_{\xi _{\perp }}(\overline{g})\,\,\,\,\,%
\text{.}  \nonumber
\end{eqnarray}

In our further consideration we will assume the existence of a proper frame
of reference in a flow. From this point of view it is possible to introduce
particular designations for some types of flows.

\section{Special types of flows}

\subsection{Inertial flow}

\begin{definition}
A flow which material points are moving on auto-parallel lines, i.e. a flow
with $\nabla _uu=a=0$ as a kinematic characteristic, is called inertial flow.
\end{definition}

Inertial flows will be considered below with respect to their relative
accelerations.

\subsection{Vortex-free (irrotational) flow}

\begin{definition}
A flow for which the vector field $u$ fulfills the condition $\omega =0$ is
called vortex-free (irrotational) flow.
\end{definition}

If we consider the explicit form for $\omega $%
\begin{equation}
\omega =h_u(k_a)h_u  \label{6.0}
\end{equation}

\noindent we can prove the following propositions:

\begin{proposition}
The necessary and sufficient condition for the existence of a contravariant
non-null vector field with vanishing rotation velocity $(\omega =0)$ is the
condition 
\begin{equation}
k_a=\frac 1e\cdot \{u\otimes [g(u)](k_a)-[g(u)](k_a)\otimes u\}\text{ ,}
\label{6.1a}
\end{equation}
\end{proposition}

\noindent or in a co-ordinate basis 
\begin{equation}
k_a^{ij}=\frac 1e\cdot g_{\overline{m}\overline{n}}\cdot u^n\cdot (u^i\cdot
k_a^{mj}-u^j\cdot k_a^{mi})\text{ \thinspace \thinspace \thinspace
\thinspace .}  \label{6.1}
\end{equation}

Proof: 1. Necessity. Form $h_u(k_a)h_u=0$, it follows that 
\begin{eqnarray*}
h_u(k_a)h_u &=&0=g(k_a)g-\frac 1e\cdot g(u)\otimes [g(u)](k_a)g-\frac
1e\cdot g(k_a)[g(u)]\otimes g(u)+ \\
&&+\frac 1{e^2}\cdot [g(u)](k_a)[g(u)]\cdot g(u)\otimes g(u)\text{ .}
\end{eqnarray*}

Since 
\[
\lbrack g(u)](k_a)[g(u)]=g_{\overline{i}\overline{m}}\cdot u^m\cdot
k_a^{ij}\cdot g_{\overline{j}\overline{n}}\cdot u^n=-g_{\overline{i}%
\overline{m}}\cdot u^m\cdot k_a^{ij}\cdot g_{\overline{j}\overline{n}}\cdot
u^n\text{ ,} 
\]

\noindent we have $[g(u)](k_a)[g(u)]=0$. Therefore, 
\[
g(k_a)g=\frac 1e\cdot \{g(u)\otimes [g(u)](k_a)g+g(k_a)[g(u)]\otimes g(u)\}%
\text{ .} 
\]

From the last expression and from the relation $\overline{g}[g(k_a)g]%
\overline{g}=k_a$, it follows that 
\begin{eqnarray*}
k_a &=&\frac 1e\cdot \{u\otimes [g(u)](k_a)+(k_a)[g(u)]\otimes u\}= \\
&=&\frac 1e\cdot \{u\otimes [g(u)](k_a)-[g(u)](k_a)\otimes u\}\text{ ,}
\end{eqnarray*}

\noindent because of $(k_a)[g(u)]=-$ $[g(u)](k_a)$. In a co-ordinate basis
we obtain (\ref{6.1}).

2. Sufficiency. From (\ref{6.1a}) we have 
\[
g(k_a)g=\frac 1e\cdot \{g(u)\otimes [g(u)](k_a)g+g(k_a)[g(u)]\otimes g(u)\}%
\text{ ,} 
\]

\noindent which is identical to $h_u(k_a)h_u=0$.

On the other hand, after direct computations, it follows that 
\[
(k_a)[g(u)]=\frac 12\cdot \{(k)[g(u)]-[g(u)](k)\}\text{ .} 
\]

Since $(k)[g(u)]=a$, we have the relation 
\[
(k_a)[g(u)]=\frac 12\cdot \{a-[g(u)](k)\}\text{ .} 
\]

Then 
\[
k_a=\frac 1{2\cdot e}\cdot \{a\otimes u-u\otimes a+u\otimes
[g(u)](k)-[g(u)](k)\otimes u\}\text{ .} 
\]

\begin{proposition}
A sufficient condition for the existence of a contravariant non-null vector
field with vanishing rotation velocity $(\omega =0)$ is the condition 
\[
k_a=0\text{ .} 
\]
\end{proposition}

Proof: If $k_a=0$, then it follows directly from $\omega =h_u(k_a)h_u$ that $%
\omega =0$.

In a co-ordinate basis $k_a$ is equivalent to the expression 
\[
u^i\,_{;l}\cdot g^{lj}-u^j\,_{;l}\cdot g^{li}=(T_{lm}\,^i\cdot
g^{lj}-T_{lm}\,^j\cdot g^{li})\cdot u^m\text{ .} 
\]

On the other side, after multiplying the last expression with $g_{\overline{j%
}\overline{k}}\cdot u^k$ and summarizing over $j$, we obtain 
\[
a^i=u^i\,_{;k}\cdot u^k=g_{\overline{j}\overline{k}}\cdot u^k\cdot
(u^j\,_{;l}-T_{lm}\,^j\cdot u^m)\cdot g^{li}=g_{\overline{j}\overline{k}%
}\cdot u^k\cdot k^{ji}\text{ ,} 
\]

\noindent or in a form 
\[
a=[g(u)](k)\text{ .} 
\]

\begin{proposition}
The necessary condition for $k_a=0$ is the condition 
\[
a=[g(u)](k)\,\,\,\text{.} 
\]
\end{proposition}

Proof: From $k_a=0$ and $(k_a)[g(u)]=\frac 12\cdot \{a-[g(u)](k)\}$, it
follows that $a=[g(u)](k)$.

\subsection{Volume-preserving (isochoric) flow}

\begin{definition}
A flow for which the vector field $u$ fulfills the conditions
\end{definition}

\begin{eqnarray}
_\nabla \theta &=&\frac 1{2\cdot e}\cdot (\nabla _ug)(u,u)\text{\thinspace
\thinspace \thinspace \thinspace \thinspace \thinspace ,}  \label{6.2} \\
_{\pounds }\theta &=&\frac 1{2\cdot e}\cdot (\pounds _ug)(u,u)\text{%
\thinspace }=\frac 1{2\cdot e}\cdot ue\text{ \thinspace \thinspace
\thinspace \thinspace ,}  \label{6.3}
\end{eqnarray}

is called volume-preserving (isochoric) flow.

From the last two conditions, it follows that 
\begin{eqnarray*}
\theta &=&\,_\nabla \theta -\,_{\pounds }\theta =\frac 1{2\cdot e}\cdot
[(\nabla _ug)(u,u)-ue]=-\frac 1e\cdot g(u,a)\text{ \thinspace \thinspace
\thinspace ,} \\
\nabla _u(d\omega ) &=&0\text{ \thinspace \thinspace \thinspace ,\thinspace
\thinspace \thinspace \thinspace \thinspace \thinspace \thinspace \thinspace 
}\pounds _u(d\omega )=0\text{ \thinspace \thinspace ,}
\end{eqnarray*}

\noindent where $d\omega $ is the invariant volume element in the
differentiable manifold $M$, considered as a model of a continuous media.
Since $(\nabla _ug)(u,u)=ue-2\cdot g(u,a)$, we have for $_\nabla \theta $
the expression 
\begin{equation}
_\nabla \theta =\frac 1{2\cdot e}\cdot [ue-2\cdot g(u,a)]\text{ \thinspace
\thinspace \thinspace .}  \label{6.4}
\end{equation}

\begin{proposition}
For an inertial $(a=\nabla _uu=0)$ and volume-preserving flow the following
relations are fulfilled 
\begin{eqnarray*}
_\nabla \theta =\,_{\pounds }\theta =\frac 1{2\cdot e}\cdot ue\text{ ,} \\
\theta =0\text{ .}
\end{eqnarray*}
\end{proposition}

The proof is trivial. It follows from the expression for $_\nabla \theta $
and $_{\pounds }\theta $ in the case of a volume-preserving (isochoric)
flow. From the condition $\theta =0$, it follows that

\begin{proposition}
An inertial and volume-preserving flow is an expansion-free flow $(\theta
=0) $.
\end{proposition}

The proof follows immediately from the above relations for $\theta $ in the
case of an inertial and volume-preserving flow.

\begin{proposition}
For volume-preserving flow and a normalized vector field $u$ [$g(u,u)=\,$%
const.$\,\neq 0$]\thinspace the following relations are valid 
\begin{eqnarray*}
_\nabla \theta =-\frac 1e\cdot g(u,a)\text{ ,} \\
_{\pounds }\theta =0\text{ .}
\end{eqnarray*}
\end{proposition}

The proof is trivial. It follows from the expression for $_\nabla \theta $
and $_{\pounds }\theta $ in the case of a volume-preserving (isochoric) flow.

\subsection{Shear-free flow}

\begin{definition}
A flow with $_\nabla \sigma =0$ and $_{\pounds }\sigma =0$ is called
shear-free flow.
\end{definition}

From the last conditions it follows that $\sigma =0$.

The shear velocity tensor $\sigma $ and the expansion velocity invariant $%
\theta $ are composed as the difference between the corresponding quantities
induced by a transport and by its corresponding dragging along $u$.

A transport along $u$ (action of $\nabla _u$) is a motion of a material
point along a line with the tangent vector $u$. A dragging along $u$ (action
of $\pounds _u$) is a motion of all material points lying in a vicinity
(determined by the vectors $\xi _{(a)}$) of the material point at the curve
with tangent vector $u$%
\begin{eqnarray}
\nabla _u\xi _{(a)} &=&(\xi _{(a),i}^k\cdot u^i+\Gamma _{ji}^k\cdot \xi
_{(a)}^j\cdot u^i)\cdot \partial _k=\xi _{(a);i}^k\cdot u^i\cdot \partial _k%
\text{\thinspace \thinspace \thinspace \thinspace \thinspace \thinspace
\thinspace \thinspace \thinspace \thinspace \thinspace \thinspace ,}
\label{6.5} \\
\pounds _u\xi _{(a)} &=&(\xi _{(a),i}^k\cdot u^i-u^k\,_{,i}\cdot \xi
_{(a)}^i)\cdot \partial _k=(\pounds _u\xi ^k)\cdot \partial _k=  \nonumber \\
&=&\nabla _u\xi _{(a)}-\nabla _{\xi _{(a)}}u-T(u,\xi _{(a)})\text{
\thinspace \thinspace \thinspace \thinspace ,}  \label{6.6}
\end{eqnarray}
\[
\nabla _u\xi _{(a)}-\pounds _u\xi _{(a)}\,=\nabla _{\xi _{(a)}}u+T(u,\xi
_{(a)})\,\,\,\,\,\,\,\text{.} 
\]

The difference between a transport along $u$ and a dragging along $u$ of a
vector field $\xi _{(a)}$ could be interpreted as a characteristic
describing the (relative) change of the vector field $u$ under the influence
of the vector field $\xi _{(a)}$. Since $\nabla _{\xi _{(a)}}u$ is related
to the friction velocity of the flow, the non-vanishing difference $\nabla
_u\xi _{(a)}-\pounds _u\xi _{(a)}$ could characterize the friction velocity
in the flow.

If we consider the structure of the relative velocity we can find the
relations: 
\begin{equation}
_{rel}v=\overline{g}[h_u(\nabla _u\xi )]=\overline{g}(h_u)(\pounds _u\xi
)+\frac le\cdot \overline{g}[h_u(a)]+\overline{g}[d(\xi )]\text{ \thinspace
\thinspace \thinspace \thinspace ,}  \label{6.7}
\end{equation}
\begin{equation}
\overline{g}(h_u)(\nabla _u\xi -\pounds _u\xi )=\frac le\cdot \overline{g}[%
h_u(a)]+\overline{g}[d(\xi )]\,\,\,\,\,\,\,\text{.}  \label{6.8}
\end{equation}

{\it Special case}: $l=g(u,\xi _{(a)}):=0$, \thinspace $\xi _{(a)}=\xi
_{(a)\perp }$. 
\begin{eqnarray}
\overline{g}(h_u)(\nabla _u\xi _{(a)\perp }-\pounds _u\xi _{(a)\perp }) &=&%
\overline{g}[d(\xi _{(a)\perp })]\text{ \thinspace \thinspace \thinspace
\thinspace \thinspace \thinspace \thinspace \thinspace ,}  \nonumber \\
(h_u)(\nabla _u\xi _{(a)\perp }-\pounds _u\xi _{(a)\perp }) &=&[d(\xi
_{(a)\perp })]\text{ }\,\,\,\,\,\,\,\,\,\,\,\text{.}  \label{6.9}
\end{eqnarray}

The orthogonal to $u$ projection of the difference $\nabla _u\xi _{(a)\perp
}-\pounds _u\xi _{(a)\perp }$ is proportional to the deformation velocity
tensor $d$. This means that $d$ is a measure for the relative deformation
induced by a transport along $u$ and a dragging along $u$. This relative
deformation could be an object of measurement because we can (locally)
measure a deformation velocity at a point of a line with respect to the
deformation velocity of its neighboring points outside the line. Usually,
the deformation induced by a dragging along $u$ is ignored by choosing the
vectors $u$ and $\xi _{(a)\perp }$ as tangent vectors to the co-ordinate
lines and vice versa, by choosing the co-ordinate lines as lines with
tangent vectors $u$ and $\xi _{(a)\perp }$. Then $\pounds _u\xi _{(a)\perp
}=-\pounds _{\xi _{(a)\perp }}u=0$ and the deformation velocity tensor
represent the deformation velocity for the special type of co-ordinates.
this is the common (canonical) method for description of deformations in the
relativistic continuous media mechanics in $V_n$-spaces ($n=4$). The
condition $\pounds _\xi u=0$ has been introduced by Ehlers \cite{Ehlers} at
the beginning of all further considerations about relativistic mechanics of
continuous media. If appropriate co-ordinates are imposed by the condition $%
\pounds _u\xi _{(a)\perp }=[u,\xi _{(a)\perp }]=0$ a relative deformation
velocity and its corresponding structures (shear, rotation, and expansion
velocities) could be considered as absolute kinematic characteristics with
respect to the given co-ordinates. The same is valid for the kinematic
characteristics related to the deformation acceleration tensor and its
corresponding structures (shear, rotation, and expansion accelerations).
This is the reason for introducing and considering of many notions of
continuous media mechanics under the condition $\pounds _u\xi _{(a)\perp }=0$
or $\pounds _u\xi =0$.

If we consider the explicit form of the shear velocity tensor (shear
velocity, shear) 
\begin{equation}
\sigma =h_u(k_s)h_u-\frac 1{n-1}\cdot \overline{g}[h_u(k_s)h_u]\cdot h_u
\label{6.10}
\end{equation}

\noindent we can prove the following propositions:

\begin{proposition}
The necessary and sufficient condition for the existence of a non-null
contravariant vector field $u$ with vanishing shear velocity ($\sigma =0$)
is the condition 
\begin{eqnarray}
k_s=\frac 1{2\cdot e}\cdot \{u\otimes a+a\otimes u+u\otimes
[g(u)](k)+[g(u)](k)\otimes u-  \nonumber \\
-\frac 1e\cdot [ue-(\nabla _ug)(u,u)]\cdot u\otimes u\}+  \nonumber \\
+\frac 1{n-1}\cdot \theta \cdot h^u\text{,}  \label{6.11}
\end{eqnarray}
\end{proposition}

\noindent or in a co-ordinate basis 
\begin{eqnarray}
h_s^{ij} &=&\frac 1{2\cdot e}\cdot \{u^i\cdot a^j+u^j\cdot a^i+u^i\cdot g_{%
\overline{m}\overline{n}}\cdot u^n\cdot k^{mj}+u^j\cdot g_{\overline{m}%
\overline{n}}\cdot u^n\cdot k^{mi}-  \nonumber \\
&&-\frac 1e\cdot [e_{,k}\cdot u^k-g_{km;n}\cdot u^n\cdot u^{\overline{k}%
}\cdot u^{\overline{m}}]\cdot u^i\cdot u^j\}+\frac 1{n-1}\cdot \theta \cdot
h^{ij}\text{ .}  \label{6.11a}
\end{eqnarray}

Proof: 1. Necessity. From $\sigma =0$, it follows that $h_u(k_s)h_u=\frac
1{n-1}\cdot \overline{g}[h_u(k_s)h_u]\cdot h_u=\frac 1{n-1}\cdot \theta
\cdot h_u$. Further, from the explicit form of $h_u$ and $k_s$, it follows
that 
\begin{eqnarray*}
h_u(k_s)h_u &=&\frac 1{n-1}\cdot \theta \cdot h_u=g(k_s)g-\frac 1e\cdot
\{g(u)\otimes [g(u)](k_s)g+g(k_s)[g(u)]\otimes g(u)\}+ \\
&&+\frac 1{e^2}\cdot [g(u)](k_s)[g(u)]\cdot g(u)\otimes g(u)\text{ , }
\end{eqnarray*}

\noindent or 
\begin{eqnarray*}
k_s &=&\frac 1e\cdot \{u\otimes [g(u)](k_s)+(k_s)[g(u)]\otimes u\}-\frac
1{e^2}\cdot [g(u)](k_s)[g(u)]\cdot u\otimes u+ \\
&&+\frac 1{n-1}\cdot \theta \cdot \overline{g}(h_u)\overline{g}\text{ .}
\end{eqnarray*}

Since $[g(u)](k_s)=(k_s)[g(u)]$, $(k_s)[g(u)]=\frac 12\cdot
\{(k)[g(u)]+[g(u)](k)\}$, $(k)[g(u)]=a$, $(k_s)[g(u)]=\frac 12\cdot
\{a+[g(u)](k)\}$, $[g(u)](k_s)[g(u)]=[g(u)](k)[g(u)]=g(u,a)=\frac 12\cdot
[ue-(\nabla _ug)(u,u)]$, $\theta =\overline{g}[h_u(k_s)h_u]=\overline{g}[%
h_u(k)h_u]$, and $\overline{g}(h_u)\overline{g}=h^u$, the explicit form of $%
k_s$ can be found as 
\begin{eqnarray*}
k_s &=&\frac 1{2\cdot e}\cdot \{u\otimes a+a\otimes u+u\otimes
[g(u)](k)+[g(u)](k)\otimes u- \\
&&-\frac 1e\cdot [ue-(\nabla _ug)(u,u)]\cdot u\otimes u\}+ \\
&&+\frac 1{n-1}\cdot \theta \cdot h^u\text{ \thinspace \thinspace \thinspace
.}
\end{eqnarray*}

2. Sufficiency. From the last expression and the above relations, it follows
that $h_u(k_s)h_u=\frac 1{n-1}\cdot \theta \cdot h_u$, and therefore $\sigma
=0$.

\begin{proposition}
A sufficient condition for the existence of a non-null vector field with
vanishing shear velocity ($\sigma =0$) and expansion velocity ($\theta =0$)
is the condition 
\[
h_u(k_s)h_u=0\text{ ,} 
\]
\end{proposition}

\noindent identical with the condition 
\begin{eqnarray*}
k_s &=&\frac 1{2\cdot e}\cdot \{u\otimes a+a\otimes u+u\otimes
[g(u)](k)+[g(u)](k)\otimes u- \\
&&-\frac 1e\cdot [ue-(\nabla _ug)(u,u)]\cdot u\otimes u\}
\end{eqnarray*}

Proof: If $h_u(k_s)h_u=0$, then $\theta =\overline{g}[h_u(k_s)h_u]=0$.
Therefore, $\sigma =h_u(k_s)h_u-\frac 1{n-1}\cdot \theta \cdot h_u=0$.

{\it Corollary}. If $h_u(k_s)h_u=0$, then 
\begin{equation}
g[k_s]=\frac 1{2\cdot e}\cdot [ue-(\nabla _ug)(u,u)]\text{ .}  \label{6.12}
\end{equation}

Proof: It follows from the above proposition that 
\begin{eqnarray*}
\theta &=&g[k_s]-\frac 1e\cdot g(u,a)=0\text{ ,} \\
g[k_s] &=&\frac 1e\cdot g(u,a)=\frac 1{2\cdot e}\cdot [ue-(\nabla _ug)(u,u)]%
\text{ .}
\end{eqnarray*}

\subsection{Rigid flow}

\begin{definition}
A flow which is isochoric (volume-preserving) and shear-free is called rigid
flow.
\end{definition}

An other definition of the notion of rigid flow could also be introduced.
Let us consider the vectors $\xi _{(a)\perp }$ as infinitesimal vectors,
determining a cross section in the flow orthogonal to the vector $u$. If the
vectors $\xi _{(a)\perp }$ are Fermi-Walker transported \cite{Manoff-6} they
will not change their lengths and angles between them. The cross section
will move along $u$ without any deformation. Therefore, the Fermi-Walker
transport determines a motion of a cross section of a flow as a rigid body.
A rigid flow is then defined as a flow with cross sections transported along
a vector $u$ by means of a Fermi-Walker transport. Since a Fermi-Walker
transport is not a priori related to the kinematic characteristics of a
flow, the last definition of a rigid flow is more general than the first
definition using the kinematic characteristics related to the relative
velocity in a flow. For a Fermi-Walker transport of type $C$ \cite{Manoff-6}
we have the relation 
\[
\overline{g}[h_u(\overline{g}))(^F\omega )-\omega ](\xi _{(a)\perp })=\frac
12\cdot h^u[(\nabla _ug)(\xi _{(a)\perp })]+ 
\]
\begin{equation}
+\overline{g}[h_u(\pounds _u\xi _{(a)\perp })]+\overline{g}[\sigma (\xi
_{(a)\perp })]+\frac 1{n-1}\cdot \theta \cdot \xi _{(a)\perp }\text{%
\thinspace \thinspace \thinspace \thinspace \thinspace \thinspace \thinspace
\thinspace ,}  \label{6.13}
\end{equation}

\noindent where $^F\omega $ is the covariant antisymmetric tensor of second
rank from the structure of the Fermi derivative [for more details see \cite
{Manoff-5}, \cite{Manoff-6}]. By the use of the relations 
\begin{eqnarray}
h^u(\nabla _ug) &=&h^u[h_u(\overline{g})(\nabla _ug)(\overline{g})h_u+ 
\nonumber \\
&&+\frac 1e\cdot [h_u(\overline{g})(\nabla _ug)(u)\otimes g(u)+g(u)\otimes
h_u(\overline{g})(\nabla _ug)(u)]+  \nonumber \\
&&+\frac 1{e^2}\cdot (\nabla _ug)(u,u)\cdot g(u)\otimes g(u)]\text{%
\thinspace \thinspace \thinspace ,}  \label{6.14}
\end{eqnarray}
\begin{eqnarray*}
h^u(h_u)\overline{g} &=&h^u\text{ \thinspace \thinspace \thinspace
\thinspace ,} \\
(h_u)\overline{g}(h_u) &=&h_u\text{\thinspace \thinspace \thinspace
\thinspace \thinspace \thinspace ,}
\end{eqnarray*}
\[
h^u(g)(u)=h_u[g(u)]=0\text{ \thinspace \thinspace \thinspace ,} 
\]

\noindent we can find a representation of the tensor $^F\omega $ in the form 
\begin{eqnarray}
^F\omega &=&h_u(\overline{g})(^F\omega )(\overline{g})h_u+  \nonumber \\
&&+\frac 1e\cdot [h_u(\overline{g})(^F\omega )(u)\otimes g(u)-g(u)\otimes
h_u(\overline{g})(^F\omega )(u)]\text{ \thinspace \thinspace \thinspace .}
\label{6.15}
\end{eqnarray}

At the same time $^F\omega $ has the following properties: 
\begin{eqnarray}
^F\omega (u) &=&h_u(\overline{g})(^F\omega )(u)\text{\ \ ,}  \label{6.16} \\
^F\omega (\xi _{(a)\perp }) &=&h_u(\overline{g})(^F\omega )(\xi _{(a)\perp
})-\frac 1e\cdot (\xi _{(a)\perp })(h_u)\overline{g}(^F\omega )(u)\cdot g(u)%
\text{ \thinspace \thinspace \thinspace \thinspace ,}  \label{6.16a}
\end{eqnarray}

\noindent where 
\begin{eqnarray*}
(\xi _{(a)\perp })(h_u) &=&g(\xi _{(a)\perp })=(\xi _{(a)\perp })g\text{
\thinspace \thinspace ,} \\
(\xi _{(a)\perp })(h_u)\overline{g}(^F\omega )(u) &=&\,^F\omega (\xi
_{(a)\perp },u)\text{ \thinspace \thinspace \thinspace \thinspace ,} \\
^F\omega (\xi _{(a)\perp }) &=&h_u(\overline{g})(^F\omega )(\xi _{(a)\perp
})-\frac 1e\cdot \,^F\omega (\xi _{(a)\perp },u)\cdot g(u)\text{ \thinspace
\thinspace .}
\end{eqnarray*}

Now $^F\omega $ could be represented in the form 
\begin{equation}
^F\omega =\,^F\omega _{\perp }+\,^F\widetilde{\omega }\text{ \thinspace
\thinspace \thinspace ,}  \label{6.17}
\end{equation}

\noindent where 
\begin{eqnarray*}
^F\omega _{\perp } &=&h_u(\overline{g})(^F\omega )(\overline{g})h_u\text{
\thinspace \thinspace \thinspace ,\thinspace \thinspace \thinspace
\thinspace } \\
^F\text{\thinspace }\omega _{\perp }(u) &=&\,-(u)(^F\omega _{\perp })=0\text{%
\thinspace \thinspace \thinspace \thinspace \thinspace \thinspace
,\thinspace \thinspace \thinspace \thinspace \thinspace \thinspace
\thinspace \thinspace }
\end{eqnarray*}
\begin{eqnarray*}
^F\widetilde{\omega } &=&\frac 1e\cdot [h_u(\overline{g})(^F\omega
)(u)\otimes g(u)-g(u)\otimes h_u(\overline{g})(^F\omega )(u)]\text{
\thinspace \thinspace ,} \\
^F\widetilde{\omega }(u) &=&-(u)(^F\widetilde{\omega })=h_u(\overline{g}%
)(^F\omega )(u)\text{\thinspace \thinspace \thinspace \thinspace \thinspace
\thinspace \thinspace \thinspace ,} \\
^F\widetilde{\omega }(\xi _{(a)\perp }) &=&-\frac 1e\cdot \,^F\omega (\xi
_{(a)\perp },u)\cdot g(u)\text{\thinspace \thinspace \thinspace \thinspace
\thinspace \thinspace \thinspace \thinspace \thinspace \thinspace .}
\end{eqnarray*}

The tensor $^F\omega $ contains in general terms not orthogonal to the
vector $u$ [$^F\omega (u)=-(u)(^F\omega )\neq 0$] in contrast to the tensor $%
\omega $ [$\omega (u)=-(u)(\omega )=0$]. Therefore, a rigid dynamic system
is either a rigid flow or a system transported by means of a Fermi-Walker
transport.

\subsubsection{Rigid flow and Fermi-Walker transports}

Ehlers \cite{Ehlers} has defined a Fermi derivative $^e\nabla _u\xi $ in the
form 
\begin{equation}
^e\nabla _u\xi _{\perp }:=\overline{g}[h_u(\nabla _u\xi _{\perp })]\text{
\thinspace \thinspace \thinspace .}  \label{6.18}
\end{equation}

If an external covariant differential operator $^e\nabla _u$ is chosen as 
\cite{Manoff-6} 
\begin{equation}
^e\nabla _u:=\nabla _u-\overline{A}_u  \label{6.19}
\end{equation}

\noindent with 
\[
\overline{A}_u:=\frac 1e\cdot [\nabla _uu\otimes g(u)-u\otimes g(\nabla _uu)]%
\text{ ,} 
\]

\noindent i.e. if $^e\nabla _u$ is chosen as 
\begin{equation}
^e\nabla _u:=\nabla _u-\frac 1e\cdot [\nabla _uu\otimes g(u)-u\otimes
g(\nabla _uu)]\,\,\,\,\,\,\text{,}  \label{6.20}
\end{equation}

\noindent then 
\begin{eqnarray}
^e\nabla _u\xi _{\perp } &=&\nabla _u\xi _{\perp }-\frac 1e\cdot [\nabla
_uu\otimes g(u)-u\otimes g(\nabla _uu)](\xi _{\perp })=  \nonumber \\
&=&\nabla _u\xi _{\perp }+\frac 1e\cdot g(\nabla _uu,\xi _{\perp })\cdot u%
\text{ ,}\,\,  \label{6.21}
\end{eqnarray}

\noindent because of $g(u,\xi _{\perp })=0$. Now, using the expression for $%
\nabla _u\xi _{\perp }$, 
\[
\nabla _u\xi _{\perp }=\frac 1e\cdot g(u,\nabla _u\xi _{\perp })\cdot u+%
\overline{g}[h_u(\nabla _u\xi _{\perp })]\text{ \thinspace \thinspace
\thinspace \thinspace \thinspace \thinspace } 
\]

\noindent we can find the form of the Fermi derivative, introduced by Ehlers 
\begin{eqnarray}
^e\nabla _u\xi _{\perp } &=&\frac 1e\cdot g(u,\nabla _u\xi _{\perp })\cdot u+%
\overline{g}[h_u(\nabla _u\xi _{\perp })]+\frac 1e\cdot g(\nabla _uu,\xi
_{\perp })\cdot u=  \nonumber \\
&=&\overline{g}[h_u(\nabla _u\xi _{\perp })]-\frac 1e\cdot (\nabla
_ug)(u,\xi _{\perp })\cdot u\text{ ,}  \label{6.22}
\end{eqnarray}

\noindent where 
\begin{eqnarray*}
\nabla _u[g(u,\xi _{\perp })] &=&u[g(u,\xi _{\perp })]=0= \\
&=&(\nabla _ug)(u,\xi _{\perp })+g(\nabla _uu,\xi _{\perp })+g(u,\nabla
_u\xi _{\perp })\text{ \thinspace \thinspace \thinspace .}
\end{eqnarray*}

For $\nabla _ug=0$ we have 
\begin{equation}
^e\nabla _u\xi _{\perp }=\overline{g}[h_u(\nabla _u\xi _{\perp })]=\,_{rel}v%
\text{\thinspace \thinspace \thinspace \thinspace \thinspace \thinspace .}
\label{6.23}
\end{equation}

The last condition is not fulfilled if $\nabla _ug\neq 0$.

The notion of Fermi-Walker transport has richer contents than usually
assumed on the basis of different heuristic viewpoints (Manoff 1998, 2000).
In the structure of a Fermi-Walker transport a covariant antisymmetric
tensor field $^F\omega $ of second rank plays an important role. On the
other side, in the deformation velocity tensor and in the relative velocity
respectively the rotation (vortex) velocity tensor $\omega $ is exactly of
the type of the tensor $^F\omega $. This fact leads to the assumption for
identification of both the tensors in (pseudo) Riemannian spaces without
torsion. Such convention \cite{Ehlers} could be unique only if we consider
the kinematics of a continuous media. If we consider in addition dynamical
models of substratum then there could exist other interpretations of the
antisymmetric tensor $^F\omega $ in the structure of a Fermi-Walker
transport. In $(\overline{L}_n,g)$- and $(L_n,g)$-spaces there is no unique
relation between the rotation velocity tensor and a Fermi-Walker transport.
This means that in general there is no need for a relation between the
rotation velocity tensor and a Fermi-Walker transport. Only if material
points in a flow are Fermi-Walker transported a relation between $^F\omega $
and $\omega $ could be established.

\subsection{Deformation-free flow}

\begin{definition}
A flow with vanishing deformation velocity tensor $d$, i.e. with $d=0$, is
called deformation-free flow.
\end{definition}

If the co-ordinates in a flow are chosen in the way that $\pounds _\xi
u=-\pounds _u\xi =0$ then from the differential geometry in $(\overline{L}%
_n,g)$-spaces we have the relation 
\begin{equation}
\nabla _u\xi -\nabla _\xi u-T(u,\xi )=0\text{ .}  \label{6.24}
\end{equation}

The parallel transports of the deviation vector $\xi $ along the velocity
vector $u$ and vice versa assure the vanishing of the vector of torsion $%
T(u,\xi )$. Therefore, the conditions 
\begin{eqnarray}
\pounds _u\xi &=&-\pounds _\xi u=0\text{ ,}  \nonumber \\
\nabla _u\xi &=&0\text{ \thinspace \thinspace ,}  \label{6.25} \\
\nabla _\xi u &=&0\text{ \thinspace \thinspace ,}  \nonumber
\end{eqnarray}

\noindent should lead to deformation-free flow of a continuous media. The
last two conditions ($\nabla _u\xi =0$, $\nabla _\xi u=0$) should by
dynamically generated.. We could speak about deformations if $\nabla _u\xi
\neq 0$ or $\nabla _\xi u\neq 0$, or if $\nabla _u\xi \neq 0$ and $\nabla
_\xi u\neq 0$. The condition $\nabla _u\xi \neq 0$ means that the deviation
vector $\xi $ changes in the time and generates changes of the distance, the
relative velocity and the relative acceleration between the material points
in the media. The condition $\nabla _\xi u\neq 0$ means that the velocity
vector $u$ changes along the co-ordinate line (if $\pounds _u\xi =-\pounds
_\xi u=0$) with tangent vector $\xi $ and this changes could be a corollary
of friction between the material points in the media.

If we consider the explicit form for $d$%
\begin{equation}
d:=h_u(k)h_u  \label{6.26}
\end{equation}

\noindent we can prove the following propositions:

\begin{proposition}
The necessary and sufficient condition for the existence of a non-null
contravariant vector field $u$ with vanishing deformation velocity $(d=0)$
is the condition 
\[
k=\frac 1e\cdot \{a\otimes u+u\otimes [g(u)](k)\}-\frac 1{2e^2}\cdot
[ue-(\nabla _ug)(u,u)]\cdot u\otimes u\text{ ,} 
\]
\end{proposition}

\noindent or in a co-ordinate basis 
\begin{eqnarray*}
k^{ij} &=&\frac 1e\cdot (a^i\cdot u^j+u^i\cdot k^{lj}\cdot g_{\overline{l}%
\overline{m}}\cdot u^m) \\
&&-\frac 1{2e^2}\cdot (e_{,k}\cdot u^k-g_{\overline{k}\overline{l};m}\cdot
u^m\cdot u^k\cdot u^l)\cdot u^i\cdot u^j\text{ .}
\end{eqnarray*}

Proof: 1. Necessity. From $d=h_u(k)h_u$, after writing the explicit form of $%
h_u$, it follows that 
\begin{eqnarray*}
d &=&g(k)g-\frac 1e\cdot g(u)\otimes [g(u)](k)g-\frac 1e\cdot
g(k)[g(u)]\otimes g(u) \\
&&+\frac 1{e^2}\cdot [g(u)](k)[g(u)]\cdot u\otimes u\text{ .}
\end{eqnarray*}

Since $(k)[g(u)]=a=\nabla _uu$, it follows further that 
\[
\lbrack g(u)](k)[g(u)]=[g(u)](a)=g(u,a)=\frac 12\cdot [ue-(\nabla _ug)(u,u)]%
\text{ .} 
\]

Therefore, 
\begin{eqnarray*}
d &=&0:\,\,\,\,\,\,\,\,\,\,\,\,\,\,\,\,\,\,\,\,\,\,\,\,\,\,\,g(k)g=\frac
1e\cdot \{g(u)\otimes [g(u)](k)g+g(a)\otimes g(u)\} \\
&&-\frac 1{e^2}\cdot g(u,a)\cdot g(u)\otimes g(u)\text{ .}
\end{eqnarray*}

From $\overline{g}(g(k)g)\overline{g}=k$, we obtain 
\begin{eqnarray*}
k &=&\frac 1e\cdot \{a\otimes u+u\otimes [g(u)](k)\}- \\
&&-\frac 1{2\cdot e^2}\cdot [ue-(\nabla _ug)(u,u)]\cdot u\otimes u\text{ .}
\end{eqnarray*}

2. Sufficiency. From the explicit form of $k$, it follows that 
\begin{eqnarray*}
g(k)g &=&\frac 1e\cdot g(u)\otimes [g(u)](k)g+\frac 1e\cdot
g(k)[g(u)]\otimes g(u)- \\
&&-\frac 1{e^2}\cdot g(u,a)\cdot g(u)\otimes g(u)\text{ }
\end{eqnarray*}

\noindent which is identical to $h_u(k)h_u=d=0$.

{\it Special case:} $\nabla _uu=a:=0$, $\nabla _\xi g:=0$ for $\forall \xi
\in T(M)$ ($U_n$-space), $ue=0:e=\,$const.$\,\neq 0$ ($u$ is a normalized,
non-null contravariant vector field). 
\[
d=0:\,\,\,\,\,\,\,\,\,\,\,\,\,\,\,\,\,\,\,\,\,\,\,\,\,\,\,k=\frac 1e\cdot
u\otimes [g(u)](k)\text{ ,} 
\]
\[
(k)[g(\xi )]=\frac 1e\cdot u\otimes [g(u)](k)[g(\xi )]=\frac 1e\cdot
[g(u)](k)[g(\xi )]\cdot u\text{ ,} 
\]
\begin{eqnarray*}
(k)[g(\xi )] &=&(u^i\,_{;l}-T_{lk}\,^i\cdot u^k)\cdot g^{lm}\cdot g_{%
\overline{m}\overline{j}}\cdot \xi ^j\cdot \partial _i=\nabla _\xi u-T(\xi
,u)= \\
&=&\nabla _u\xi -\pounds _\xi u\text{ ,}
\end{eqnarray*}
\begin{equation}
_{rel}v=\overline{g}[h_u(\nabla _u\xi )]=-\,\overline{g}(h_u)(\pounds _\xi u)%
\text{ for }\forall \xi \in T(M)\text{ \thinspace \thinspace .}  \label{6.27}
\end{equation}

\begin{proposition}
A sufficient condition for the existence of a non-null contravariant vector
field with vanishing deformation velocity ($d=0$) is the condition 
\[
k=0\text{ ,} 
\]
\end{proposition}

\noindent equivalent to the condition 
\[
\nabla _\xi u=T(\xi ,u)\text{ \thinspace \thinspace \thinspace for }\forall
\xi \in T(M)\text{ ,} 
\]

\noindent or in a co-ordinate basis 
\[
k^{ij}=0:\,\,\,\,\,\,\,\,\,\,\,\,\,\,\,\,\,\,\,u^i\,_{;j}=T_{jk}\,^i\cdot u^k%
\text{ .} 
\]

Proof: From $k=0$ and $(k)[g(\xi )]=\nabla _\xi u-T(\xi ,u)$ for $\forall
\xi \in T(M)$, it follows that $\nabla _\xi u-T(\xi ,u)=0$ or in a
co-ordinate basis $u^i\,_{;j}-T_{jk}\,^i\cdot u^k=0$. In this case $\pounds
_\xi u=\nabla _\xi u-\nabla _u\xi -T(\xi ,u)=-\nabla _u\xi $.

{\it Corollary}. A deformation-free contravariant vector field $u$ with $k=0$
is an auto-parallel contravariant vector field.

Proof: It follows immediately from the condition $\nabla _\xi u=T(\xi ,u)$
and for $\xi =u$ that $\nabla _uu=a=0$.

\begin{proposition}
The necessary condition for the existence of a deformation-free
contravariant vector field $u$ with $k=0$ is the condition 
\[
[R(u,v)]\xi =[\pounds \Gamma (u,v)]\xi \text{ \thinspace \thinspace
\thinspace \thinspace \thinspace \thinspace \thinspace \thinspace \thinspace
for\thinspace \thinspace \thinspace \thinspace }\forall \xi ,v\in T(M)\text{
,} 
\]
\end{proposition}

\noindent or in a co-ordinate basis 
\[
R^k\,_{ilj}\cdot u^l=\pounds _u\Gamma _{ij}^k\text{ .} 
\]

Proof: By the use of the explicit form of the curvature operator $R(u,v)$
acting on a contravariant vector field $\xi $%
\[
\lbrack R(u,v)]\xi =\nabla _u\nabla _v\xi -\nabla _v\nabla _u\xi -\nabla
_{\pounds _uv}\xi \text{ , \thinspace \thinspace \thinspace \thinspace
\thinspace \thinspace \thinspace \thinspace \thinspace \thinspace \thinspace
\thinspace \thinspace }\xi \text{, }v\text{, }u\in T(M)\text{ ,} 
\]

\noindent and the explicit form of the deviation operator $\pounds \Gamma
(u,v)$ acting on a contravariant vector field $\xi $%
\[
\lbrack \pounds \Gamma (u,v)]\xi =\pounds _u\nabla _v\xi -\nabla _v\pounds
_u\xi -\nabla _{\pounds _uv}\xi 
\]

\noindent we obtain under the condition $\nabla _\xi u=T(\xi ,u)$
(equivalent to the condition $\pounds _u\xi =\nabla _u\xi $) 
\begin{eqnarray*}
\lbrack \pounds \Gamma (u,v)]\xi &=&\pounds _u\nabla _v\xi -\nabla _v\pounds
_u\xi -\nabla _{\pounds _uv}\xi = \\
&=&\nabla _u\nabla _v\xi -\nabla _{\nabla _v\xi }u-T(u,\nabla _v\xi )-\nabla
_v\nabla _u\xi -\nabla _{\pounds _uv}\xi = \\
&=&\nabla _u\nabla _v\xi -\nabla _v\nabla _u\xi -\nabla _{\pounds _uv}\xi
-[\nabla _{\nabla _v\xi }u+T(u,\nabla _v\xi )]= \\
&=&[R(u,v)]\xi -[\nabla _{\nabla _v\xi }u+T(u,\nabla _v\xi )]\text{ .}
\end{eqnarray*}

Since 
\[
\nabla _{\nabla _v\xi }u+T(u,\nabla _v\xi )=0\text{ \thinspace \thinspace
\thinspace \thinspace \thinspace \thinspace \thinspace for\thinspace
\thinspace \thinspace \thinspace \thinspace \thinspace \thinspace \thinspace 
}\forall v,\xi \in T(M)\text{ , } 
\]

\noindent we have 
\[
\lbrack R(u,v)]\xi =[\pounds \Gamma (u,v)]\xi \text{ \thinspace \thinspace
\thinspace \thinspace \thinspace \thinspace for\thinspace \thinspace
\thinspace \thinspace \thinspace \thinspace }\forall v,\xi \in T(M)\text{
\thinspace \thinspace . } 
\]

The last condition appears as the integrability condition for the equation
for $u$%
\[
\nabla _\xi u=T(\xi ,u)\text{ \thinspace \thinspace \thinspace \thinspace
for \thinspace \thinspace \thinspace \thinspace }\forall \xi \in T(M)\text{
\thinspace \thinspace .} 
\]

\begin{proposition}
A deformation-free contravariant non-null vector field $u$ with $k=0$ is an
auto-parallel non-null shear-free ($\sigma =0$), rotation-free ($\omega =0$)
and expansion-free ($\theta =0$) contravariant vector field with vanishing
deformation acceleration ($A=0$) \cite{Manoff-9}.
\end{proposition}

Proof: If $k=k^{ij}\cdot \partial _i\otimes \partial _j=0$ and $\nabla
_uu=a=0$, then $k_s=k^{(ij)}\cdot \partial _i.\partial _j=\frac 12\cdot
(k^{ij}+k^{ji})\cdot \partial _i.\partial _j=0$, and $k_a=k^{[ij]}\cdot
\partial _i\wedge \partial _j=\frac 12\cdot (k^{ij}-k^{ji})\cdot \partial
_i\wedge \partial _j=0$. Therefore, $\sigma =h_u(k_s)h_u-\frac 1{n-1}\cdot 
\overline{g}[h_u(k_s)h_u]\cdot h_u=0$, $\theta =\overline{g}[h_u(k_s)h_u]=0$%
, and $\omega =h_u(k_a)h_u=0$. From the explicit form of the deformation
acceleration $A$ (see below), it follows that $A=0$.

From the identity for the Riemannian tensor $R^i\,_{jkl}$%
\begin{eqnarray}
R^i\,_{jkl}+R^i\,_{ljk}+R^i\,_{klj} &\equiv
&T_{jk}\,^i\,_{;l}+T_{lj}\,^i\,_{;k}+T_{kl}\,^i\,_{;j}+  \nonumber \\
&&+T_{jk}\,^m\cdot T_{ml}\,^i+T_{lj}\,^m\cdot T_{mk}\,^i+T_{kl}\,^m\cdot
T_{mj}\,^i\text{ ,}  \label{6.28}
\end{eqnarray}

\noindent after contraction with $g_i^l$ (equivalent to the action of the
contraction operator $S=C$) and summation over $l$ we obtain 
\begin{eqnarray}
R_{jk}-R_{kj}+R^i\,_{ijk} &\equiv
&T_{jk}\,^i\,_{;i}+T_{ij}\,^i\,_{;k}-T_{ik}\,^i\,_{;j}+  \label{6.29} \\
&&+T_{jk}\,^m\cdot T_{mi}\,^i+T_{ij}\,^m\cdot T_{mk}\,^i-T_{ik}\,^m\cdot
T_{mj}\,^i\text{ .}  \nonumber
\end{eqnarray}

If we introduce the abbreviations 
\begin{equation}
_aR_{jk}:=\frac 12\cdot (R_{jk}-R_{kj})\text{ , \thinspace \thinspace
\thinspace \thinspace \thinspace \thinspace \thinspace \thinspace \thinspace
\thinspace \thinspace \thinspace }T_{ji}\,^i:=T_j\text{ ,\thinspace
\thinspace \thinspace \thinspace \thinspace \thinspace \thinspace \thinspace
\thinspace \thinspace \thinspace \thinspace }  \label{6.30}
\end{equation}

\noindent where \thinspace $T_{ik}\,^i=-T_{ki}\,^i=-T_k$, then the last
expression for $R_{jk}$ can be written in the form 
\begin{eqnarray}
2\cdot \,_aR_{jk} &\equiv
&-R^i\,_{ijk}+T_{jk}\,^i\,_{;i}+T_{ij}\,^i\,_{;k}-T_{ik}\,^i\,_{;j}+ 
\nonumber \\
&&+T_{jk}\,^m\cdot T_m+T_{ij}\,^m\cdot T_{mk}\,^i-T_{ik}\,^m\cdot T_{mj}\,^i%
\text{ .}  \label{6.31}
\end{eqnarray}

Therefore, $_aR_{ij}\cdot u^j$ can be written in the form 
\begin{eqnarray}
2\cdot \,_aR_{ij}\cdot u^j+R^i\,_{ijk}\cdot u^j &=&T_{k;j}\cdot
u^j-T_{j;k}\cdot u^j+T_{jk}\,^i\,_{;i}\cdot u^j+  \nonumber \\
&&+T_m\cdot T_{jk}\,^m\cdot u^j+T_{ij}\,^m\cdot u^j\cdot
T_{mk}\,^i-T_{ik}\,^m\cdot T_{mj}\,^i\cdot u^j\text{ .}  \label{6.32}
\end{eqnarray}

From the other side, from $u^i\,_{;j}=T_{jl}\,^i\cdot u^l$ and $%
a^i=u^i\,_{;j}\cdot u^j=0$, we have 
\begin{equation}
u^i\,_{;j;k}=T_{jl}\,^i\,_{;k}\cdot u^l+T_{jm}\,^i\cdot T_{kl}\,^m\cdot u^l%
\text{ ,}  \label{6.33}
\end{equation}
\begin{eqnarray}
u^i\,_{;j;k}-u^i\,_{;k;j} &=&-u^l\cdot R^i\,_{ljk}+T_{jk}\,^m\cdot
T_{ml}\,^i\cdot u^l=  \nonumber \\
&=&T_{jl}\,^i\,_{;k}\cdot u^l+T_{jm}\,^i\cdot T_{kl}\,^m\cdot u^l-  \nonumber
\\
&&-T_{kl}\,^i\,_{;j}\cdot u^l-T_{km}\,^i\cdot T_{jl}\,^m\cdot u^l\,\text{ ,}
\label{6.34}
\end{eqnarray}
\begin{eqnarray}
u^l\cdot R^i\,_{ljk} &=&T_{jk}\,^m\cdot T_{ml}\,^i\cdot
u^l+T_{kl}\,^i\,_{;j}\cdot u^l-T_{jl}\,^i\,_{;k}\cdot u^l+  \nonumber \\
&&+T_{km}\,^i\cdot T_{jl}\,^m\cdot u^l-T_{jm}\,^i\cdot T_{kl}\,^m\cdot u^l%
\text{ \thinspace \thinspace ,}  \label{6.35}
\end{eqnarray}
\begin{equation}
R_{lj}\cdot u^l=-T_{l;j}\cdot u^l-T_{jl}\,^i\,_{;i}\cdot u^l-T_m\cdot
T_{jl}\,^m\cdot u^l\text{ \thinspace \thinspace \thinspace ,}  \label{6.36}
\end{equation}
\begin{equation}
R_{lj}\cdot u^l\cdot u^j=\,_sR_{lj}\cdot u^l\cdot u^j=I=-T_{l;j}\cdot
u^l\cdot u^j=-(T_i\cdot u^i)_{;j}\cdot u^j=\dot{\theta}_1\text{ . }
\label{6.37}
\end{equation}

By the use of the decompositions $R_{ij}=\,_aR_{ij}+\,_sR_{ij}$, $%
R_{ij}\cdot u^j=\,_aR_{ij}\cdot u^j+\,_sR_{ij}\cdot u^j$, and the above
expression for $R_{lj}\cdot u^l$, we can find the following relations 
\begin{equation}
2\cdot \,_aR_{jk}\cdot u^j=T_{k;j}\cdot u^j-T_{j;k}\cdot u^j-2\cdot
T_{kj}\,^i\,_{;i}\cdot u^j-2\cdot T_m\cdot T_{kj}\,^m\cdot u^j\text{ ,}
\label{6.38}
\end{equation}
\begin{equation}
R^i\,_{ijk}\cdot u^j=(T_{kj}\,^i\,_{;i}+T_m\cdot T_{kj}\,^m+T_{ij}\,^m\cdot
T_{mk}\,^i-T_{ik}\,^m\cdot T_{mj}\,^i)\cdot u^j\text{ ,}  \label{6.39}
\end{equation}
\begin{equation}
2\cdot \,_sR_{jk}\cdot u^j=-(T_{j;k}+T_{k;j})\cdot u^j\text{ }.  \label{6.40}
\end{equation}

It follows that in a $(\overline{L}_n,g)$-space the projections of the
symmetric part of the Ricci tensor on the non-null contravariant vector
field $u$ with $k=0$ is depending on the covariant derivatives of $T_i$
(respectively on the covariant derivatives of the torsion $T_{ik}\,^l$) and
not on the torsion $T_{ik}\,^l$ itself.

\subsection{Conformal flow}

\begin{definition}
A flow for which $\pounds _ug=\lambda \cdot g$ \thinspace \thinspace [or $%
\pounds _u\overline{g}=-\lambda \cdot \overline{g}$] with $\lambda
=(1/n)\cdot \overline{g}[\pounds _ug]$ is called conformal flow.
\end{definition}

\begin{proposition}
For a conformal flow the following relations are valid 
\[
_{\pounds }\theta =-\frac{n-1}2\cdot \lambda \text{ \thinspace \thinspace
\thinspace , \thinspace \thinspace \thinspace \thinspace \thinspace
\thinspace \thinspace \thinspace \thinspace }_{\pounds }\sigma =0\text{
,\thinspace \thinspace \thinspace \thinspace \thinspace \thinspace
\thinspace }\sigma =\,_\nabla \sigma \text{,\thinspace \thinspace \thinspace
\thinspace \thinspace \thinspace } 
\]
\[
\theta =\,_\nabla \theta -\,_{\pounds }\theta =\,_\nabla \theta +\frac{n-1}%
2\cdot \lambda \text{ \thinspace \thinspace \thinspace ,} 
\]
\[
\pounds _u(d\omega )=\frac n2\cdot \lambda \cdot d\omega \text{ \thinspace
\thinspace .} 
\]
\end{proposition}

The proof follows from the explicit forms of $\pounds _u\overline{g}$, $%
_{\pounds }\theta $, and $\pounds _u(d\omega )$, $\pounds _u\overline{g}=-%
\overline{g}(\pounds _ug)\overline{g}$, $h_u[\overline{g}]=n-1$, under the
definition of a conformal flow.

{\it Special case:} $\overline{U}_n$- and $\overline{V}_n$-spaces. 
\[
_\nabla \sigma =\sigma =0\text{ \thinspace \thinspace \thinspace ,
\thinspace \thinspace \thinspace \thinspace \thinspace \thinspace \thinspace
\thinspace }_\nabla \theta =0\text{ \thinspace \thinspace \thinspace
\thinspace \thinspace ,\thinspace \thinspace \thinspace \thinspace
\thinspace \thinspace }\theta =\text{\thinspace }\frac{n-1}2\cdot \lambda 
\text{ \thinspace \thinspace \thinspace \thinspace \thinspace .\thinspace
\thinspace \thinspace } 
\]

\subsection{Isometric flow}

\begin{definition}
A flow for which $\pounds _ug=0$ is called isometric flow.
\end{definition}

\begin{proposition}
For an isometric flow the following relations are valid 
\begin{eqnarray*}
_{\pounds }\sigma &=&0\text{ \thinspace \thinspace ,\thinspace \thinspace
\thinspace \thinspace \thinspace \thinspace }_{\pounds }\theta =0\text{
,\thinspace \thinspace \thinspace \thinspace \thinspace \thinspace
\thinspace \thinspace \thinspace \thinspace }\pounds _u(d\omega )=0\text{
\thinspace \thinspace ,} \\
\sigma &=&\,_\nabla \sigma \text{ \thinspace \thinspace ,\thinspace
\thinspace \thinspace \thinspace \thinspace \thinspace \thinspace \thinspace
\thinspace }\theta =\,_\nabla \theta \text{ \thinspace \thinspace .}
\end{eqnarray*}
\end{proposition}

\section{Conclusion}

In this paper the notion of relative velocity and its kinematic
characteristics are introduced and considered. On an analogous basis, the
notion of friction velocity and its kinematic characteristics in a
continuous media are also introduced. The deformation and friction velocity
tensors are found. Special types of flows show that some notions of
classical continuous mechanics and hydrodynamics could be generalized
without difficulties for continuous media mechanics and hydrodynamics in $(%
\overline{L}_n,g)$-spaces.

\end{document}